\documentclass[article,twocolumn,twocolappendix]{aastex6}

  \usepackage{amsmath}
  \usepackage{tabularx}
  \usepackage{color}
  \usepackage{graphicx}
  \usepackage{rotating}

\begin{document}


\title{On the redshift of TeV BL Lac objects \\
                    }



\author{Simona Paiano\altaffilmark{1,2},  Marco Landoni\altaffilmark{2}, Renato Falomo\altaffilmark{1}, Aldo Treves\altaffilmark{3}, Riccardo Scarpa\altaffilmark{4}, Chiara Righi\altaffilmark{2,3}}

\email{simona.paiano@oapd.inaf.it}




\altaffiltext{1}{INAF, Osservatorio Astronomico di Padova, Vicolo dell'Osservatorio 5 I-35122 Padova (PD) - ITALY}
\altaffiltext{2}{INAF, Osservatorio Astronomico di Brera, Via E. Bianchi 46 I-23807 Merate (LC) - ITALY}
\altaffiltext{3}{Universit\`a degli Studi dell'Insubria, Via Valleggio 11 I-22100 Como - ITALY}
\altaffiltext{4}{Instituto de Astrofisica de Canarias, C/O Via Lactea, s/n E38205 - La Laguna (Tenerife) - ESPANA}

\begin{abstract}

We report results of a spectroscopic campaign carried out at the 10 m
Gran Telescopio Canarias for a sample of 22 BL Lac objects detected
(or candidates) at TeV energies, aimed to determine or constrain their redshift. 
This is of fundamental importance for the interpretation of their emission models, for population studies and also mandatory to study the interaction of high energy photons with the extragalactic background light using TeV sources.
High signal-to-noise optical spectra in the range 4250~-~ 10000~$\textrm{\AA}$ were obtained to search for
faint emission and/or absorption lines from both the host galaxy or the
nucleus.  
We determine a new redshift for PKS~1424+240 (z~$=$~0.604) and a tentative one for 1ES~0033+595 (z~$=$~0.467).  
We are able to set new spectroscopic redshift lower limits for other three sources on the basis of Mg II and Ca II intervening
absorption features: BZB~J1243+3627 (z~$>$~0.483), BZB~J1540+8155 (z~$>$~0.672), and BZB~0J2323+4210 (z~$>$~0.267).  
We confirm previous redshift estimates for four blazars: S3~0218+357 (z~$=$~0.944), 1ES~1215+303 (z~$=$~0.129), W~Comae (z~$=$~0.102), and MS~1221.8+2452 (z~$=$~0.218).  
For the remaining targets, in seven cases (S2~0109+22, 3C~66A, VER~J0521+211, S4~0954+65, BZB~J1120+4214, S3~1227+25, BZB~J2323+4210), we do not validate the proposed redshift.  
Finally for all sources of still unknown redshift, we set a lower limit based on the minimum  equivalent width of absorption features
expected from the host galaxy.

\end{abstract}



\section{Introduction} \label{sec:intro}

Blazars are luminous emitters over the whole electromagnetic spectrum
up to TeV energies. They are highly variable and polarized and are
often dominated, especially during outbursts, by the gamma-ray
emission.  The standard paradigm for these sources is that they owe
their extreme physical behavior to the presence of a relativistic jet
closely aligned with the observer’s direction, a model that
explains most of the peculiar properties of these sources: superluminal
motion, rapid variability, huge radio brightness temperature,
etc. From the optical point of view, blazars showing very weak lines or
a completely featureless spectra are named BL Lac objects (BLLs) \citep[see e.g. the review of][]{falomo2014}.

Compared to other AGN, the featureless spectrum of BLLs is due to
the extreme dominance of the non-thermal
emission over the stellar emission of the host galaxy, 
which make the assessment of their redshift very difficult.
\citep{sbarufatti2005b, sbarufatti2006a, sbarufatti2006b,
  sbarufatti2008, landoni2012, shaw2013a, landoni2013, landoni2014,
  massaro2014, massaro2015a, landoni2015, crespo2016}.  

The knowledge of the distance is, however, crucial to understand the nature of these
sources, the physical mechanism responsible for their extremely
energetic emission, their intrinsic luminosity, and cosmic evolution.
Furthermore, in the case of TeV BLLs, the simple knowledge of the redshift
converts these sources into powerfull probe of the Extragalactic
Background Light (EBL) through $\gamma-\gamma$ absorption, also 
improving our understanding of supersymmetric particles thought to be produced in their
ultra-relativistic jet \citep[see
  e.g.][]{tavecchio2015}. Thus, 
  we undertook a spectroscopic observational
campaign of a sample of TeV (or TeV candidate) BLLs with unknown or
uncertain redshift to be observed at the 10.4m Gran Telescopio
CANARIAS (GTC), in order to improve our knowledge of the redshift 
of  TeV BLLs, a possibly unique test-bench for ultra-high energy
fundamental physics.

The first results of this program were presented in
\citet{landoni2015} for S4 0954+65 and \citet{paiano2016} for S2
0109+22.  In this paper we report results for 22 additional BLLs: 15 of them
detected at TeV energies, while 7 being good TeV candidates
\citep{massaro2013}.

In Section \ref{sec:sample} we outline the selection criteria of our
sample and discuss their main properties.  In Section \ref{sec:data}
we present the data collection and the reduction procedure.  In
Section \ref{sec:results} we show the optical spectra of each object,
underlying their main features, and discuss their redshift.  In
Section \ref{sec:notes} we give detailed notes on individual objects
and finally in Section \ref{sec:discu} we summarize and discuss the
results.

In this work we assume the following cosmological parameters: H$_0=$
70 km s$^{-1}$ Mpc$^{-1}$, $\Omega_{\Lambda}$=0.7, and
$\Omega_{m}$=0.3.

\section{The sample} \label{sec:sample}

We selected all BLLs that are detected at Very High Energy band (VHE;
$>$100 GeV) from the online reference catalog of TeV sources
(TeVCAT\footnote{http://tevcat2.uchicago.edu/}) with unknown or
uncertain redshift and that are observable from La Palma ($\delta>$
-20$^\circ$).  For objects with uncertain redshift we choose sources
with contrasting redshift values reported in literature or with
measurements from optical spectra of low signal-to-noise ratio.  This
selection yields 18 targets and we obtain observations for 15 of them
(see Tab.\ref{tab:table1}) that represent about 70\% of the whole
sample of TeV blazars with uncertain or unknown redshift.

In addition we selected BLLs from a sample of 41
objects\footnote{half of them have unknown or uncertain redshift}
proposed as TeV emitters by \citet{massaro2013} on the basis of the
combined IR and X-ray properties of BLLs reported in the ROMA-BZCAT
catalog \citep{bzcat2009}, satisfying the criteria of uncertain
redshift and observability.  This selection produced 12 TeV candidates
and we obtained spectra for 7 of them (see Tab.\ref{tab:table1}) that
represent $\sim$60\% of the unknown/uncertain TeV candidate emitters
proposed by \citet{massaro2013}.
1424+240

\section{Observations and data reduction} \label{sec:data}

Observations were obtained between February 2015 and August 2016 in
Service Mode at the GTC using the low resolution spectrograph OSIRIS
\citep{cepa2003}. The instrument was configured with the two grisms
R1000B and
R1000R\footnote{http://www.gtc.iac.es/instruments/osiris/osiris.php},
in order to cover the spectral range 4000-10000 $\textrm{\AA}$, and
with a slit width~$=$~1'' yielding a spectral resolution
$\lambda$/$\Delta\lambda$~$=$~800.

For each grism three individual exposures were obtained (with exposure time
ranging from 300 to 1200 seconds each, depending on the source magnitude),
that were then combined into a single average image, in order to
perform optimal cleaning of cosmic rays and CCD cosmetic defects. 
Detailed information on the
observations are given in Tab.\ref{tab:table2}.

Data reduction was carried out using IRAF\footnote{IRAF (Image
Reduction and Analysis Facility) is distributed by the National
Optical Astronomy Observatories, which are operated by the
Association of Universities for Research in Astronomy, Inc., under
cooperative agreement with the National Science Foundation.} and
adopting the standard procedures for long slit spectroscopy with bias
subtraction, flat fielding, and bad pixel correction.
Individual spectra were cleaned of cosmic-ray contamination using the L.A.Cosmic algorithm \citep{lacos}.

Wavelength calibration was performed using the spectra of Hg, Ar,
Ne, and Xe lamps and providing an accuracy of 0.1~$\textrm{\AA}$ over
the whole spectral range.  Spectra were corrected for atmospheric
extinction using the mean La Palma site extinction
table\footnote{https://www.ing.iac.es/Astronomy/observing/manuals/}.
Relative flux calibration was obtained from the observations of
spectro-photometric standard stars secured during the same nights of
the target observation.  For each object the spectra obtained with the
two grisms were merged into a final spectrum covering the whole
desired spectral range.

Thanks to the availability of a direct image of the target, which is obtained at GTC as
part of  target acquisition, the spectra could be flux calibrated.
The calibration  was assessed using the zero point
provided by the GTC-OSIRIS
webpage\footnote{http://www.gtc.iac.es/instruments/osiris/media/zeropoints.html}.
For half of our sample it was also possible to use stars with known flux from the SDSS survey to
double check the flux calibration. We found no
significant difference on average between the two methods within  $\sim$~0.1
mag.
The final spectra were then calibrated to have the flux at 6231~$\textrm{\AA}$ equal to the photometry found for the targets (see
Tab. \ref{tab:table2}).
Finally each spectrum has been dereddened, applying the extinction law
described in \citet{cardelli1989} and assuming the E(B-V) values taken
from the NASA/IPAC Infrared Science
Archive \footnote{https://irsa.ipac.caltech.edu/applications/DUST/}.

\section{Results} \label{sec:results}

The optical spectra of the targets are presented in Fig. \ref{fig:spectra}.
In order to emphasize weak emission and/or absorption features, we
show also the normalized spectrum. This was obtained by dividing the
observed calibrated spectrum by a power law continuum fit of the
spectrum, excluding the telluric absorption bands (see
Tab. \ref{tab:table3}).
These normalized spectra were used to evaluate the SNR in a number of
spectral regions.  On average the SNR ranges from 150-200 at 4500~$\textrm{\AA}$ and 8000~$\textrm{\AA}$ respectively, to a peak of 320
at 6200~$\textrm{\AA}$. See details in Tab. \ref{tab:table3} and all
these spectra can be accessed at the website
http://www.oapd.inaf.it/zbllac/.

\subsection{Search for emission/absorption features}

All spectra were carefully inspected to find emission and absorption
features.  When a possible feature was identified, we determined its
reliability checking that it was present on the three individual
exposures (see Sec. \ref{sec:data} for details).  We were able to
detect spectral lines for 9 targets.  In particular we observe [O~III]
5007~$\textrm{\AA}$ weak emission in the spectra of 1ES~1215+303, W~Comae, MS~1221.8+2452 and PKS~1424+240, [O~II] 3727~$\textrm{\AA}$ in
1ES~0033+595, 1ES~1215+303 and PKS~1424+240, the Ca~II 3934,3968~$\textrm{\AA}$ doublet absorption system and the G-band 4305~$\textrm{\AA}$ absorption line in MS~1221.8+2452, a strong emission of
Mg~II 2800~$\textrm{\AA}$ in S3~0218+357 and intervening absorption
systems due to Mg~II 2800 $\textrm{\AA}$ in BZB~J1243+3627 and BZB~J1540+8155 and, finally the Ca~II 3934,3968~$\textrm{\AA}$ doublet in
the spectrum of BZB~J2323+4210.  Details in Fig. \ref{fig:spectraCU}
and Tab. \ref{tab:line}. 
The spectrum of 7 additional targets is found completely featureless
even though a redshift is
reported in literature.  Details about the optical spectra and
redshift estimates for each objects of our sample are given in
Sec. \ref{individual}.

\subsection{Redshift lower limits}

Based on the assumption that all BLLs are hosted by a massive
elliptical galaxy \citep[e.g.][]{falomo2014} one should be able to
detect faint absorption features from the starlight provided that the
SNR and the spectral resolution are sufficiently high.  In the case of
no detection of spectral features it is possible to set a lower limit
to the redshift based on the minimum Equivalent Width (EW) that can be
measured in the spectrum.

The minimum measurable equivalent eidth (EW$_{min}$) was set
according to the scheme outlined by \citet{sbarufatti2006b,
  sbarufatti2006a}, though in a more elaborated procedure (see
Appendix A).  In brief from the normalized spectrum (see
Fig. \ref{fig:spectra}) we computed the nominal EW adopting a running
window of 15~$\textrm{\AA}$ for five intervals of the spectra that
avoid the prominent telluric absorption features (see
Tab. \ref{tab:table3}).
The procedure yields for each given interval a distribution of EW and
we took as minimum measurable EW three times the standard deviation of
the distribution (see details in Appendix A).

Five different intervals were considered because the SNR changes with
wavelength.  The range of EW$_{min}$ is reported in
Tab. \ref{tab:table3} and we give a lower limit on z assuming a
standard average luminosity for the host galaxy M$_{R}$~$=$~-22.9 (or
M$_{R}$~$=$~-21.9 in parenthesis).

\section{Notes for individual sources } \label{sec:notes}
\label{individual}

\begin{itemize}
\item[] \textbf{BZB J0035+1515}:
The source was first discovered by \citet{fischer1998} and catalogued
as BLL on the basis of its featureless optical spectrum.
A more recent optical spectrum, obtained as part of the SDSS survey, exhibits
no features (although the automatic procedure suggests some tentative
values, also included in NED).
Also \citet{shaw2013a} found a featureless spectrum.

We confirm the featureless nature of the spectrum from 4200 to 9000
$\textrm{\AA}$ and from our high SNR we obtain an EW$_{min}$ of 0.09 -
0.18 $\textrm{\AA}$, which correspond to a redshift lower limit
of z~$>$~0.55

\item[] \textbf{1ES 0033+595}: 
  \citet{perlman1996} identified this
  Einstein Slew Survey source as a BLL finding a featurless optical
  spectrum (although a tentative redshift z~$=$~0.086 was derived by
  Perlman et al. as mentioned in \citet{falomo1999}).  In
  \citet{scarpa1999} the HST images of this object shows two
  unresolved sources, ``A'' and ``B'', separated by 1''.58 and with
  magnitude $m_{R}$=17.95$\pm$0.05 and 18.30$\pm$0.05 mag
  respectively.  On the basis of radio coordinates the authors
  identified the source B as the most probable BLL counterpart and A as a possible
  star.

In our spectrum the two sources are partially resolved (see
Fig. \ref{fig:0033image}) and we perform a de-blending during the
extraction process in order to obtain two separated spectra for the
targets.  The spectrum of the object ``A'' (Fig. \ref{fig:Gstar}) shows the typical 
stellar absorption lines of G stars, confirming the previous classification. 
For the ``B'' source we obtain a spectrum with a SNR$\sim$100 (see
Fig. \ref{fig:spectra}) and although there is some contamination of
the spectrum by the companion, the non detection of H$_{\alpha}$
indicates that the ``B'' object has an extragalactic nature and it is
the blazar counterpart as proposed by \citet{scarpa1999}.  We found an emission
feature at 5468 $\textrm{\AA}$ of EW~$=$~0.4 $\textrm{\AA}$ (see
Fig. \ref{fig:spectraCU}). This feature is detected in all three individual
spectra and therefore we consider it a secure detection. If identified as [OII] 3727$\textrm{\AA}$ emission, a tentative redshift
of z~$=$~0.467 can be provided.

Finally, comparing our photometry with \citet{scarpa1999}, we obtain
the same value for the A object, while for the object B we obtain a
magnitude difference of 1.2 with respect to the previous one reinforcing 
the classification of this source as a BLL.

\begin{figure}[htbp]  
   \includegraphics[width=0.5\textwidth]{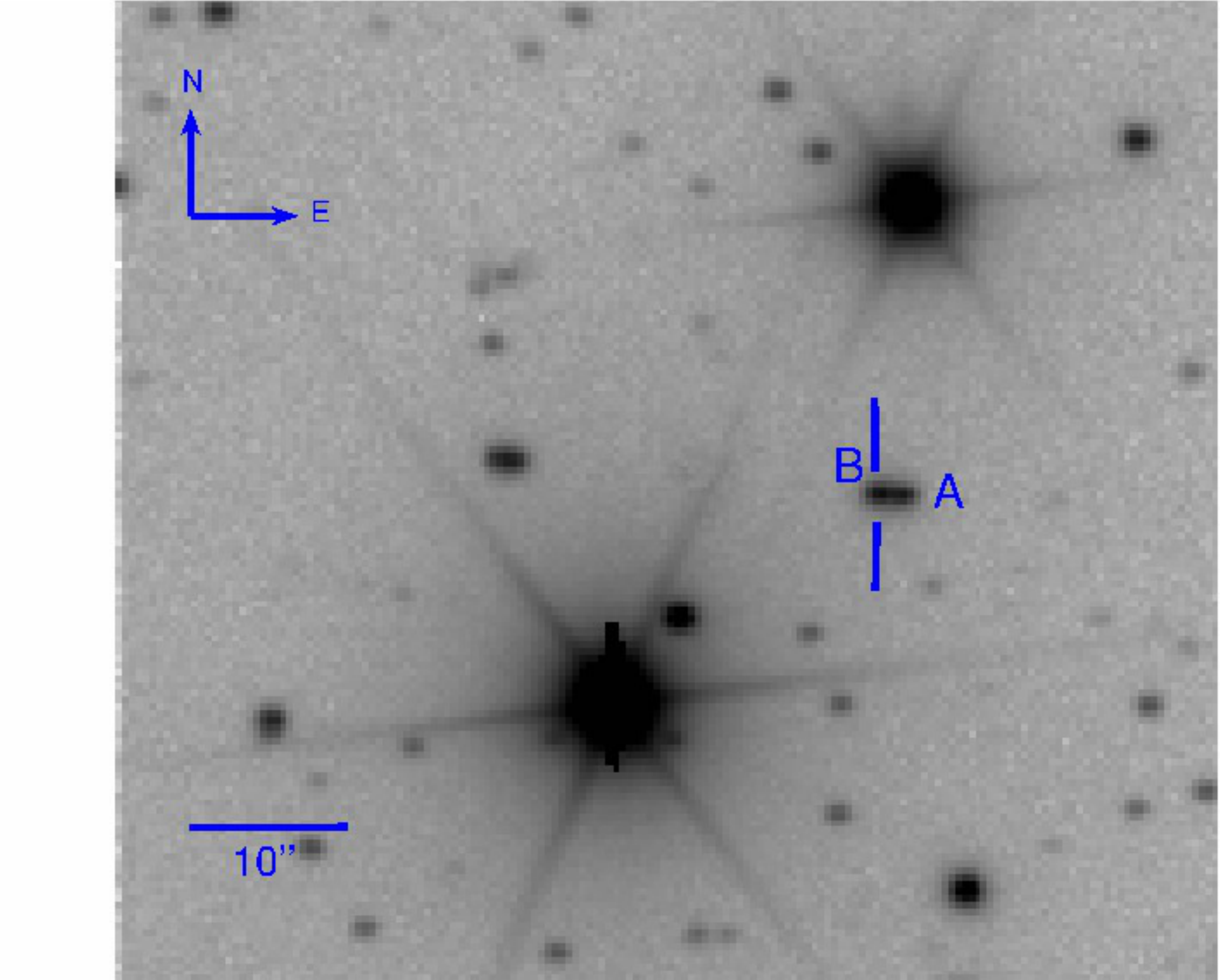}
   \caption{r-band optical image of the sky region around the BL Lac
     object 1ES 0033+595 obtained at the GTC. The source flagged as
     ``A'' is a foreground star and the BLL is the source labelled as
     ``B''. }
    \label{fig:0033image}
\end{figure}

\begin{figure}[htbp]  
   \includegraphics[width=0.4\textwidth, angle=-90]{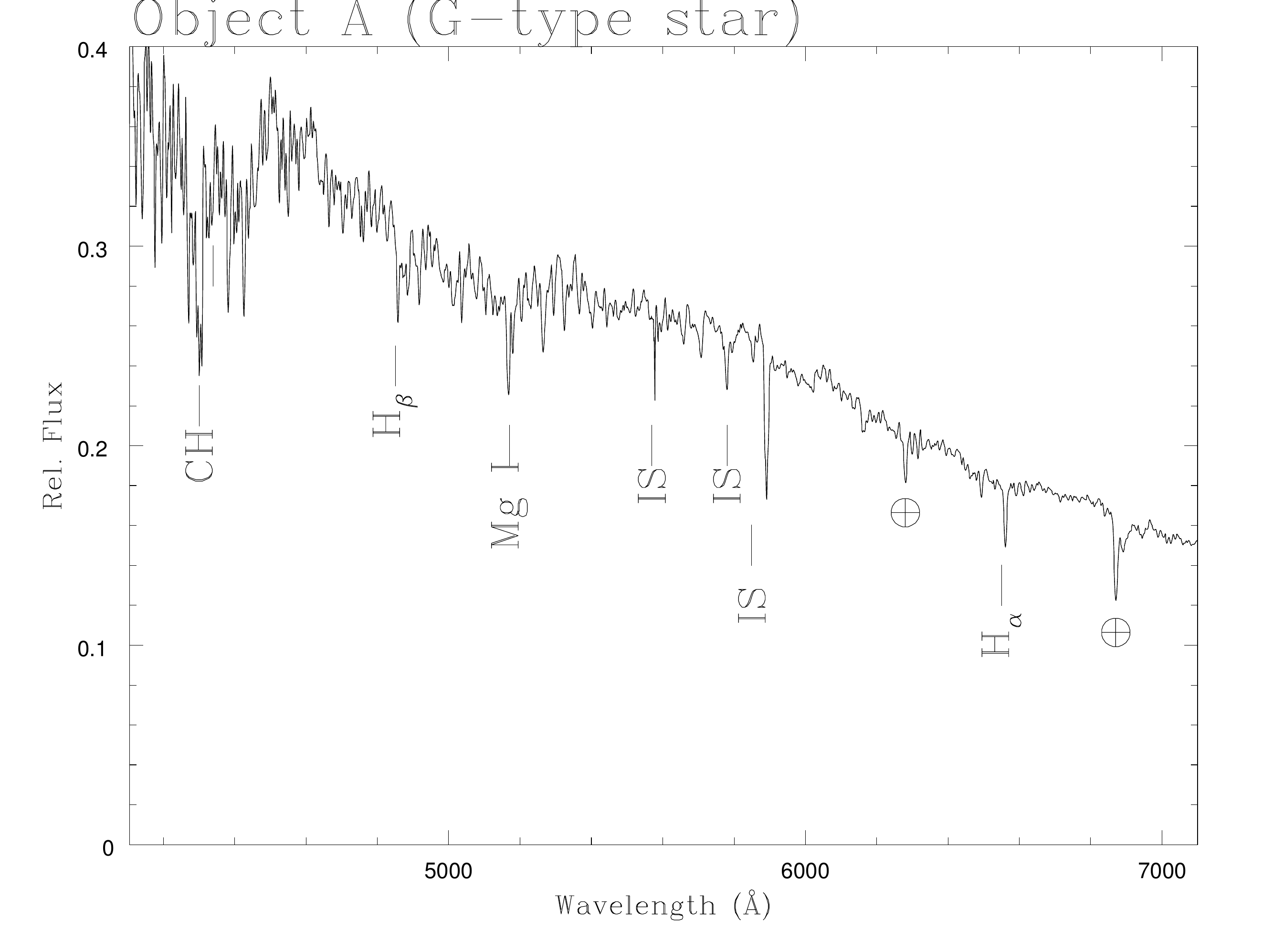}
   \caption{GTC spectrum of the companion source, labelled as ``A''
     (see Fig.\ref{fig:0033image}, of the BL Lac object 1ES
     0033+595. Absorption lines due to CH (4299 $\textrm{\AA}$),
     hydrogen (4342 $\textrm{\AA}$, 4863 $\textrm{\AA}$, 6565
     $\textrm{\AA}$), and Mg I (5176 $\textrm{\AA}$) are clearly
     detected. Telluric bands are indicated by $\oplus$. This object
     can be classified as a G-type star. }
   \label{fig:Gstar}
\end{figure}

\item[] \textbf{RGB J0136+391}:
The first identification of this source as BLL was proposed by
\citet{laurent1998} showing a featureless optical spectrum.  The same
result was found in \citet{wei1999, piranomonte2007} and
\citet{shaw2013a}.  A lower limit on the redshift of z~$>$~0.40 was
set on the basis of high quality i-band images obtained at the Nordic
Optical Telescope (NOT) (\citet{nilsson2012}).

We found our high SNR ($\sim$ 200-480) optical spectrum completely featureless, only allowing
to set a lower limit to the redshift of z~$>$~0.27.

\item[] \textbf{S3 0218+357}: 
This source was discovered to be a gravitational lens by
\citet{patnaik1993} who detected two similar radio sources with
$\sim$0.33'' separation and an Einstein ring with the same diameter.
Optical counterparts of the two radio-sources were observed and
detected by HST images \citep{jackson2000, york2005}.
An optical spectrum of the source was obtained by \citet{browne1993}
who detect absorption features of Ca II and Mg II attributed to the
lens galaxy at z~$=$~0.684. They also claim the detection of very weak
emission lines of [OII] 3727$\textrm{\AA}$ and [OIII]
5007$\textrm{\AA}$.  In addition they suggest the presence of weak
emission feature of Mg II 2800$\textrm{\AA}$ attributed to the blazar
proposing a redshift of 0.936.  The redshift of the lens galaxy was
confirmed through 21 cm HI absorption by \citet{carilli1993}.
\citet{cohen2003} obtained a high quality spectrum which confirms the
absorption features and clearly detect a strong broad emission line at
5470 $\textrm{\AA}$ that identified as Mg II 2800 $\textrm{\AA}$
yields a redshift of z~$=$~0.944 for the blazar.  In addition these
authors claimed the detection of emission lines of [OII], H$_{\beta}$
and [OIII] at z~$=$~0.684 attributed to the lens galaxy. Moreover they
claimed the detection of weak H$_{\beta}$ and [OIII] emission in the
red noisy spectrum, also attributed to the blazar at z~$=$~0.944

We obtain an optical spectrum ranging from 4500 to 10000
$\textrm{\AA}$ with a SNR in the range 25-50.  We confirm the
detection of Mg II and Ca II absorption lines at z~$=$~0.684, and in
addition we clearly detect an absorption line at 9920 $\textrm{\AA}$
identified as Na I 5892 $\textrm{\AA}$ at the redshift of the lens.
We do not detect the emissions line [OII], H$_{\beta}$ and [OIII]
\citep[claimed by][]{cohen2003}. We note that some of these latter
features occur inside the telluric absorptions of the O$_{2}$ and
H$_{2}$O.  We clearly detect the strong broad emission line at 5480
$\textrm{\AA}$ (EW=35 $\textrm{\AA}$, FWHM=4700 km/s) that if
attributed to Mg II 2800$\textrm{\AA}$, yields the redshift of
z~$=$~0.954.  We stress that in our spectrum we do not detect the
claimed emissions H$_{\beta}$ and [OIII] attributed to the blazar by
\citet{cohen2003}. We note again that these features are placed in a
spectral region that is heavily contaminated by strong H$_{2}$O
atmospheric absorption.  Therefore we conclude that the redshift of
the S3 0218+357 is still tentative since it is based on only one line. 
If confirmed, this source is the most distant
blazar detected at frequencies $>$100 GeV \citep{ahnen2016}.

It is worth to note that the shape of the continuum exhibits a marked
decline towards the blue region that is rather unusual for this type
of source. This could be due to significant intrinsic extinction or
caused by absorption in the lens.

\item[] \textbf{3C 66A}: 
\citet{wills1974} identified this strong radio source as a BLL because
of its featureless optical spectrum.  \citet{miller1978} proposed a
redshift of z~$=$~0.444, on the basis on one emission line attributed
to Mg II 2800 $\textrm{\AA}$. A value considered by the authors
as tentative and highly uncertain.
No other optical spectroscopy was done for thirty years. More recently \citet{finke2008} showed an optical 
spectrum in the range from 4200 to 8500 $\textrm{\AA}$ with no detectable optical features. 
The featureless spectrum was also confirmed by \citet{shaw2013a}.

Our high SNR ($\sim$ 200) GTC spectrum is also featureless.
Based on our procedure of redshift lower limits estimated by
\textit{EW$_{min}$}, due to the relatively bright source we can set
only a modest lower limit of z~$>$~0.10. We are not able to confirm
the Mg II emission proposed by \citet{miller1978} because it is out of our spectral range. However at z~$=$~0.444 we would expect to observe H$_{\beta}$ emission line at 7020 $\textrm{\AA}$, where we do not
detect any line with EW~$>$~0.2$\textrm{\AA}$. We conclude that the
redshift of this source is still unknown.

\item[] \textbf{VER J0521+211}: 
On the basis on a weak emission line at 5940$\textrm{\AA}$ attributed
to [N II] 6583$\textrm{\AA}$, \citet{shaw2013a} proposed this source to be at
z~$=$~0.108.  This feature was not confirmed by
\citet{archambault2013} that reports a featureless spectrum.

We do not confirm the redshift of the source, which therefore is still unknown, setting a lower limit of z~$>$~0.18.

\item[] \textbf{1ES 0647+250}: An optical spectrum of this source,
  with modest SNR was found featureless by \citet{schachter1993}, a result
  later confirmed by a better spectrum obtained with the Keck
  telescope by \citet{shaw2013a}.
A relatively high redshift can be supported by the absence of
detection of the host galaxy from high quality image by
\citet{kotilainen2011}.

Our GTC higher SNR ($\sim$ 200) spectrum confirms this featureless behaviour with absorptions 
at around 4400 $\textrm{\AA}$ and 4880 $\textrm{\AA}$ due to Diffusing Interstellar bands (DIBs) 
and at $\sim$ 6500  $\textrm{\AA}$ due to water vapor.
On the basis of our spectrum we set a lower limit of z~$>$~0.29.

\item[] \textbf{S5 0716+714}: This is a bright (V$\sim$15) and highly
  variable \citep{bach2007} source for which several attempts
  to detect the redshift failed (\citep{stickel1993, rector2001, finke2008, shaw2013a}).
From optical images \citet{sbarufatti2005a} set a lower limit of z~$>$~0.5 and \citet{nilsson2008} provided 
an imaging redshfit of z $\sim$ 0.3 based on the marginal detection of the host galaxy.
Finally we note that \citet{danforth2013}, based on the distribution of the 
absorption systems, set a statistical upper limit z $\lesssim$ 0.3 .

We obtained a featurless optical spectrum during a high state of the
source (r~$=$~13.6) and we can set a redshift lower limit of z~$>$~0.10.

\item[] \textbf{BZB J0915+2933}:
\citet{wills1976} showed a continuous optical spectrum for the source and classified it as a BLL.
The featureless behaviour was also found by \citet{white2000} and by \citet{shaw2013a}.

Through our high SNR optical spectrum, we confirm the featureless spectrum and set a lower limit to the redshift of z~$>$~0.13.

\item[] \textbf{BZB J1120+4212}:
This object (also known as RBS 0970) is a point-like radio source detected by various X-ray surveys \citep[see e.g.][]{giommi2005}. 
Optical spectral classification of the source as BLL was proposed by
\citet{perlman1996} on the basis of the quasi-featureless spectrum.
They claim the detection of starlight absorption features at
z~$=$~0.124. However, based on the spectrum reproduced in their Fig. 4,
the reliability of this features is quite uncertain.  This redshift is
not confirmed in other spectra obtained by \citet{white2000} and
\citet{massaro2014}.  Also the spectrum obtained by SDSS
(J112048.06+421212.4) appears to us featureless.

Our spectrum with SNR $\sim$ 100-190 is featureless and we set a lower limit of z~$>$~0.28.

\item[] \textbf{1ES 1215+303}:
\citet{bade1998} reported a redshift z~$=$~0.130 for this target, but
no information about the detected lines are given.  On the contrary
\citet{white2000} showed an optical spectrum claiming a redshift of
0.237, although it appears featurless from their figure.

A more recent spectrum (SNR~$=$~60) obtained by \citet{ricci2015} was also found featurless.
The target was clearly resolved in HST exposures (\citep{scarpa2000}) revealing a massive elliptical host galaxy,
suggesting the source is at low redshift.

Given these different redshift values, we secured a high quality
optical spectrum (SNR $\sim$ 300) in which we detect two emission lines:
[OII] 3727 $\textrm{\AA}$, [OIII] 5007 $\textrm{\AA}$ at z~$=$~0.131
(see also Table \ref{tab:line}) confirming the low redshift previously reported.

\item[] \textbf{W Comae}: 
\citet{weistrop1985} provided an optical
spectrum and estimated a redshift of z~$=$~0.102 based on the
detection of [OIII] 5007$\textrm{\AA}$ and H$_{\alpha}$ emission
lines.
This redshift was not confirmed by \citet{finke2008}, though their spectra cover only the range from 3800 to 5000 $\textrm{\AA}$.
In addition the spectrum obtained by the SDSS (J122131.69+281358.4) proposes a redshift of z~$=$~1.26.
In 2003 the host galaxy of W Comae was resolved by \citet{nilsson2003}. 

From our (SNR$\sim$220) optical spectrum we confirm the detection of
[O III] 5007$\textrm{\AA}$ and H$_{\alpha}$ emission lines at
z~$=$~0.102. In addition we detect at the same redshift the absorption
lines due to Ca II (3934, 3968 $\textrm{\AA}$) doublet, G-band 4305
$\textrm{\AA}$, and Mg I 5175 $\textrm{\AA}$ from the host galaxy.

\item[] \textbf{MS 1221.8+2452}:
A tentative redshift of z~$=$~0.218 was proposed by \citet{morris1991}
and \citet{rector2000}.  Imaging studies of this source were able to
resolve the host galaxy and are consistent with the low redshift of
the target \citep{falomo1999, scarpa2000}.

We detect the Ca II doublet and G-band 4305~$\textrm{\AA}$ absorption
lines at z= 0.218 and we find emission lines at $\sim$~7995 and $\sim$~8020~$AA$ that if confirmed could be attribuited to H$_{\alpha}$ and N~II~ 6583~$\textrm{\AA}$.

\item[] \textbf{S3 1227+255}:
\citet{nass1996} reported z~$=$~0.135 but no information on the detected spectral lines were provided. 
In spite of the alleged low redshift, high quality images failed to detect the host galaxy \citet{nilsson2003}. 
\citet{shaw2013a} did not confirm this redshift and no spectral features were found.

Our optical spectrum (SNR $\sim$ 250) is featureless down to EW~$=$~0.1-0.2 $\textrm{\AA}$. 
Therefore we do not confirm the literature redshift and we set a redshift lower limit of z~$>$~0.10.

\item[] \textbf{BZB J1243+3627}:
\citet{white2000} reported a featureless spectrum for this source.
An absorption feature of Mg II 2800$\textrm{\AA}$ at $\lambda \sim$
4160 $\textrm{\AA}$ was detected in the SDSS spectrum suggesting a
redshift of z $\ge$ 0.485 \citet{plotkin2010}.  This redshift limit
appears consistent with the marginal detection of the host galaxy by
\citet{meisner2010} who estimated z $\sim$ 0.50.

From our spectrum (SNR $\sim$ 330), we confirm the intervening
absorption system due to Mg II 2800$\textrm{\AA}$ and the remaining
part of the spectrum is completely featureless. The spectroscopic
redshift lower limit is thus z~$>$~0.483.

\item[] \textbf{BZB J1248+5820}:
The source was classified as a BLL by \citet{fleming1993} and no redshift was available.  
The featureless nature of the spectrum is reported in
\citep{henstock1997, plotkin2008} and \citet{shaw2013a}.  Note that
NED report z~$=$~0.847 based on the SDSS DR3 spectrum, although this
is not confirmed by SDSS DR13 analysis.
\citet{scarpa2000} failed to detect the host galaxy from HST images.

Our high SNR spectrum is featureless and we can determine a lower limit to the redshift of z~$>$~0.14.

\item[] \textbf{PKS 1424+240}:
The source was classified as BLL by \citet{fleming1993} and
a featureless spectra was reported by \citet{marcha1996, white2000} and \citet{shaw2013a}.
\citet{furniss2013}, from the Ly$_\beta$ and Ly$_\gamma$ absorptions
observed in the far-UV spectra from HST/COS (\textit{Hubble Space Telescope}/Cosmic Origins Spectrograph) spectra, reported a lower
limit z$>$0.6035.
This is consistent with the non detection of the host galaxy in HST images \citet{scarpa2000}.

In our high SNR $\sim$ 350 optical spectrum, we detect two
faint emission lines at 5981 and 8034 $\textrm{\AA}$ (see
Fig.\ref{fig:spectraCU}) due to [O II] 3727$\textrm{\AA}$ and [O
  III] 5007$\textrm{\AA}$.  The redshift corresponding to this identification is
0.6047, suggesting that the absorber at that redshift limit is
associated to the BLL.  Note also that in the
environment of the target there is a group of galaxies at z $\sim$
0.60 suggesting it is associated to the BLL \citep{rovero2016}.

\item[] \textbf{BZB J1540+8155}:
The source was identified as BLL by \citet{schachter1993}. The optical
spectra obtained by \citet{perlman1996} failed to detect emission or
absorption features.  The host galaxy was not detected by HST images
\citep{scarpa2000} posing the source at relatively high redshift.

In our GTC spectrum we detect an intervening absorption doublet at $\sim$~4680~$\textrm{\AA}$ that we identify as MgII 2800 $\textrm{\AA}$
absorption yielding a spectroscopic redshift lower limit of z $>$
0.673.  No intrinsic emission or absorption lines are found.  The
spectroscopic redshift limit is consistent with our redshift limits
determined by the absence of detection of host galaxy features.

\item[] \textbf{RGB J2243+203}:
\citet{laurent1998} presents the first optical spectrum of this source, found
featureless. Similarly the spectrum obtained by \citet{shaw2013a} is
featureless but the authors claimed the detection of an absorption
line at $\sim$ 3900 $\textrm{\AA}$ identified as Mg II
(2800$\textrm{\AA}$). If confirmed this would imply a redshift z~$>$~0.395.

Our spectrum, that does not cover the 3900$\textrm{\AA}$ region, is very
featureless from 4100$\textrm{\AA}$ to 9000$\textrm{\AA}$ with the
limits on the emission or absorption lines of EW$_{min}$ $<$ 0.2. This
corresponds to a lower limit of z~$>$~0.22.

\item[] \textbf{BZB J2323+4210}: 
From a poor optical spectrum \citet{perlman1996} claimed the detection
of two starlight absorption features identified as Mg I
(5175$\textrm{\AA}$)and Na I (5892 $\textrm{\AA}$) and they proposed
at redshift z~$=$~0.059.  We disprove this redshift because the Na I
absorption coincides with the telluric absorption at 6280
$\textrm{\AA}$.  \citet{shaw2013a} does not confirm the above
redshift either.

Our high SNR ($\sim$200) spectrum is characterized by a power law
emission ( $F_{\lambda}\varpropto\lambda^{\alpha}$; $\alpha$~$=$~-
1.2). We clearly detect an absorption doublet at $\sim$5000
$\textrm{\AA}$ (EW $\sim$ 0.25 $\textrm{\AA}$) and an absorption line
at 7465 $\textrm{\AA}$. We identify these features as Ca II 3934, 3968
$\textrm{\AA}$, and Na I 5892$\textrm{\AA}$ absorption lines at
z~$=$~0.267.  If these lines were ascribed to the starlight of the
host galaxy we would expect to observe some modulation imprinted on the continuum,
which are not present. 
Moreover this redshift appears remarkable
inconsistent with the lower limits (z$_{lim}$~$>$~0.65) derived from
the non detection of the host galaxy.

We further note these absorption features are rather narrow (FWHM
$\sim$ 10 $\textrm{\AA}$) compared to typical Ca II line width from
galaxies and are indicative of interstellar absorptions.  Indeed at
$\sim$ 8.5 and 12 '' (SE) from the target (see Fig.\ref{fig:2323note})
there are two spiral galaxies with halo gas which could be responsible
of the absorption features observed.  At z~$=$~0.267 the projected
separation between the target and these galaxies is $\sim$ 40 kpc.

We conclude that the redshift of BZB~J2323+4210 is still unknown and we set a 
spectroscopic lower limit of z~$>$~0.267 and a lower limit based on the host galaxy feature of z~$>$~0.65.

The case has some analogy with that of the BLL MH~2136-428
\citep{landoni2014}, where narrow absorption lines appear in the
spectrum due to the halo of an interloping bright galaxy.

\begin{figure}[htbp]  
   \includegraphics[width=0.47\textwidth]{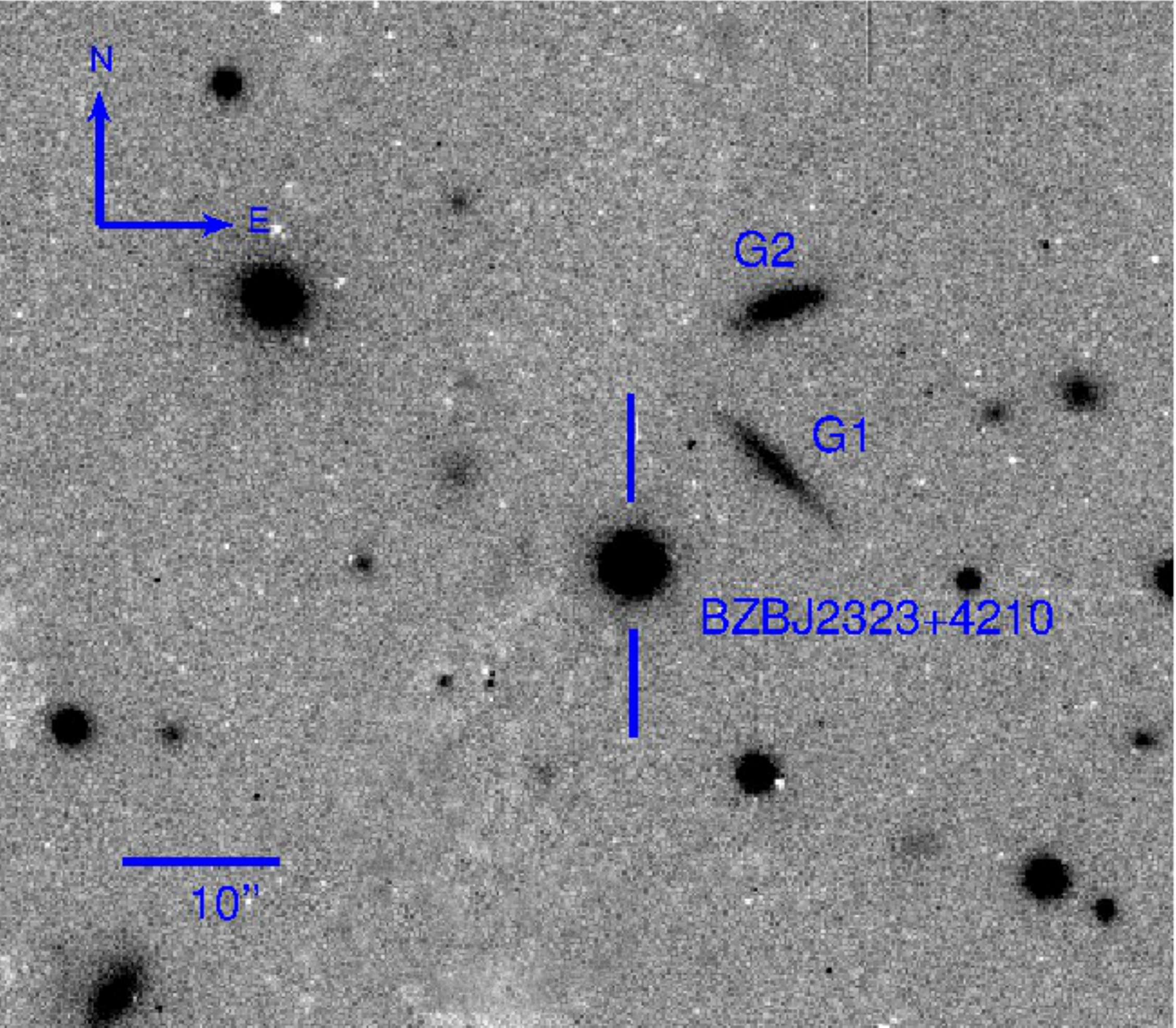}
   \caption{Optical R-image of the BL Lac object BZB~J2323+4210 taken at the NOT telescope \citep{falomo1999}. Two spiral galaxies, labelled as G1 and G2, are present in the field of view of the BL Lac object at a distance of $\sim$8.5'' and $\sim$12.0'' respectively.} 
   \label{fig:2323note}
\end{figure}

\end{itemize}

\newpage
\section{Discussion } \label{sec:discu}

We secured high SNR spectra in the range 4200-9500~$\textrm{\AA}$ for
a sample of 22 BLLs, selected for being TeV emitters or good
candidates based on the their IR properties.  Most of these sources
either had unknown redshift or the value was rather uncertain.  From
the new spectroscopy we are able to determine the redshift for 5
objects (S3~0218+357, 1ES~1215+303, W~Comae, MS~1221.8+2452, and PKS~1424+240).  For PKS~1424+240, one of the farthest BLL detected in the
TeV regime, no previous estimate of the redshift was available.  For
three objects, BZB~J1243+3627 with an uncertain redshift and BZB~J1540+8155, and BZB~J2323+4210 with previously unknown redshift, we
found intervening absorptions that allow us to set spectroscopic lower
limits.  For the remaining 13 sources we found that in spite of the
high SNR their spectrum is featureless. We can set limits to any
emission or absorption features to 0.05-0.50~$\textrm{\AA}$ depending
on the SNR of the targets and the wavelength region.  For seven of
these targets there was a previous tentative redshift that we do not
confirm from our observations.  The main reasons for this difference
are: old spectra have poor SNR, wrong source identification, very
tentative line identifications and redshift given without information
on the detected spectral features (no spectrum shown).  It is worth to
note that unfortunately these unconfirmed, likely wrong values (also
appearing in NED) are often used to derive physical properties of the
sources.

On the basis of the assumption that all the objects with pure
featureless spectra are hosted by a massive elliptical galaxy as is
the case of most (virtually all) BLLs, we have then determined a lower
limit of their redshift from the minimum detectable EW of some
absorption lines of the host galaxy (see Appendix for
details). Depending on the brightness of the observed nuclei
(r~$=$~13.6 to 19.9 ) we can set redshift lower limits for these
objects from z~$=$~0.1 to z~$=$~0.55 (see Tab. \ref{tab:table3}).

In addition to the lower limits of the redshift for objects with
featureless spectra we can also estimate an average upper limit to the
redshift of the sample of observed targets based on the number of Mg
II 2800 $\textrm{\AA}$ intervening absorption systems observed in our
spectra.  Given our observed spectral range we are potentially able to
detect Mg II 2800 $\textrm{\AA}$ intervening absorption lines (of EW
$\gtrsim$ 0.2 $\textrm{\AA}$ ) for absorbers that are at z~$>$~0.5.
Excluding the objects that are at z $<$ 0.5 (4 sources) we observe 2
absorption systems of Mg II over 18 targets.  To evaluate an average
upper limit to the redshift of these sources we compute the expected
number of Mg II intervening absorptions as a function of the
redshift. To this aim we assume the cumulative incidence rates of Mg
II absorbers derived for a very extended sample of QSO spectra
obtained by SDSS \citep{zhu2013}.  It turns out that the average
maximum redshift of sample is inferred to be z $\sim$ 0.65. At higher
redshift we would expect to detect many more absorption systems. For
instance if the average redshift of these sources were z~$=$~1 we
should detect $\sim$ 10 Mg II absorption systems in the spectra of 18 targets.

The relatively low upper limit of the redshift derived above together
with the lack of detection of absorption lines from the host galaxies
suggests that these targets have a high Nucleus-to-Host galaxy ratio
(N/H).  For each object with a featureless continuum we have derived a
lower limit to the redshift on the basis of the assumption that the
source is hosted by a massive early type galaxy (see Appendix for
details) and at a given limit of detectable EW of an absorption
feature. We can now associate a minimum N/H to these redshift lower
limits (see Tab. \ref{tab:table3}).  It turns out that some objects in
our sample have a N/H~$>$~10 (assuming the targets are hosted by a
“standard galaxy” \citep[e.g.][]{falomo2014}.
This is significantly higher than the typical value (N/H $\sim$ 1 )
for BLLs for which the host galaxy is directly imaged by HST, and it
is similar to that estimated for unresolved sources at relatively high
(z~$=$~0.5 - 1.0) redshift \citep[see e.g.][]{urry2000}.  Since on
average our targets are likely at moderate redshift (see above) the
high N/H is suggestive of a particularly beamed nuclear emission.
How strong could be the flux from the nucleus compared with that of
its host galaxy ?  For a Doppler factor $\delta$ extremely high, we
could have N/H $\sim$ 1000. The detection of the spectral features of
the host galaxy therefore will require observations with very high SNR
and adequate spectral resolution.  This appears feasible only in the
ELT era \citep[see e.g.][]{landoni2014}.


\begin{table*}
\caption{THE SAMPLE OF TEV BLLAC AND TEV-CANDIDATES}\label{tab:table1}
\centering
\begin{tabular}{lcclccll}
\hline 
Object name     & RA          & $\delta$   &  CLASS     &        V      &  $E(B-V)$     &       z$_{literature}$       & Reference \\
                & (J2000)     & (J2000)    &            &               &               &               &  \\
\hline
BZB J0035+1515	& 00:35:14.70 & 15:15:04.0 &  TeVc	&	16.9	&	0.062	&       ?       &       \\
1ES 0033+595	& 00:35:52.60 & 59:50:05.0 &  HBL	&	19.5	&	1.386	&	? 	&	\\
S2 0109+22*	& 01:12:05.08 & 22:44:39.0 &  IBL	&	15.7	&	0.034	&	0.265 ?	&	\citet{healey2008} \\
RGB J0136+391	& 01:36:32.50 & 39:06:00.0 &  HBL	&	15.8	&	0.068	&	?	&	\\
S3 0218+357	& 02:21:05.50 & 35:56:14.0 &  HBL**	&	20.0	&	0.061	&	0.944	&	\citet{cohen2003}\\
3C 66A		& 02:22:39.60 & 43:02:08.0 &  IBL	&	15.0	&	0.075	&	0.444 ?	&	\citet{miller1978}\\
VER J0521+211	& 05:21:45.90 & 21:12:51.0 &  IBL	&	17.5	&	0.604	&	0.108 ?	&	\citet{shaw2013a}\\
1ES 0647+250	& 06:50:46.50 & 25:03:00.0 &  HBL	&	15.7	&	0.087	&	?	&	\\
S5 0716+714	& 07:21:53.40 & 71:20:36.0 &  IBL	&	15.5	&	0.027	&	?	&	\\
BZB J0915+2933	& 09:15:52.40 & 29:33:24.0 &  TeVc	&	15.8	&	0.021	&	?	&	\\
S4 0954+65***   & 09:58:47.20 & 65:33:55.0 &  LBL       &       17.0    &       0.106   &       0.367 ? &       \citet{lawrence1986} \\
BZB J1120+4212	& 11:20:48.00 & 42:12:12.0 &  TeVc	&	17.3	& 	0.001	&	0.124 ?	&	\citet{perlman1996}\\
1ES 1215+303	& 12:17:52.10 & 30:07:01.0 &  HBL	&	15.8	&	0.020	&	0.13  ?	&	\citet{bade1998}\\
W Comae  	& 12:21:31.70 & 28:13:59.0 &  IBL	&	15.4	& 	0.021	&	0.102 ?	&	\citet{weistrop1985}\\
MS 1221.8+2452	& 12:24:24.20 & 24:36:24.0 &  HBL	&	16.7	&	0.019	&	0.218 ?	&	\citet{morris1991}\\
S3 1227+255	& 12:30:14.10 & 25:18:07.0 &  IBL	&	14.7	&	0.017	&	0.135 ?	&	\citet{nass1996}\\
BZB J1243+3627	& 12:43:12.70 & 36:27:44.0 &  TeVc	&	16.2	&	0.010	&	?	&	 \\ 
BZB J1248+5820	& 12:48:18.80 & 58:20:29.0 &  TeVc	&	15.4	&	0.011	&	?	&	  \\
PKS 1424+240	& 14:27:00.40 & 23:48:00.0 &  HBL	&	14.6	&	0.050	&	?	&	  \\
BZB J1540+8155	& 15:40:15.80 & 81:55:06.0 &  TeVc	&	17.6	&	0.044	&	?	&	\\
RGB J2243+203	& 22:43:54.70 & 20:21:04.0 &  HBL	&	16.0	&	0.042	&	?	&	\\
BZB J2323+4210	& 23:23:52.10 & 42:10:59.0 &  TeVc	&	17.0	&	0.134	&	0.059 ?	&	\citet{perlman1996} \\
\hline
\end{tabular}
\tablenotetext{}{
\raggedright
\footnotesize \texttt{Col.1}: Name of the target; \texttt{Col.2}: Right Ascension; \texttt{Col.3}: Declination; \texttt{Col.4}: Class of the source: High-synchrotron peaked BL Lac (HBL), Intermediate-synchrotron peaked BL Lac (IBL), Low-synchrotron peaked BL Lac (LBL), TeV Candidate BL Lac (TeVc); \texttt{Col.5}: V-band magnitudes taken from NED; \texttt{Col.6}: $E(B-V)$ taken from the NASA/IPAC Infrared Science Archive (https://irsa.ipac.caltech.edu/applications/DUST/); \texttt{Col.7}: Redshift; \texttt{Col.8}: Reference to the redshift.}
\tablenotetext{}{
\raggedright
\footnotesize * Details for S2 0109+22 are reported in \citet{paiano2016}, ** Gravitationally lensed system , *** =  Details for S4 0954+65 are reported in \citet{landoni2015} } 
\end{table*}

\newpage

\begin{table*}
\caption{LOG OBSERVATIONS OF TEV SOURCES AND TEV-CANDIDATES OBTAINED AT GTC}\label{tab:table2}
\centering
\begin{tabular}{llll|llll}
\hline
&  \multicolumn{3}{c}{Grism B}    & \multicolumn{3}{c}{Grism R} &  \\
\hline
Obejct          & t$_{Exp}$ (s)  &       Date            & Seeing & t$_{Exp}$ (s) &         Date         & Seeing & r  \\
 & & & & & & & \\
\hline
BZB J0035+1515	&	2100	   &	2015 Sept 30    & 0.6 &      	1800	 &	2015 Oct 01	& 0.6	 &	17.00	\\
1ES 0033+595	        &	3600	   &	2015 Sept 18	& 1.3 & 	         2700 &	2015 Sept 25	& 0.9	 &	17.80	\\
S2 0109+22 	        &	750	   &	2015 Sept 19	& 1.8 & 	           750 &	2015 Sept 19	& 1.8	 &	15.20	\\
RGB J0136+391	&	900	   &	2015 Sept 28	& 0.9 &		  600	 &	2015 Sept 28	& 0.9	 &	15.80	\\
S3 0218+357	        &	3600	   &	2015 Feb 05	& 0.9 &		8700 &	2015 Feb 05	& 1.2	 &	19.90	\\
3C 66A	       	        &	750	   &	2015 Sept 09	& 0.8 &		  210	 &	2015 Sept 06	& 0.8	 &	14.70	\\
VER J0521+211	&	900	   &	2015 Sept 21	& 0.8 &		1050 &	2015 Sept 21	& 0.8	 &	16.40	\\
1ES 0647+250	        &	1500	   &	2015 Sept 22	& 1.4 &		1200	 &	2015 Sept 22	& 1.4	 &	15.80	\\
S5 0716+714	        &	210	   &	2015 Nov  30	& 1.6 &		  210 &	2015 Nov 30	& 1.6	 &	13.60	\\
BZB J0915+2933	&	750	   &	2015 Dec 24	& 2.0 &		  450	 &	2015 Jun 06	& 2.0	 &	15.90	\\
S4 J0954+65            &       300     &   2015 Feb 28     & 1.0 &               450  &     2015 Feb 28     & 1.0 &       15.5    \\
BZB J1120+4212	&	3000   &	2016 Jun 24	& 1.5 &   	        3600	 &	2015 Jul 01	& 0.7	 &	16.10 	\\
1ES 1215+303	        &	900	   &	2015 May 20	& 1.5 &		  900	 &	2015 May 20	& 1.5	 &	14.50	\\
W Comae 	        &	1800	   &	2015 Jun 30	& 1.4 &		1800  &	2015 Jun 30	& 1.4	 &	15.50	\\
MS 1221.8+2452	&	3000	   &	2015 May 31	& 1.3 &		3000	 &     2015 May 31	& 1.2	 &	16.70	\\
S3 1227+255	        &	450	   &	2015 Dec 25	& 1.5 &		  500	 &	2015 Dec 25	& 1.5	 &	14.90	\\
BZB J1243+3627	&	1350	   &	2015 May 21	& 1.2 &		1350	 &	2015 May 21	& 1.2	 &	15.60	\\
BZB J1248+5820	&	600	   &	2015 Dec 25	& 2.2 &		  900	 &	2015 Dec 25	& 2.2	 &	15.70	\\
PKS 1424+240	        &	450	   &	2015 Jun 30	& 1.0 &		  450	 &	2015 Jun 30	& 1.0	 &	14.20	\\
BZB J1540+8155	&	900	   &	2015 Jun 23	& 1.0 &		  900	 &	2015 Jun 23	& 1.0	 &	17.30	\\
RGB J2243+203	&	600	   &	2015 Sept 19	& 2.0 &		  750	 &	2015 Sept 19	& 2.0	 &	16.20	\\
BZB J2323+4210	&	3000	   & 	2016 Aug 07	& 1.3 &		3600	 &	2015 Feb 28	& 0.7	 &	17.50	\\
\hline
\hline
\end{tabular}
\tablenotetext{}{
\raggedright
\footnotesize \texttt{Col.1}: Name of the target; \texttt{Col.2}: Total integration time with the Grism B; \texttt{Col.3}: Date of Observation with Grism B; \texttt{Col.4}: Seeing during the observation with the Grism B; \texttt{Col.5}: Total integration time with the Grism R; \texttt{Col.6}: Date of Observation with Grism R; \texttt{Col.7}: Seeing during the observation with the Grism R; \texttt{Col.8}: r' mag measured on the acquisition images.}
\end{table*}

\newpage

\begin{table*}
\caption{PROPERTIES OF THE OPTICAL SPECTRA OF OBSERVED SOURCES}\label{tab:table3}
\centering
\begin{tabular}{lcrclcl}
\hline
OBJECT           & $\alpha$  &   SNR       &   EW$_{min}$   &   z$_{lim}$         &  z             &  N/H$_{lim}$   \\
\hline
BZB J0035+1515   &   -1.3    &   183-275   &   0.09-0.18   &   0.55 (0.32)      &                & 11  \\ 
1ES 0033+595       &     *     &    40-135   &   0.27-0.52   &   0.53 (0.10)      &  0.467$^{e}$          & 5   \\  
S2 0109+22           &   -1.0    &   167-375   &   0.07-0.16   &   0.35 (0.15)      &                & 20  \\ 
RGB J0136+391    &   -1.5    &   196-482   &   0.08-0.15   &   0.27 (0.14)      &                & 6   \\ 
S3 0218+357         &     *     &    5-20     &    *          &       *            &   0.944$^{e}$   & 3  \\
3C 66A                   &   -1.1    &   118-314   &   0.10-0.22   &   0.10 (*)         &                & 2 \\   
VER J0521+211    &   -0.9    &    82-221   &   0.15-0.37   &   0.18 (0.10)      &                & 1  \\   
1ES 0647+250      &   -1.3    &   115-294   &   0.09-0.21   &   0.29 (0.12)      &                & 7  \\   
S5 0716+714        &   -0.8    &   180-346   &   0.04-0.14   &   0.10 (* )        &                & 4   \\   
BZB J0915+2933  &   -1.1    &    89-241   &   0.14-0.34   &   0.13 (* )        &                & 1   \\   
S4 J0954+65        &   -0.9    &   50-120    &   0.15-0.20   &   0.45 (0.27)      &                & 25  \\
BZB J1120+4212  &   -1.6    &   100-190   &   0.12-0.23   &   0.28 (0.12)      &                & 5  \\
1ES 1215+303     &   -1.0    &   205-375   &   0.09-0.14   &   0.14 (*)         &  0.129$^{e}$    & 4 \\ 
W Comae             &   -0.6    &   180-260   &   0.09-0.17   &   0.19 (0.10)      &  0.102$^{e,g}$    & 1 \\   
MS 1221.8+2452  &   -1.2    &   115-199   &   0.13-0.23   &   0.34 (0.15)      &  0.218$^{e,g}$    & 2  \\   
S3 1227+25          &   -0.8    &   124-397   &   0.09-0.24   &   0.10 (*)         &                & 2  \\   
BZB J1243+3627  &   -1.3    &   208-465   &   0.05-0.15   &   0.28 (0.10)      &  $>$0.48$^{a}$ & 29 \\   
BZB J1248+5820  &   -0.9    &    76-225   &   0.12-0.29   &   0.14 (*)         &                & 2   \\   
PKS 1424+240     &   -1.1    &   254-436   &   0.04-0.10   &   0.10 (*)         &  0.604$^{e}$    & 184 \\   
BZB J1540+8155  &   -1.3    &    97-211   &   0.15-0.28   &   0.56 (0.22)      &  $>$0.67$^{a}$ & 14 \\   
RGB J2243+203   &   -1.1    &   109-178   &   0.15-0.22   &   0.22 (0.10)      &                & 3   \\   
BZB J2323+4210  &   -1.2    &   160-315   &   0.07-0.17   &   0.73 (0.65)      &  $>$0.267$^{a}$ & 1  \\
\hline
\end{tabular}
\tablenotetext{}{
\raggedright
\footnotesize \texttt{Col.1}: Name of the target; \texttt{Col.2}: Optical spectral index derived from a Power Law fit in the range 4250-10000; \texttt{Col.3}: range of SNR of the spectrum; \texttt{Col.4}: Range of the minimum equivalenth width (EW$_{min}$) derived from different regions of the spectrum (see text), \texttt{Col.5}: Lower limit (3$\sigma$ level) of the redshift by assuming BL Lac host galaxy with $M_{R}$ = -22.9 (-21.9), in parenthesis we give the redshift lower limit assuming a host galaxy one magnitude fainter. An asterisk indicates that the redshift limit is out of observed range for the case of fainter host galaxy (see Appendix), \texttt{Col.6}: Spectroscopic redshift; the superscript letters are: \textit{e} = emission line, \textit{g} = host galaxy absorption, \textit{a}= intervening absorption ; \texttt{Col.7}: Lower limit of the Nucleus-Host galaxy Ratio (N/H) in r-band considering the whole flux of the host galaxy.}
\end{table*}

\newpage

\begin{table*}
\caption{CORRESPONDENCE BETWEEN THE WAVELENGTH RANGE, ABSORPTION LINES AND REDSHIFT RANGE}\label{tab:table4}
\centering
\begin{tabular}{ccc}
\hline
Wavelength Range  & Absorption Line & Redshift Range\\
\hline
4250 - 5000  & CaII & 0.08 - 0.27\\
5000 - 6200  & CaII & 0.27 - 0.58\\
6400 - 6800  & CaII & 0.63 - 0.73\\
\hline
7800 - 8100  & MgI  & 0.51 - 0.57\\
8400 - 8900  & MgI  & 0.63 - 0.72\\
\hline
\end{tabular}
\tablenotetext{}{
\raggedright
\footnotesize \texttt{Col.1}: Wavelength range of the optical spectrum; \texttt{Col.2}: Host galaxy absorption line used; \texttt{Col.3}: Redshift range corrisponding to the wavelength range}
\end{table*}

\begin{table*}
\caption{MEASUREMENTS OF SPECTRAL LINES}\label{tab:line}
\centering
\begin{tabular}{lccll}
\hline
OBJECT            &  $\lambda_{obs}$    &    EW (observed)    &     Line ID    &   z$_{line}$   \\
                  &  $\AA$             &     $\AA$           &                &               \\
\hline
1ES 0033+595      & 5468       & 0.40      & [OII] 3727      &  0.467      \\     
S3 0218+357       & 5470       & 46.8      & Mg II 2800      & 0.954     \\
1ES 1215+303      & 4214       & 0.16      & [O II] 3727      &  0.131  \\  
                  & 5661       & 0.11      & [O III] 5007     &  0.131  \\ 
W Comae           & 4747       & 0.15      & G-band 4305      &  0.102  \\   
                  & 5520       & 0.65      & [O III] 5007     &  0.102  \\  
                  & 5704       & 0.30      & Mg I 5175        &  0.102  \\  
                  & 6496       & 0.31      & Na I  5892       &  0.102  \\
MS 1221.8+2452    & 4794       & 0.18      & Ca II 3934       &  0.218   \\   
                  & 4834       & 0.16      & Ca II 3968       &  0.218   \\   
                  & 5244       & 0.20      & G-band 4305      &  0.218   \\   
                  & 7995       & 0.35      & H$_{\alpha}$ 6563  &  0.218   \\   
                  & 8022       & 0.40      & N II 6583        &  0.218   \\   
BZB J1243+3627    & 4150       & 0.90      & Mg II 2800       &  $>$0.483   \\  
PKS 1424+240      & 5981       & 0.05      & [O II]  3727     &  0.604  \\  
                  & 8035       & 0.10      & [O III] 5007     &  0.604  \\ 
BZB J1540+8155    & 4680       & 0.60      & Mg II 2800       &  $>$0.672   \\  
BZB J2323+4210    & 4987       & 0.30      & Ca II 3934       &  $>$0.267  \\
                  & 5031       & 0.22      & Ca II 3968       &  $>$0.267  \\
                  & 7470       & 0.35      & Na I  5892       &  $>$0.267  \\
\hline
\end{tabular}
\tablenotetext{}{
\raggedright
\footnotesize \texttt{Col.1}: Name of the target; \texttt{Col.2}: Barycenter of the detected line; \texttt{Col.3}: Measured equivalent width; \texttt{Col.4}: Line identification; \texttt{Col.5}: Spectroscopic redshift.}
\end{table*}


\begin{figure*}
  \includegraphics[width=0.4\textwidth, angle=-90]{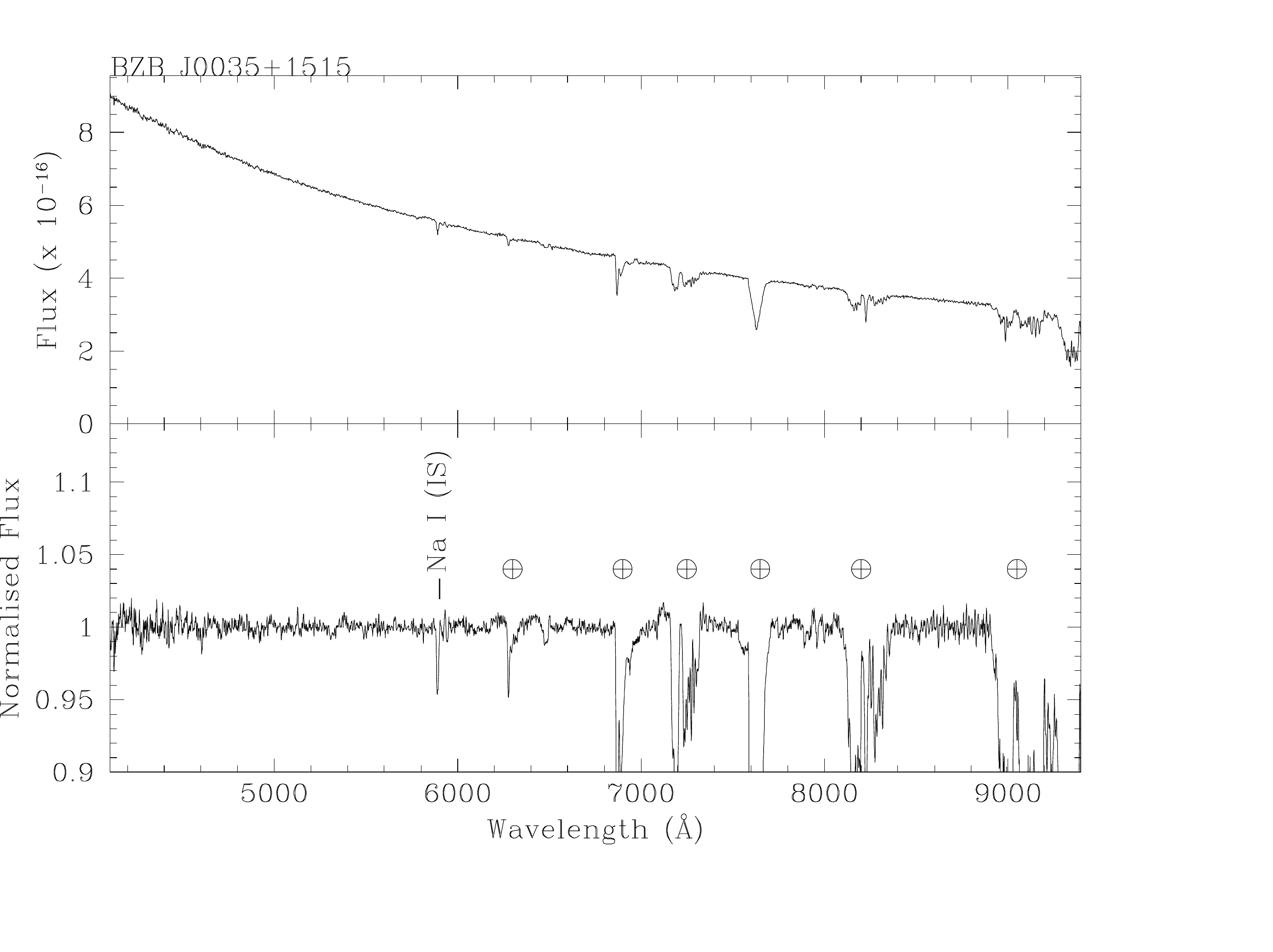}
   \includegraphics[width=0.4\textwidth, angle=-90]{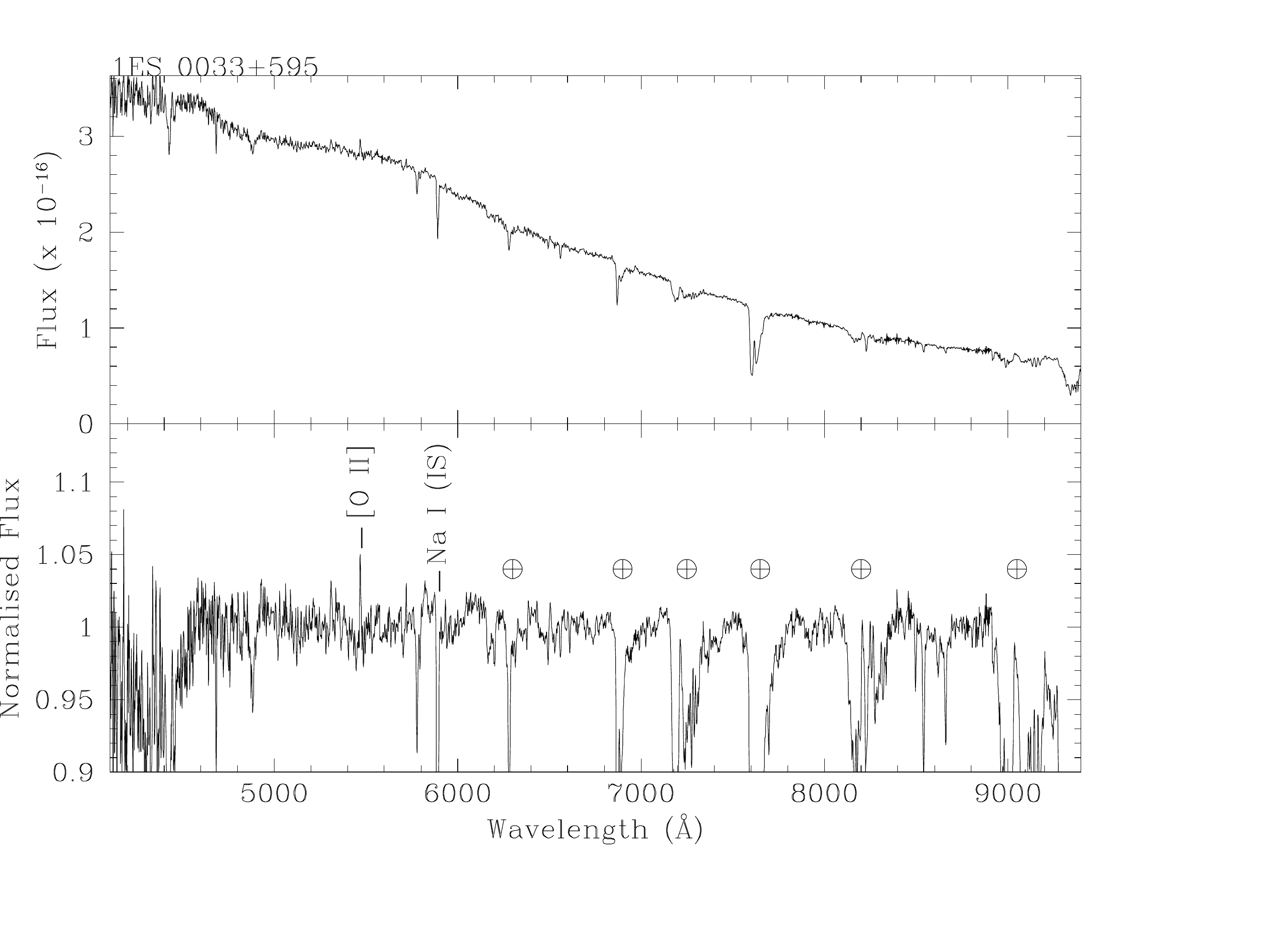} 
   \includegraphics[width=0.4\textwidth, angle=-90]{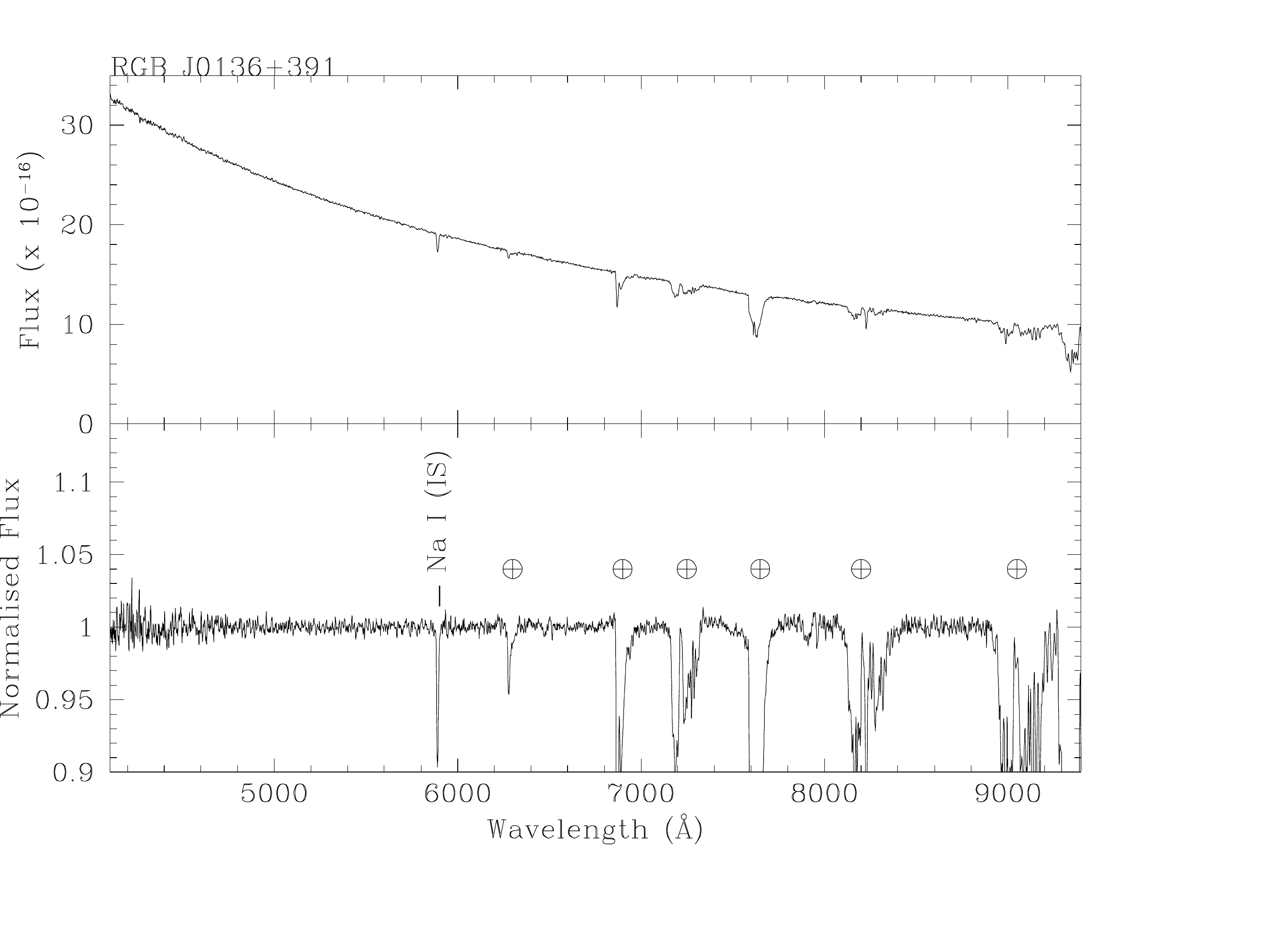}
   \includegraphics[width=0.4\textwidth, angle=-90]{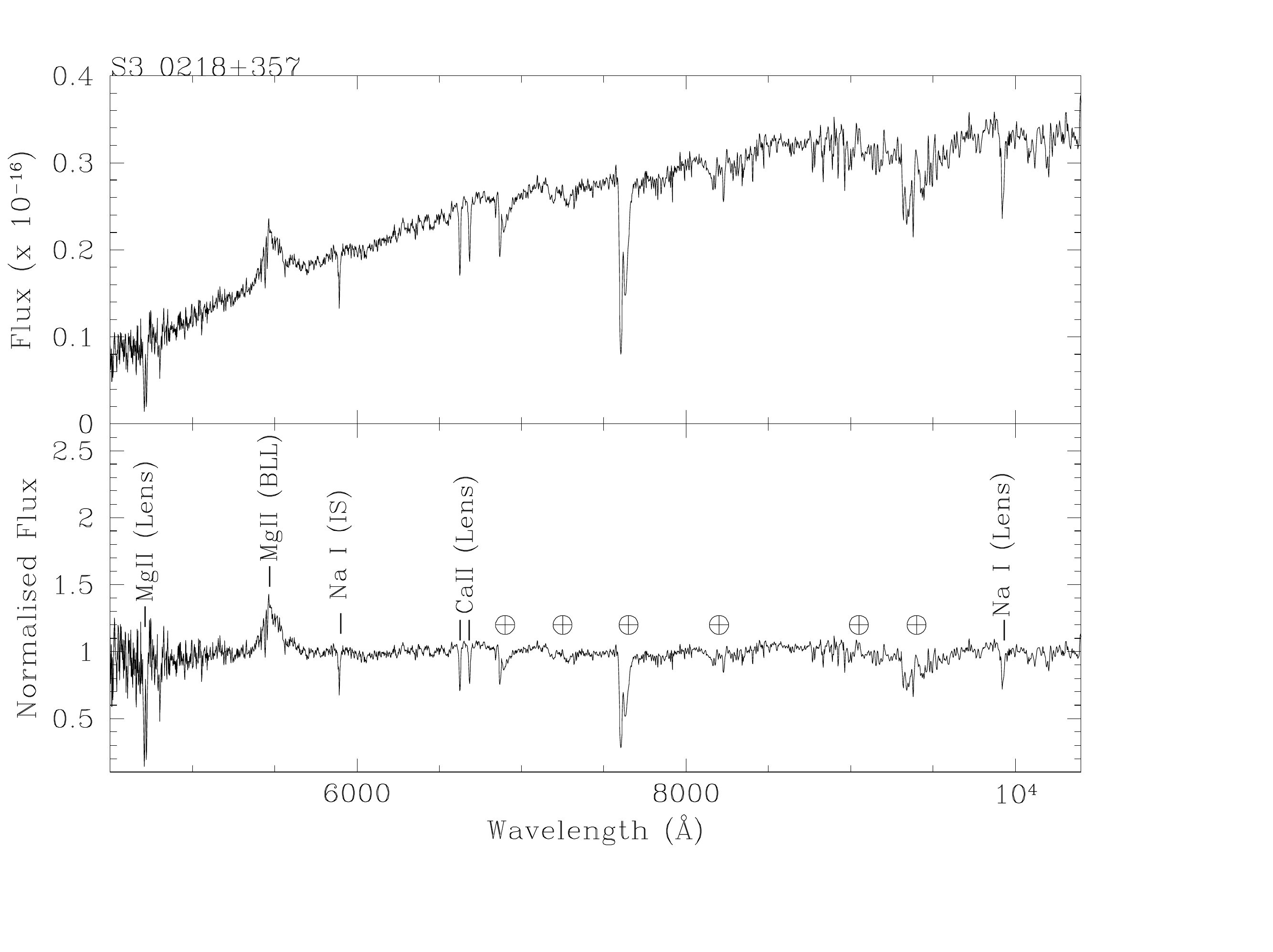}
   \includegraphics[width=0.4\textwidth, angle=-90]{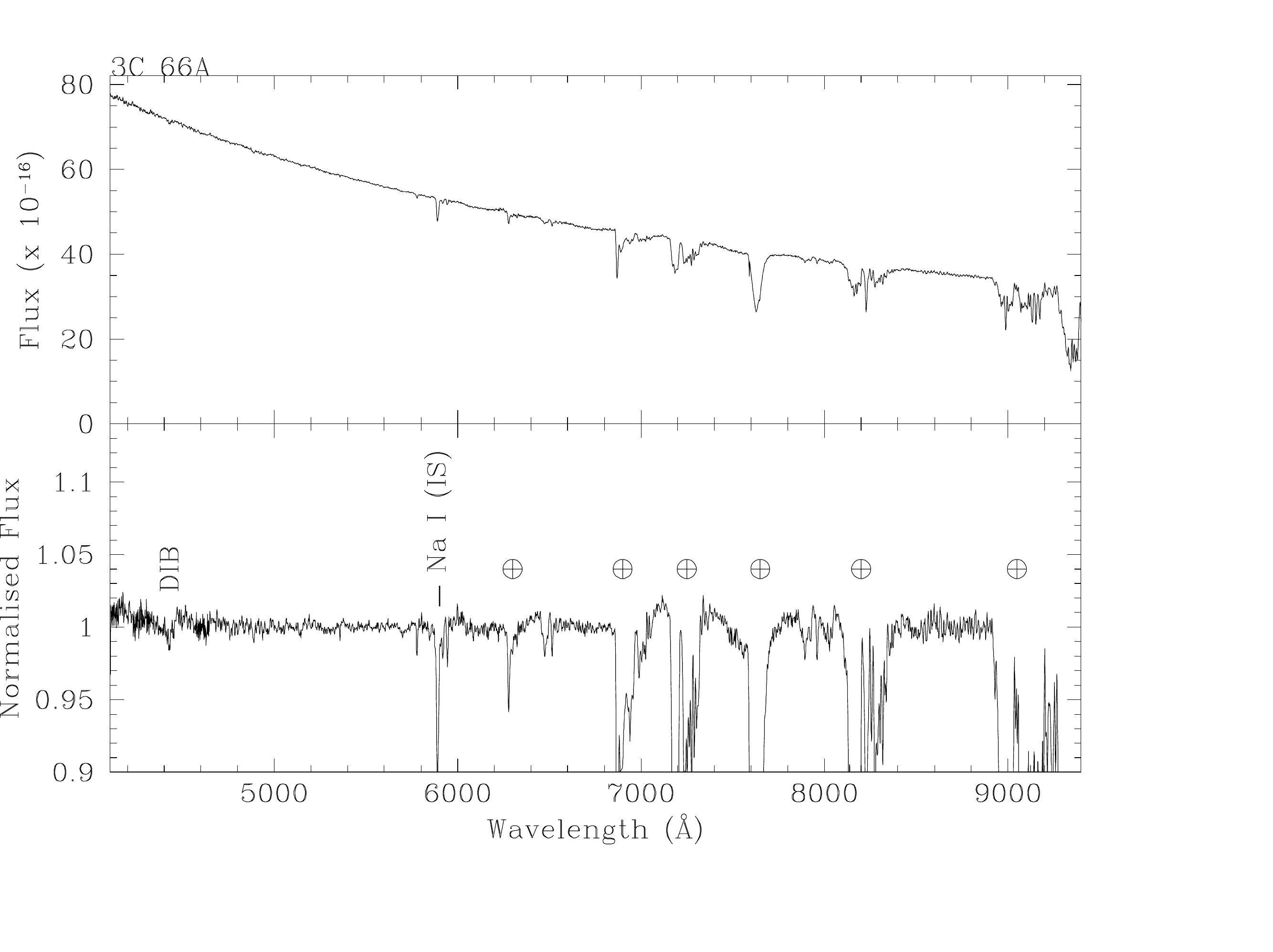}
   \includegraphics[width=0.4\textwidth, angle=-90]{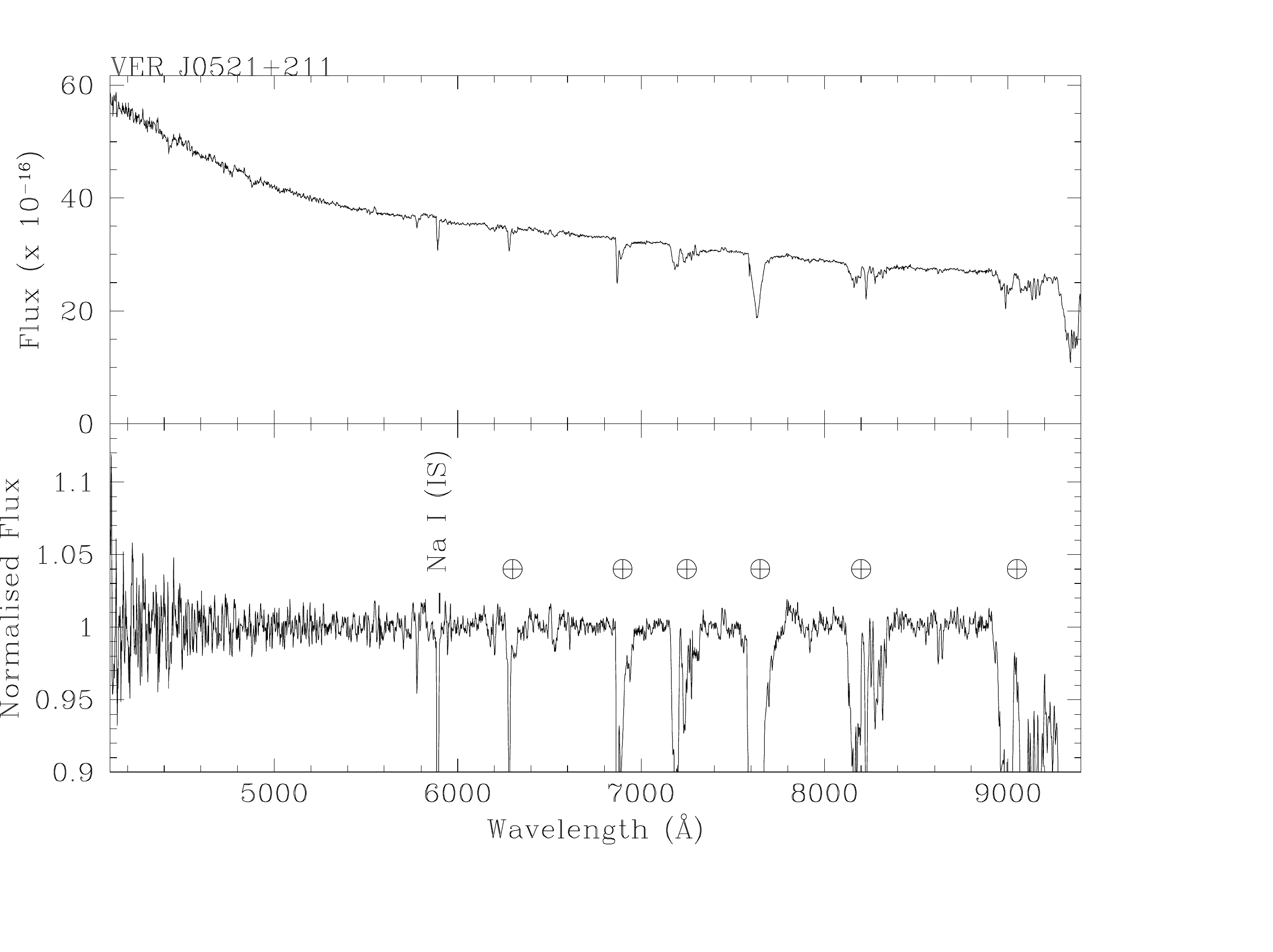}
\caption{Spectra of the TeV sources and TeV-candidates obtained at GTC. \textit{Top panel}: Flux calibrated and dereddered spectra. \textit{Bottom panel}: Normalized spectra. The main telluric bands are indicated by $\oplus$, the absorption features from interstellar medium of our galaxies are labelled as IS (Inter-Stellar)}
\label{fig:spectra}
\end{figure*}

\setcounter{figure}{3}
\begin{figure*}
 \includegraphics[width=0.4\textwidth, angle=-90]{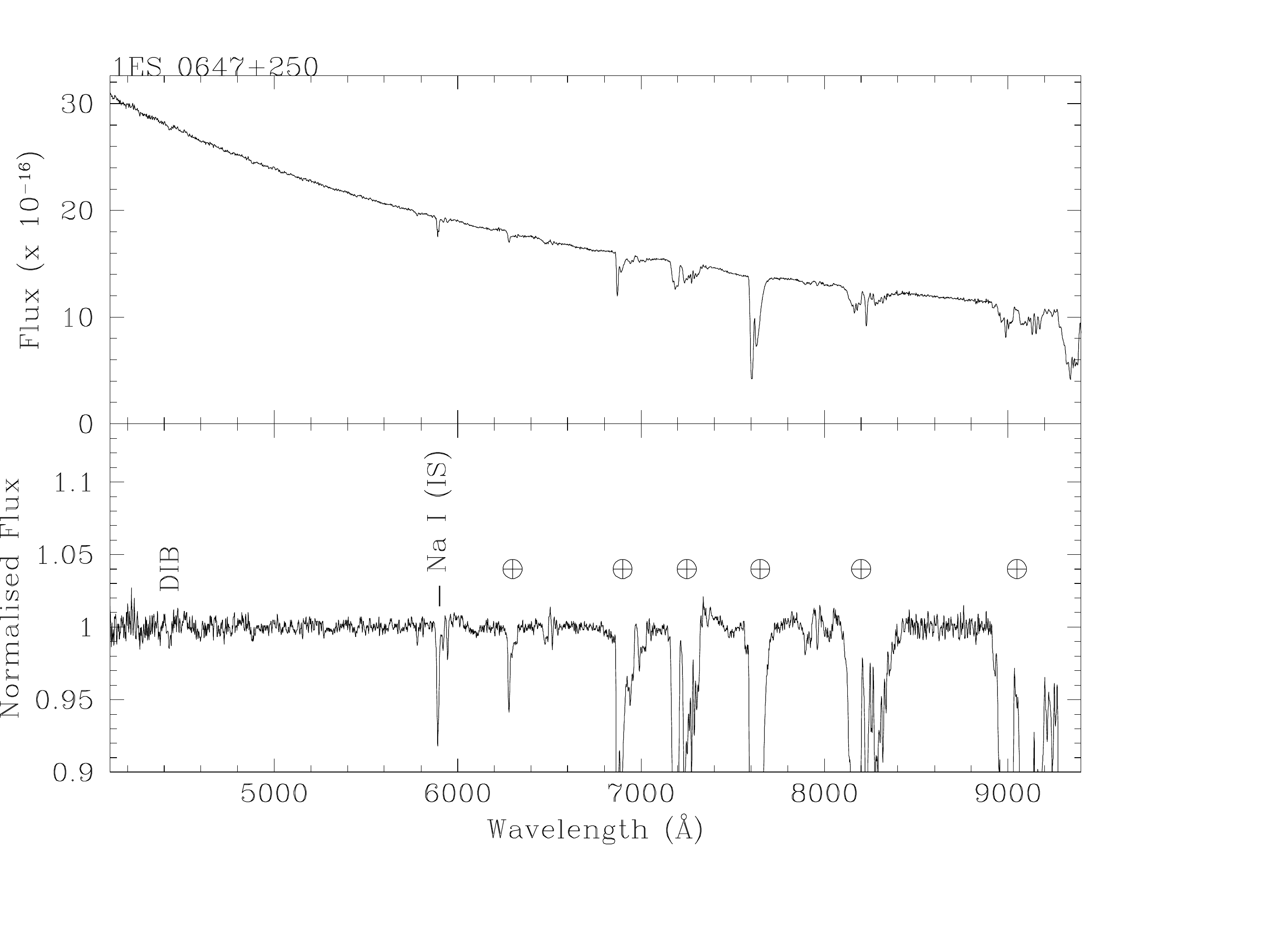}
   \includegraphics[width=0.4\textwidth, angle=-90]{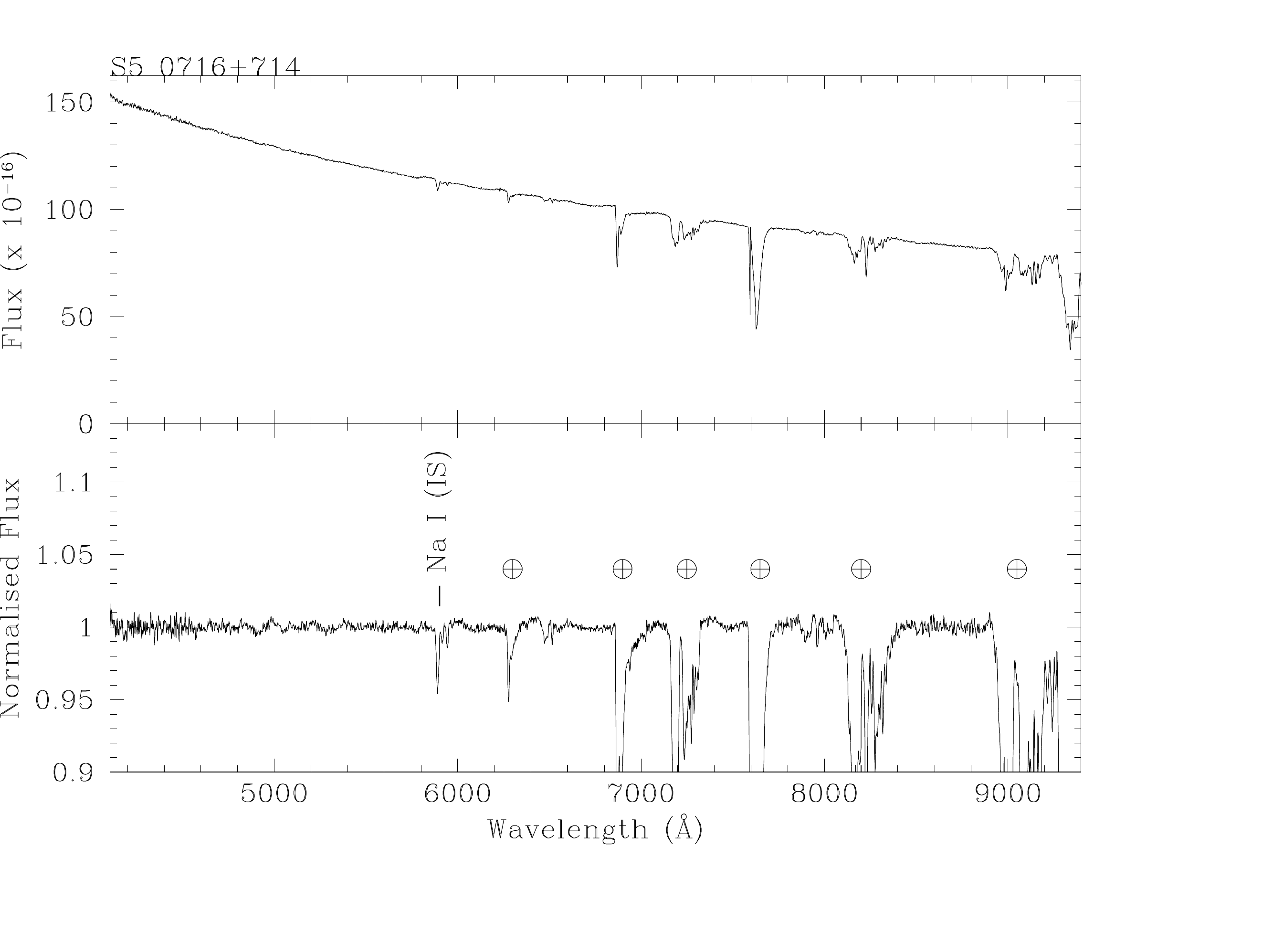}
   \includegraphics[width=0.4\textwidth, angle=-90]{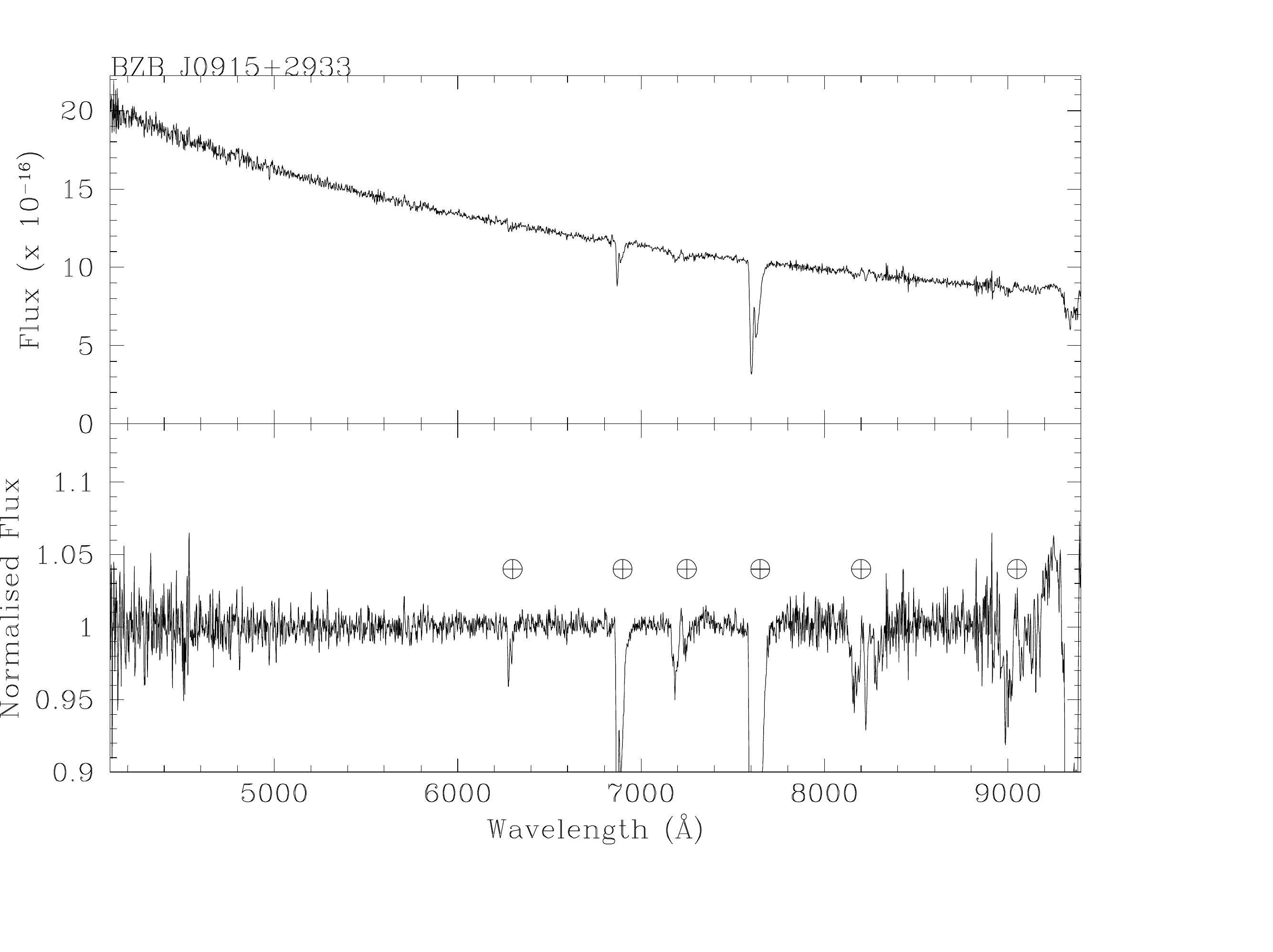}
   \includegraphics[width=0.4\textwidth, angle=-90]{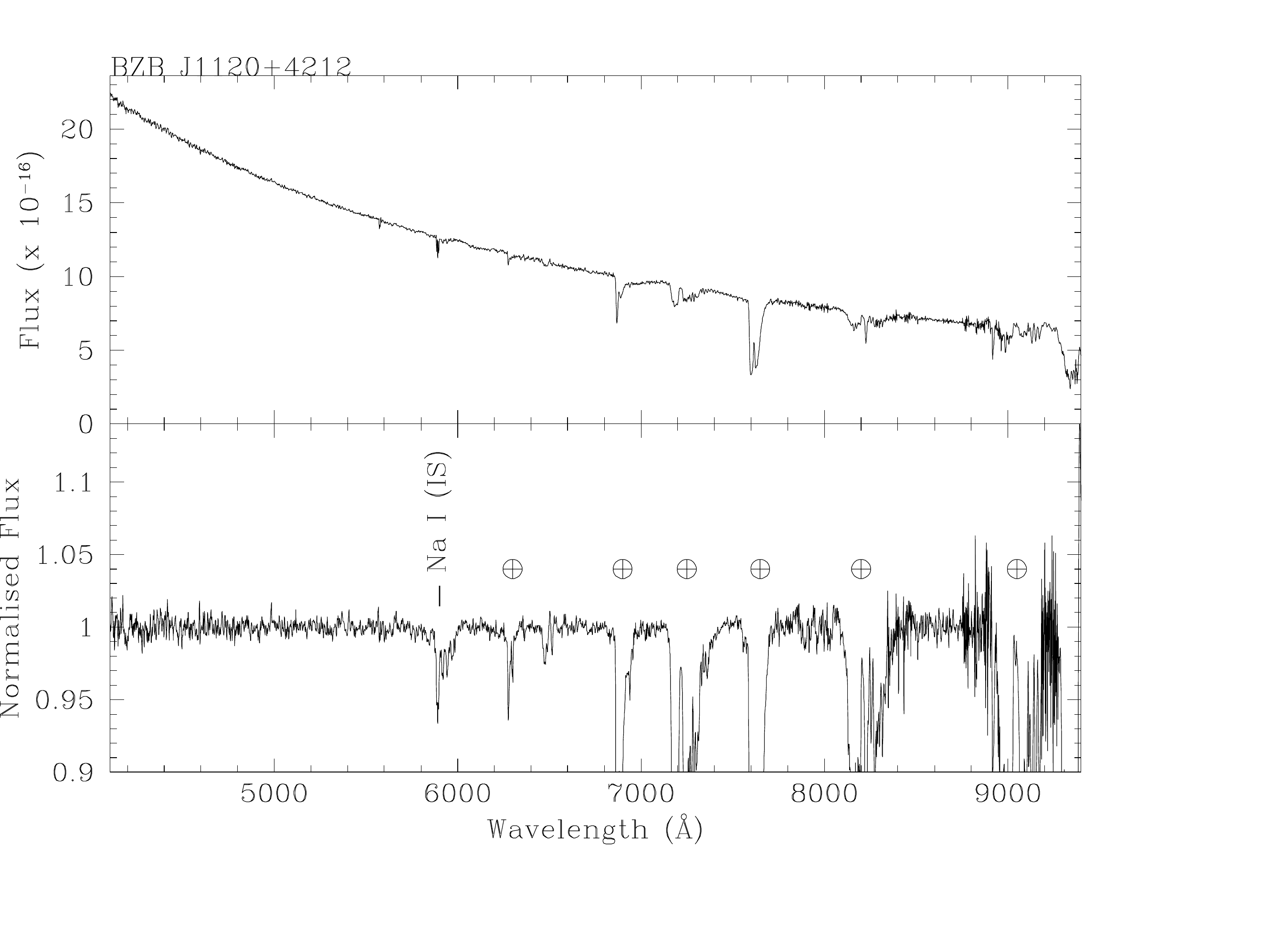}
   \includegraphics[width=0.4\textwidth, angle=-90]{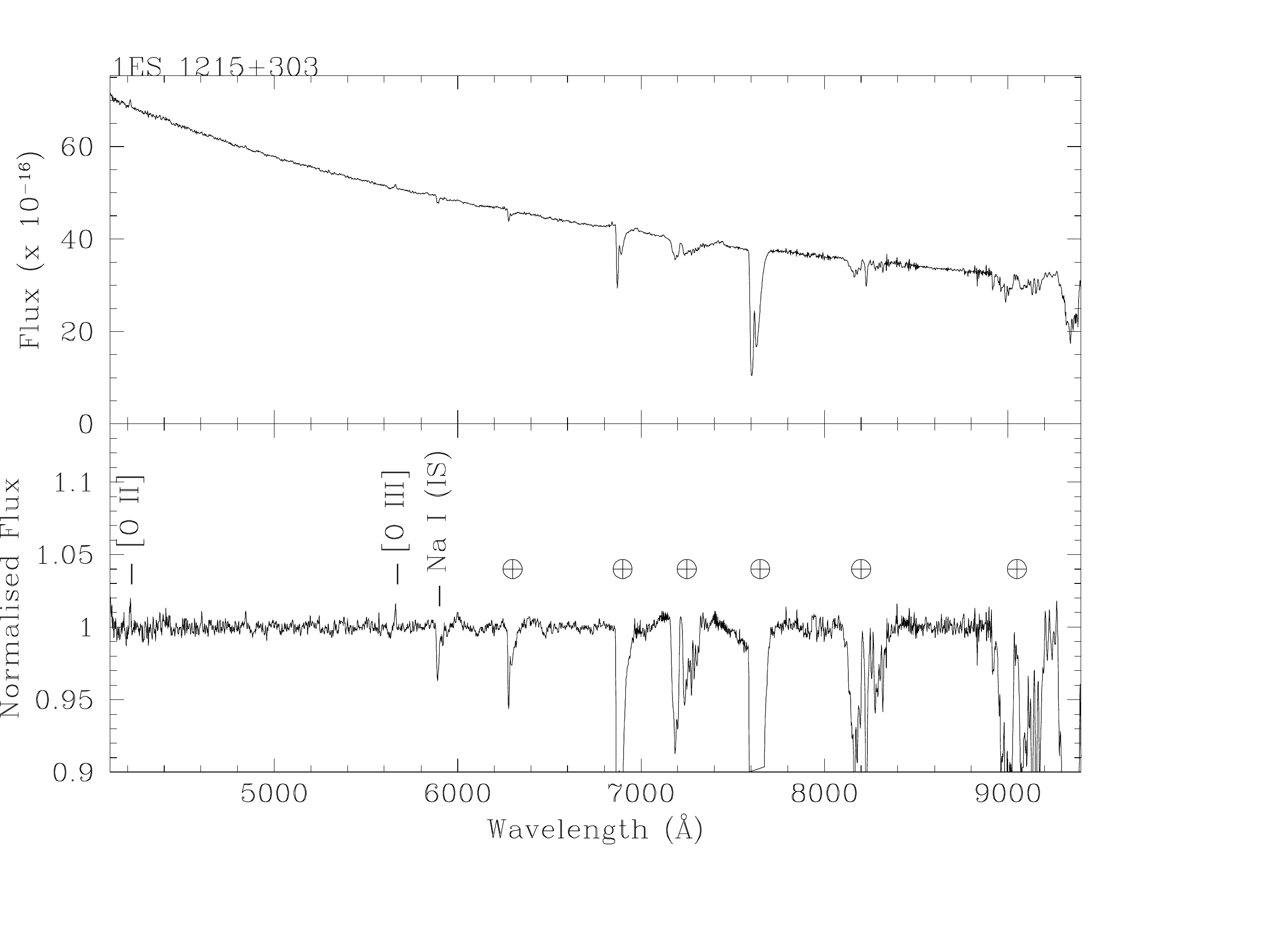}
   \includegraphics[width=0.4\textwidth, angle=-90]{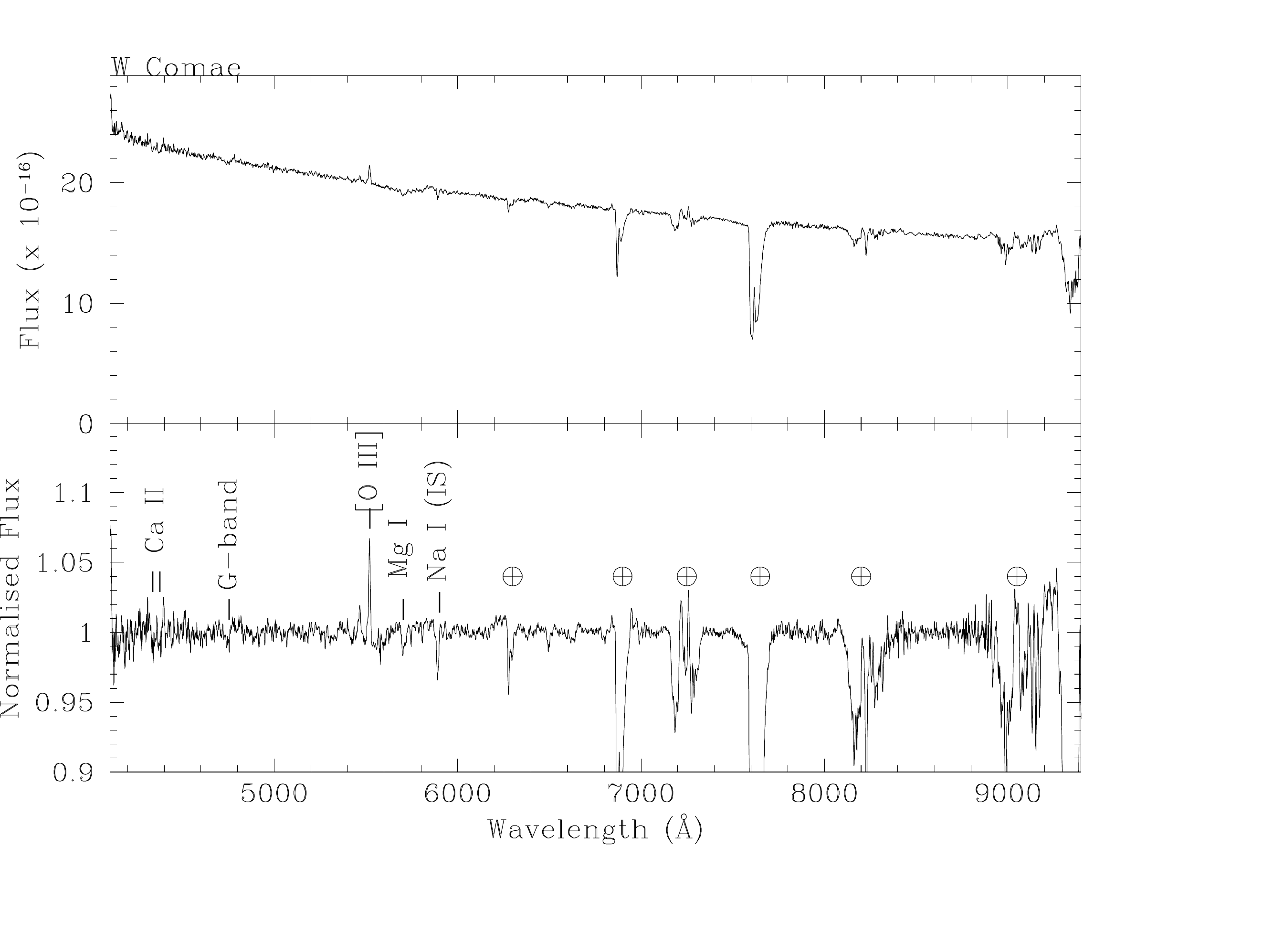}
\caption{Continued from Fig. \ref{fig:spectra}.}
\end{figure*}

\setcounter{figure}{3}
\begin{figure*}
\includegraphics[width=0.4\textwidth, angle=-90]{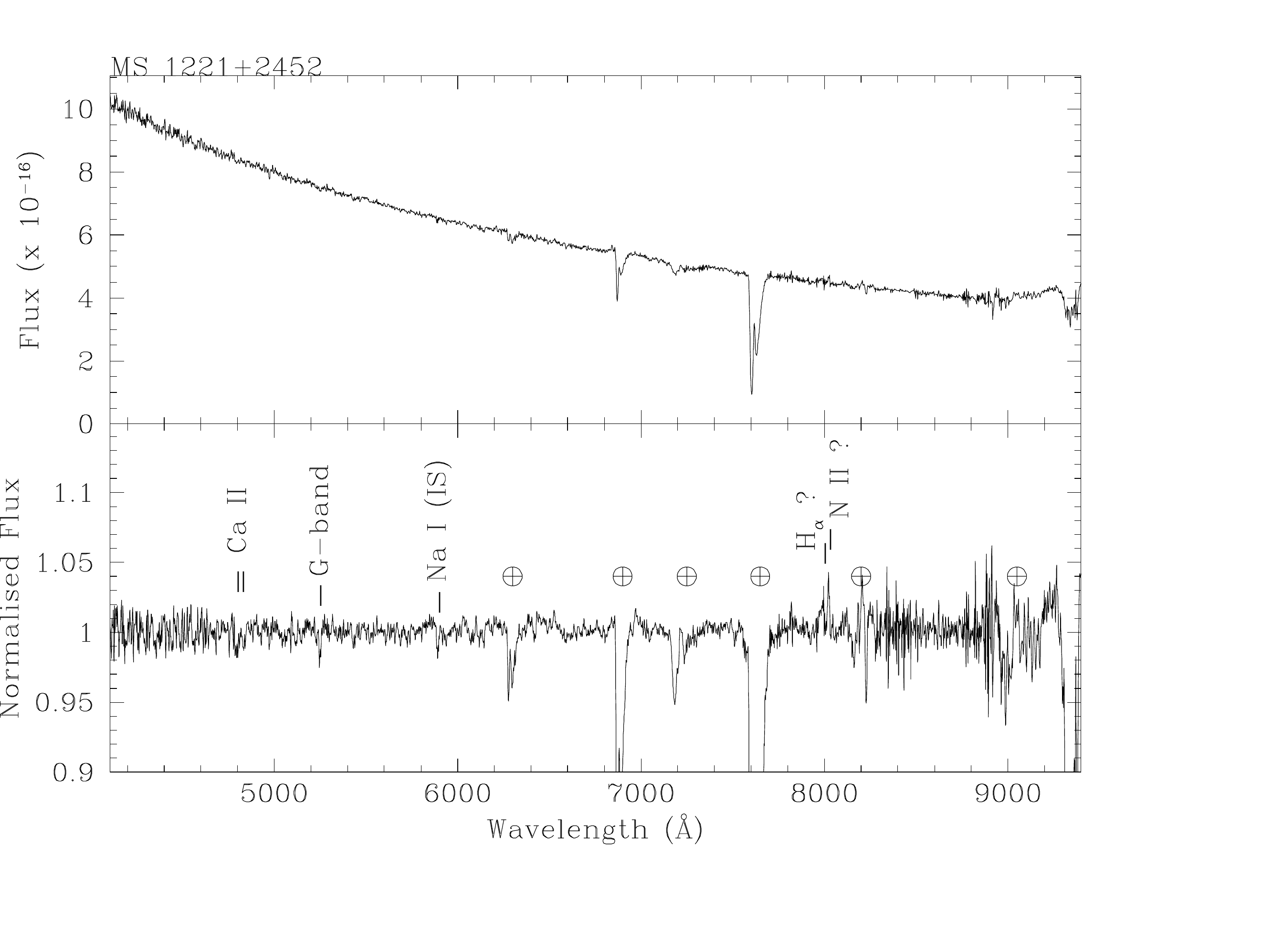}
   \includegraphics[width=0.4\textwidth, angle=-90]{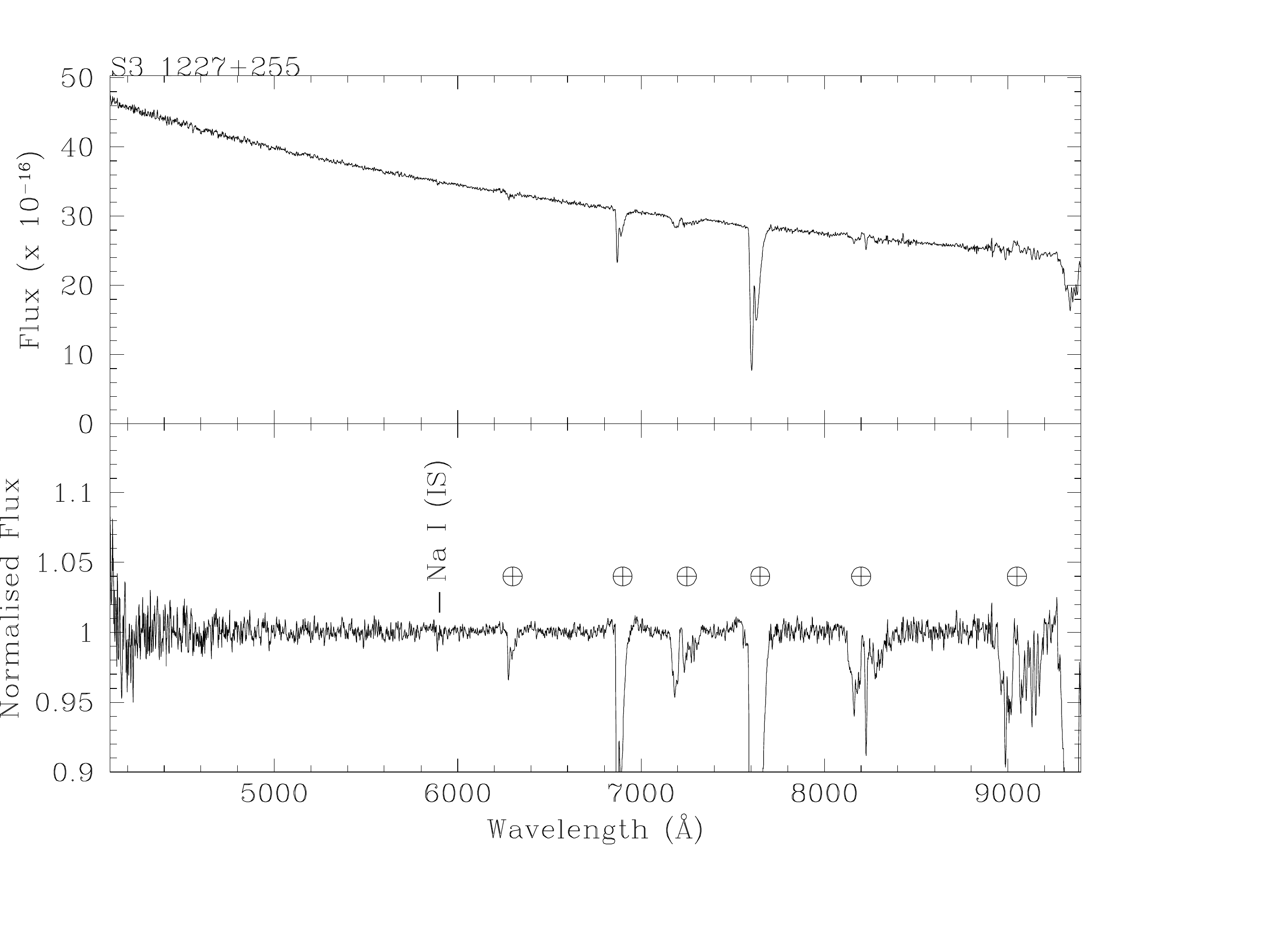}
   \includegraphics[width=0.4\textwidth, angle=-90]{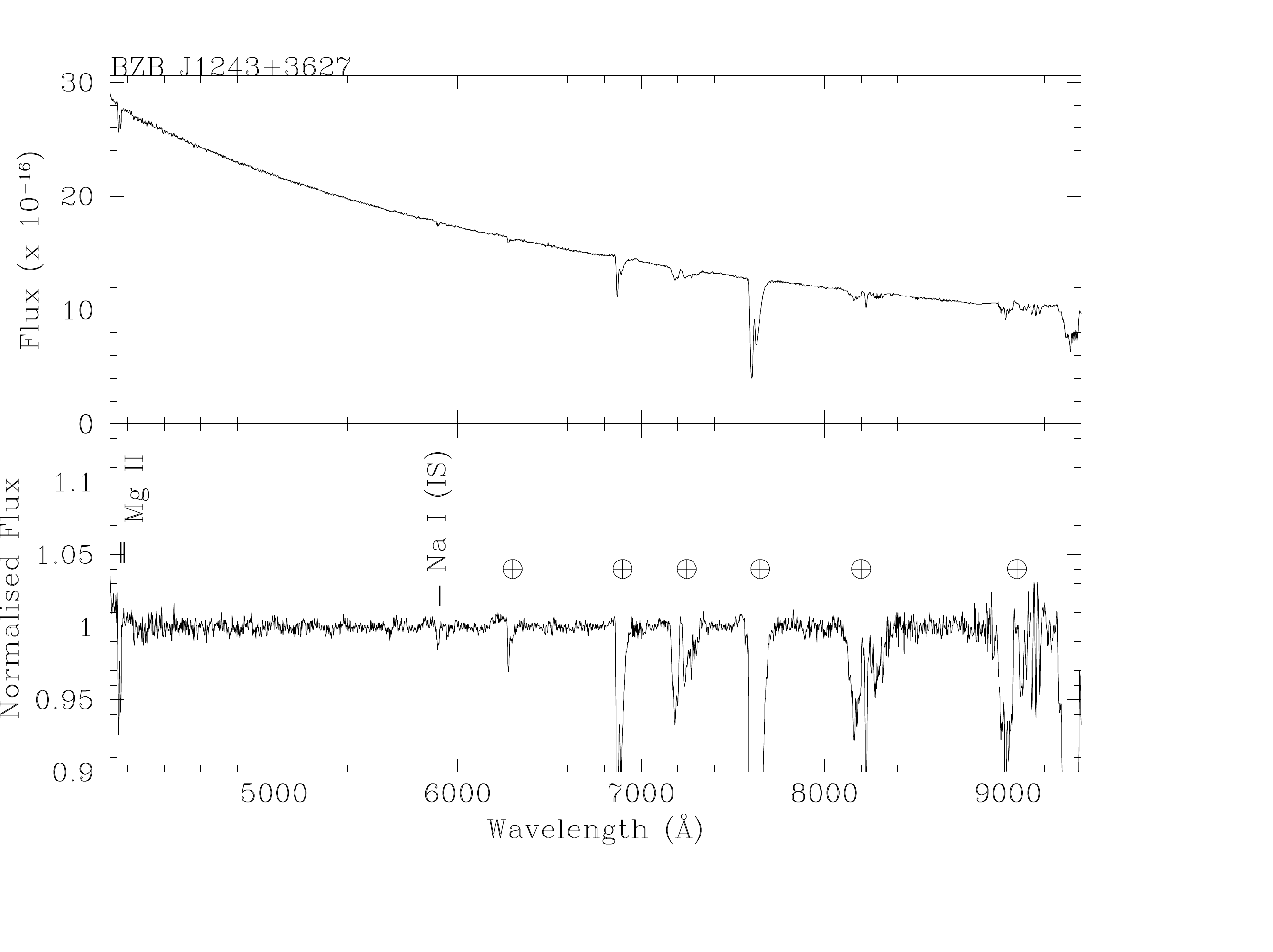}
   \includegraphics[width=0.4\textwidth, angle=-90]{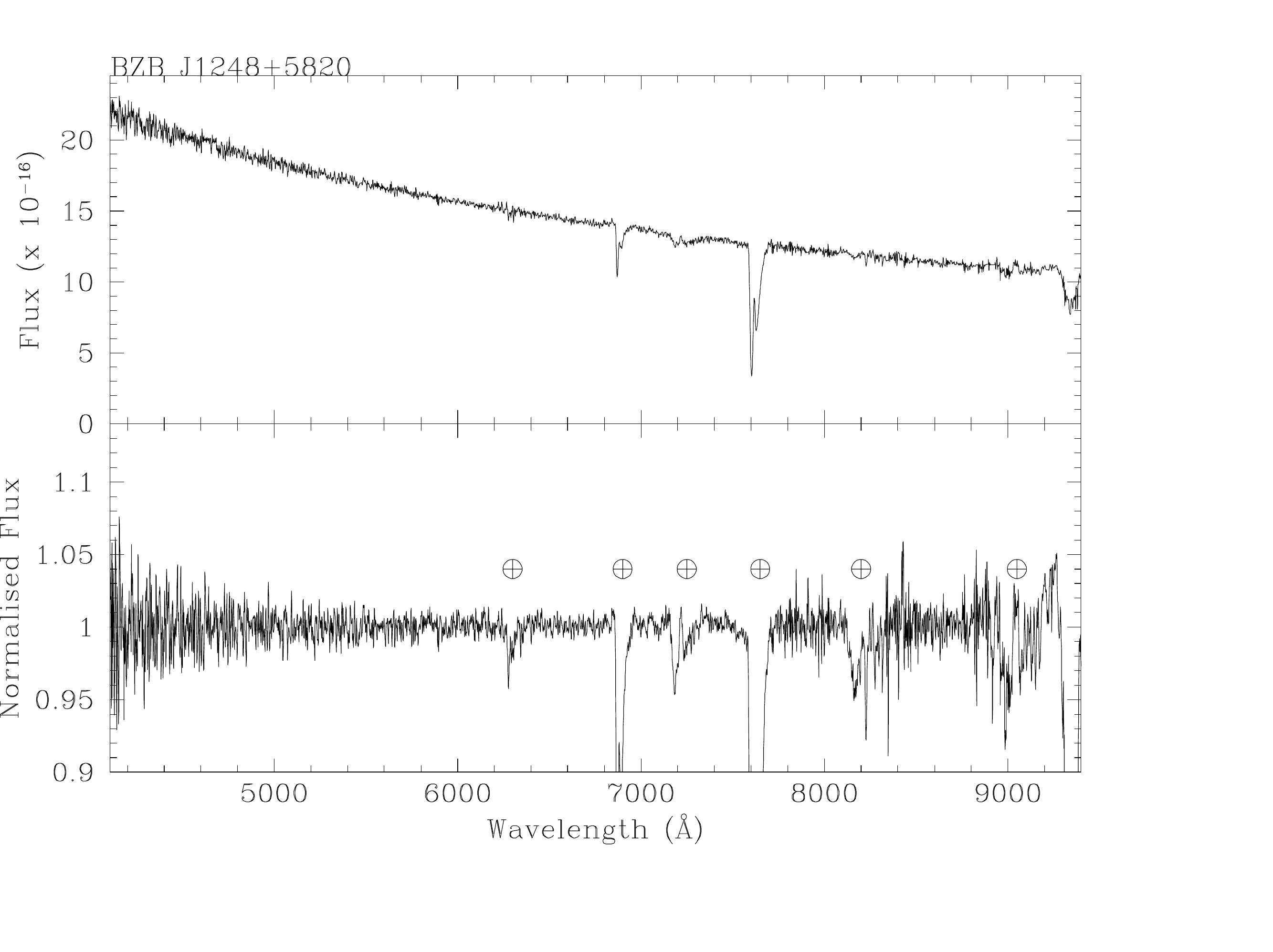}
   \includegraphics[width=0.4\textwidth, angle=-90]{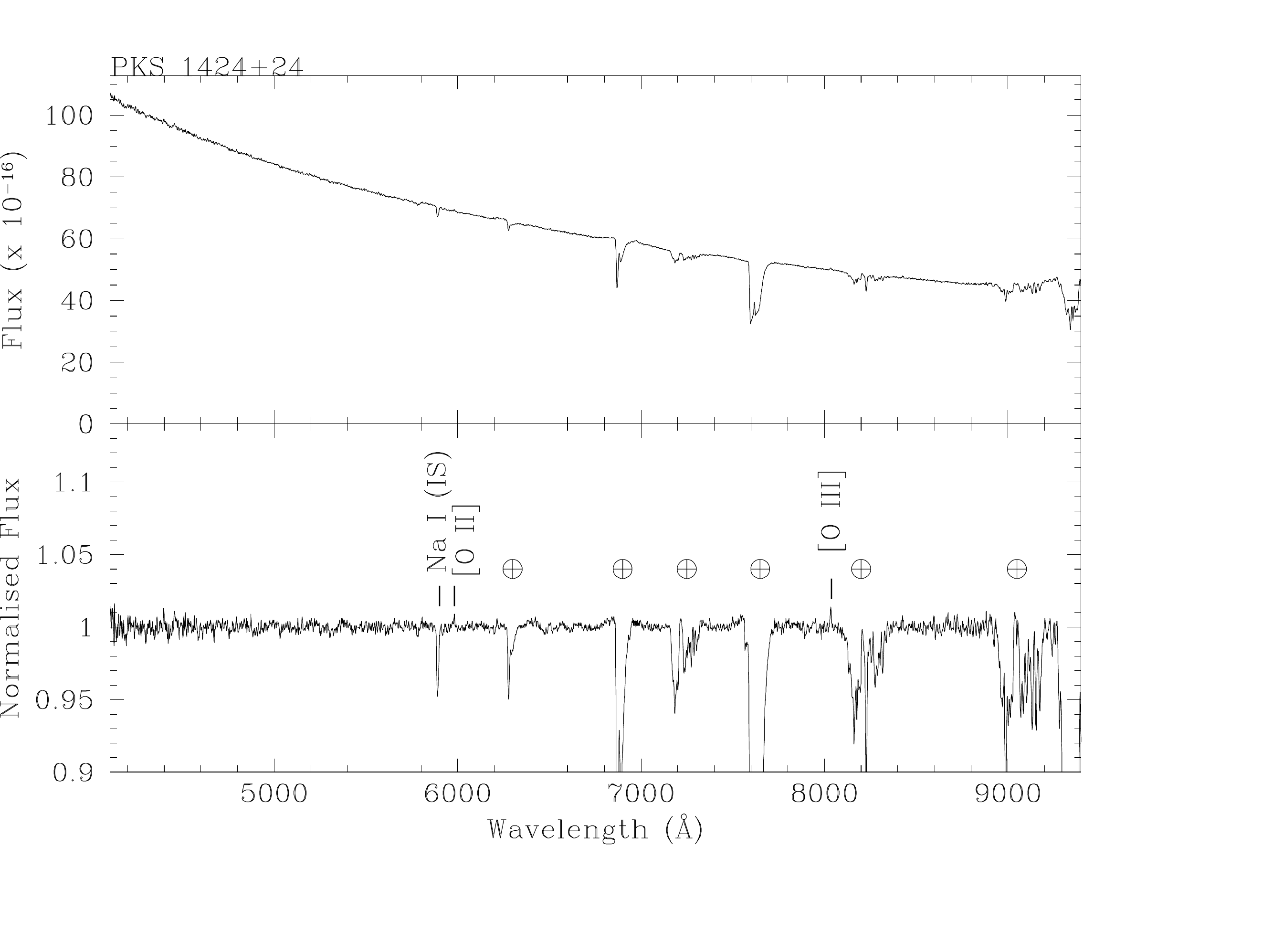}
 \includegraphics[width=0.4\textwidth, angle=-90]{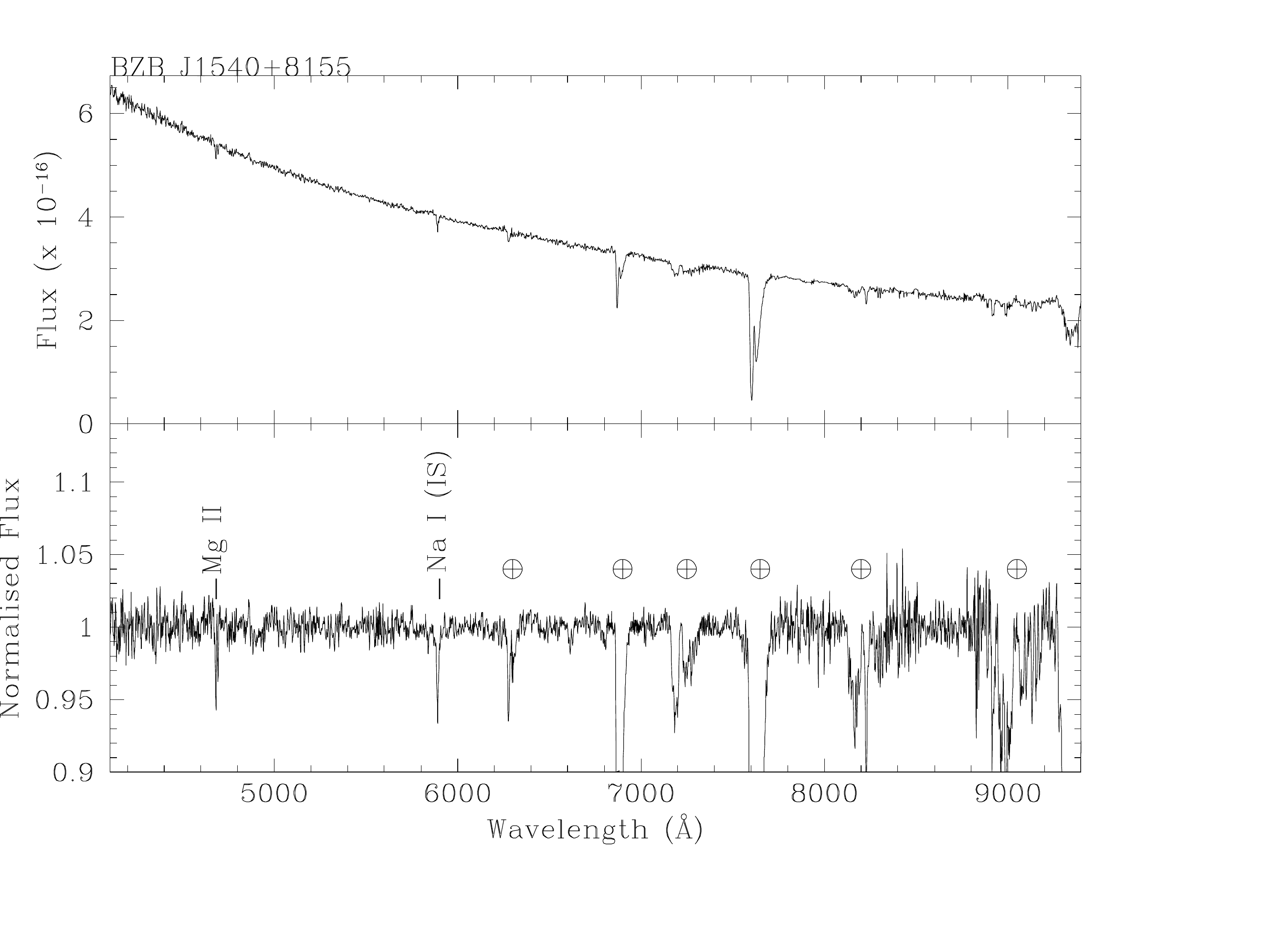}
\caption{Continued.}
\end{figure*}

\setcounter{figure}{3}
\begin{figure*}
 \includegraphics[width=0.4\textwidth, angle=-90]{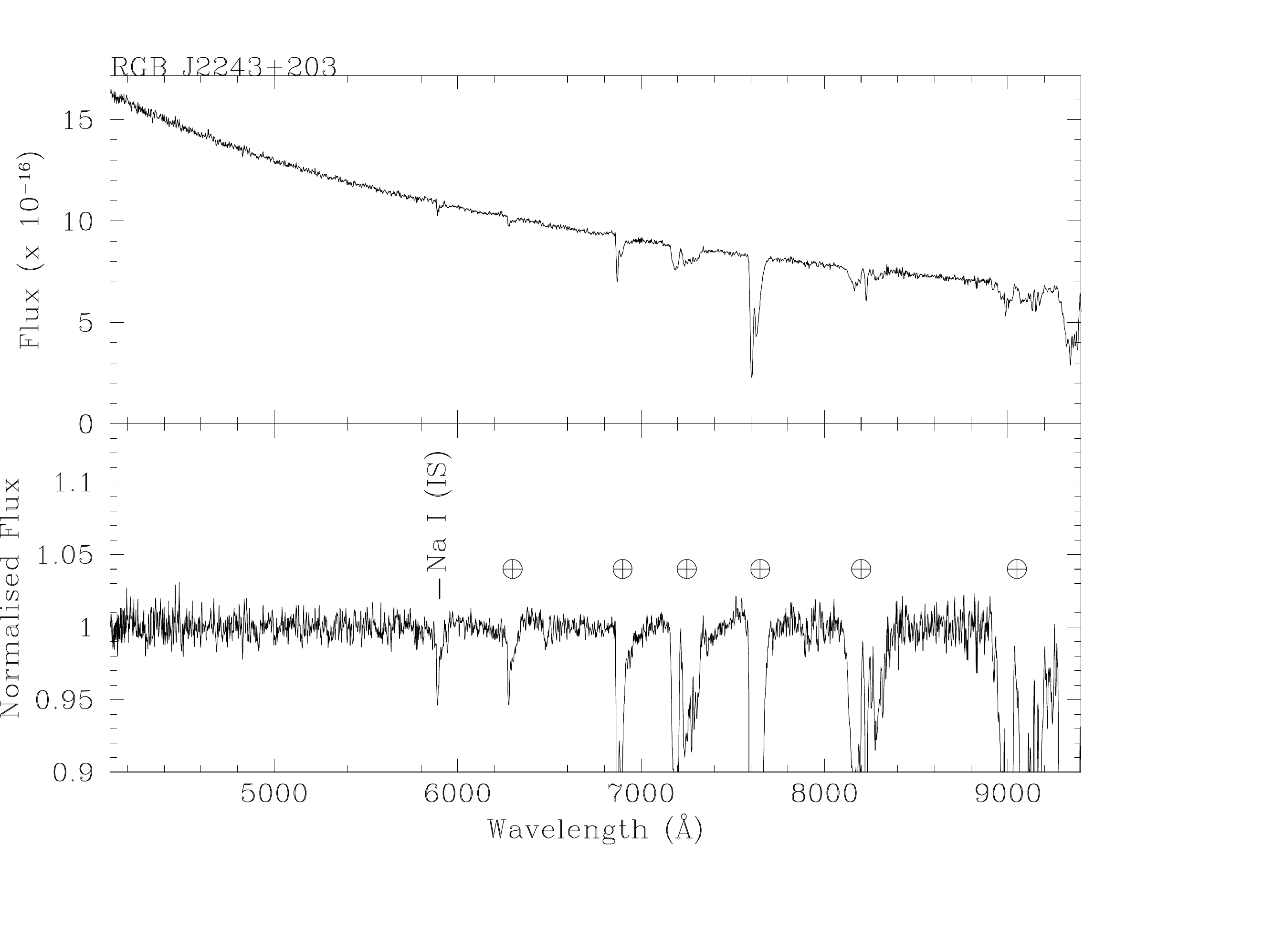}
   \includegraphics[width=0.4\textwidth, angle=-90]{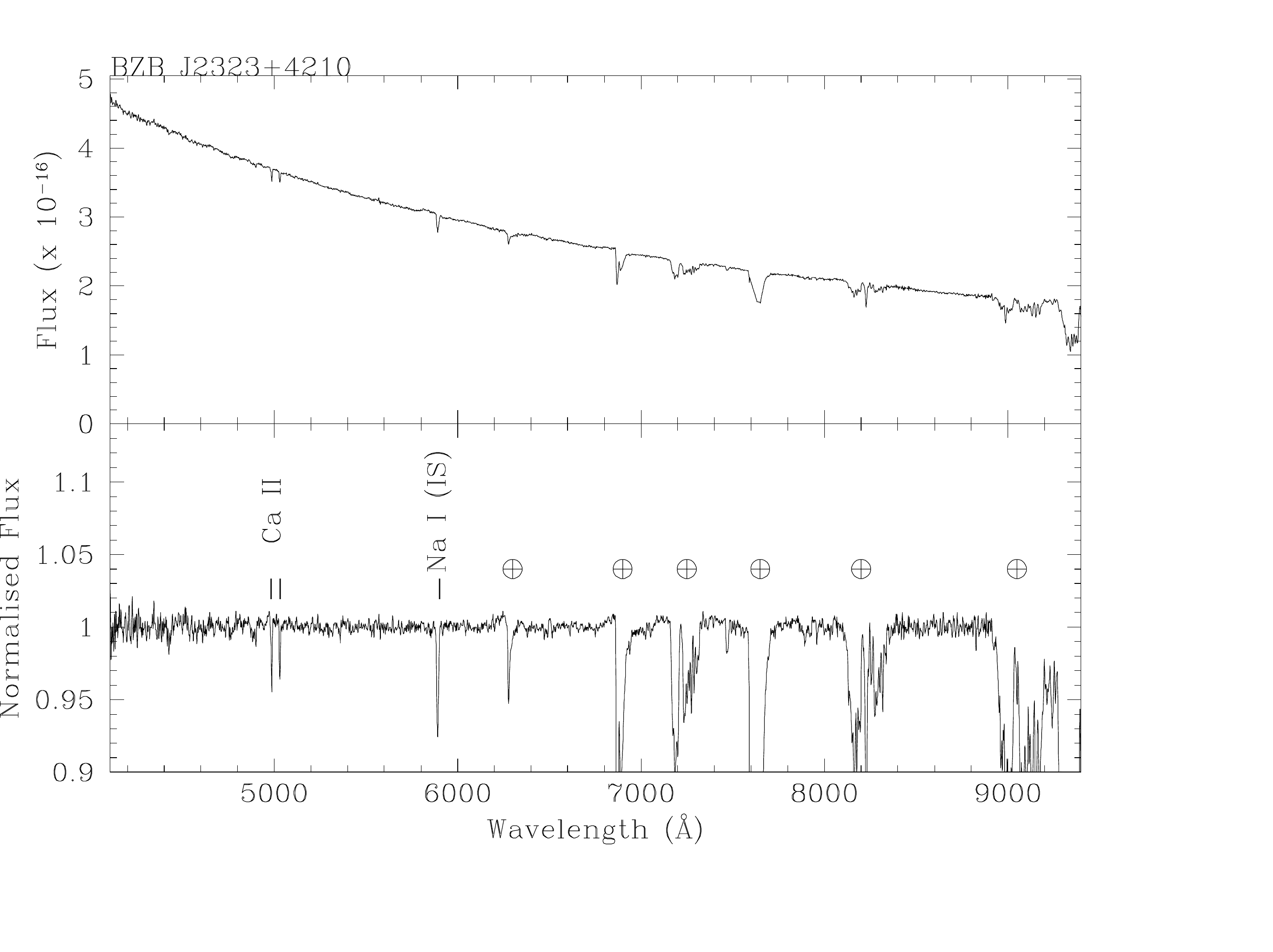}
  \caption{Continued.}
\end{figure*}

\newpage

\begin{figure*}
\includegraphics[width=0.4\textwidth, angle=-90]{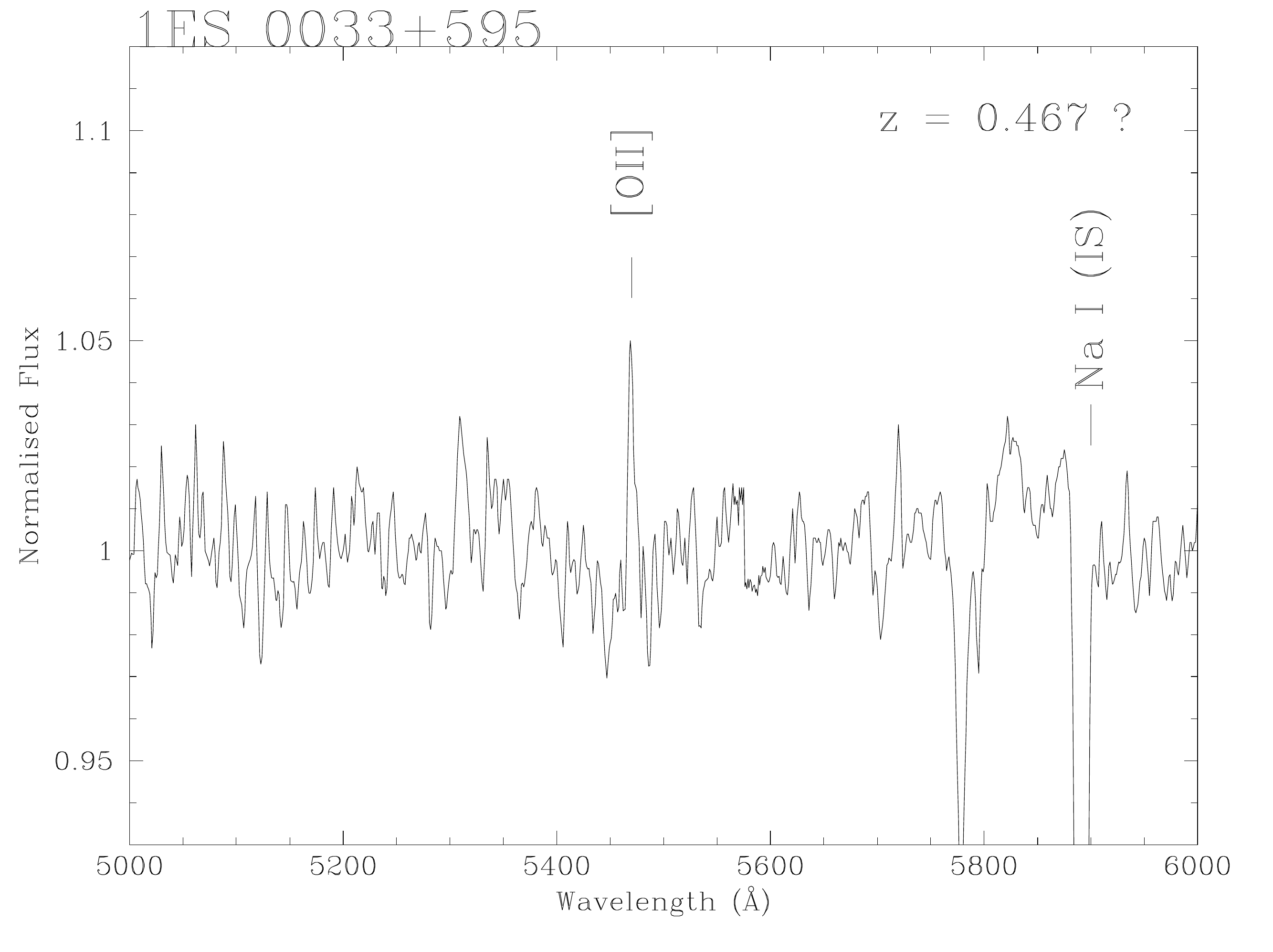}
   \includegraphics[width=0.4\textwidth, angle=-90]{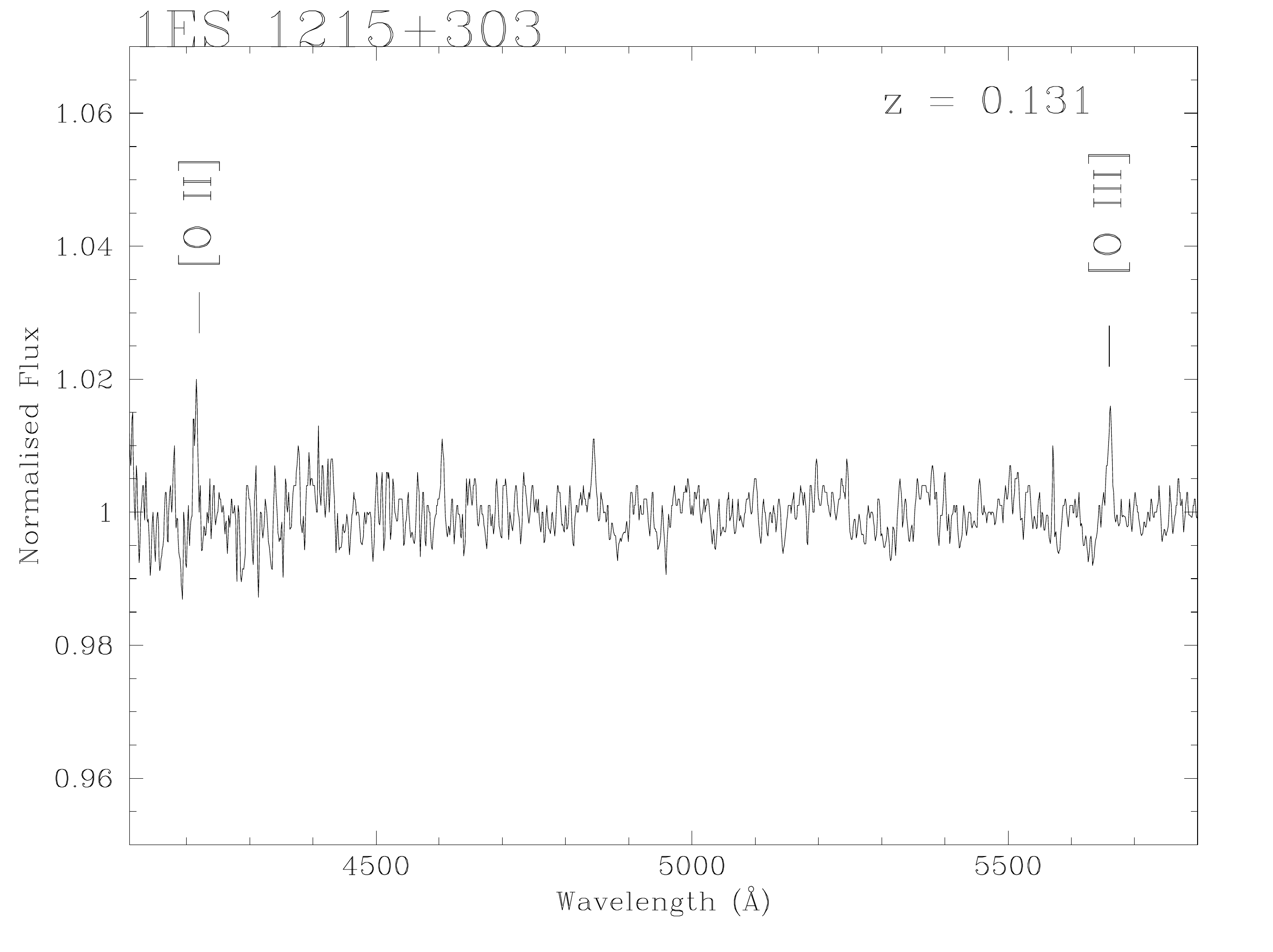}
   \includegraphics[width=0.4\textwidth, angle=-90]{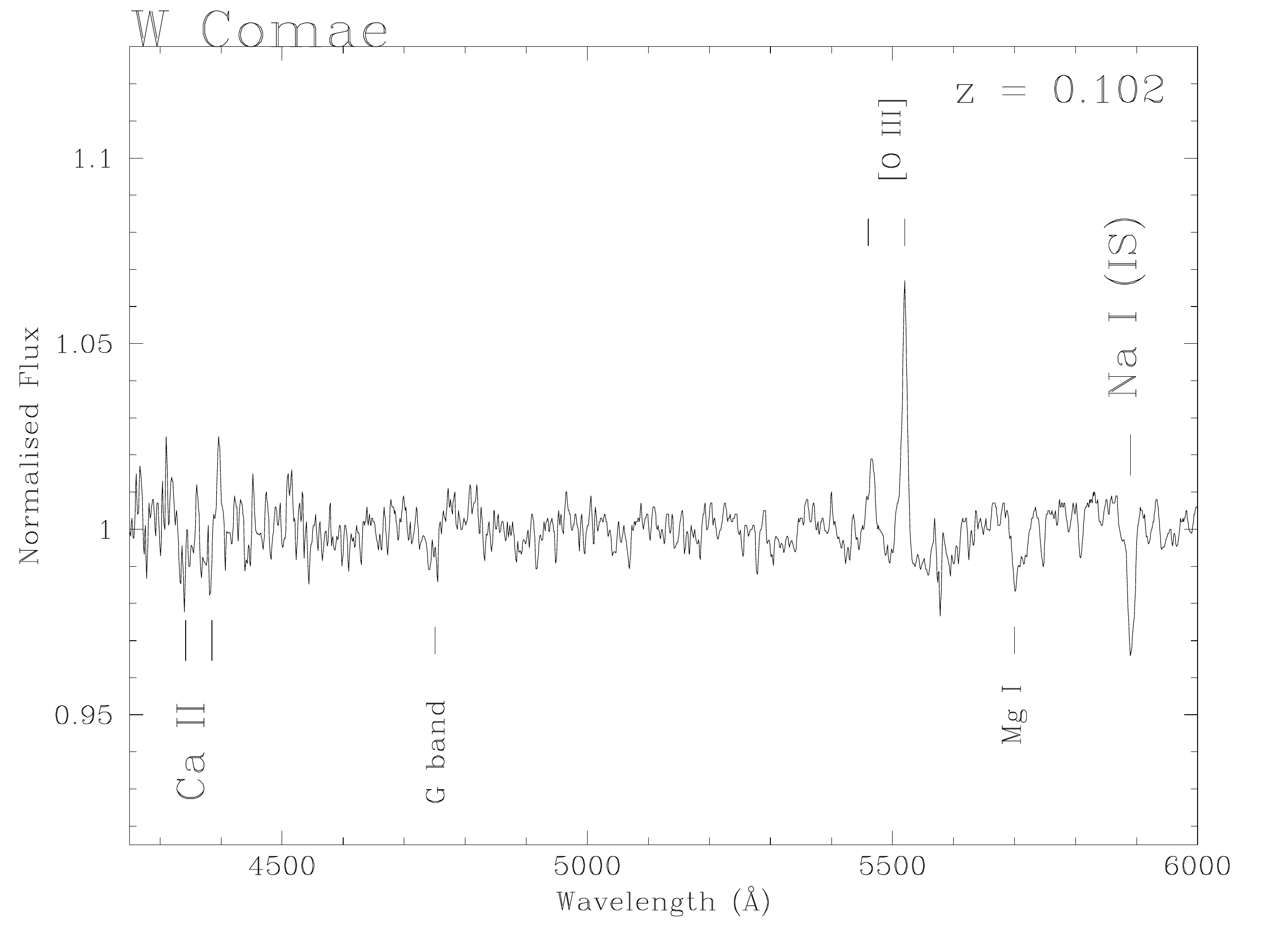} 
   \includegraphics[width=0.4\textwidth, angle=-90]{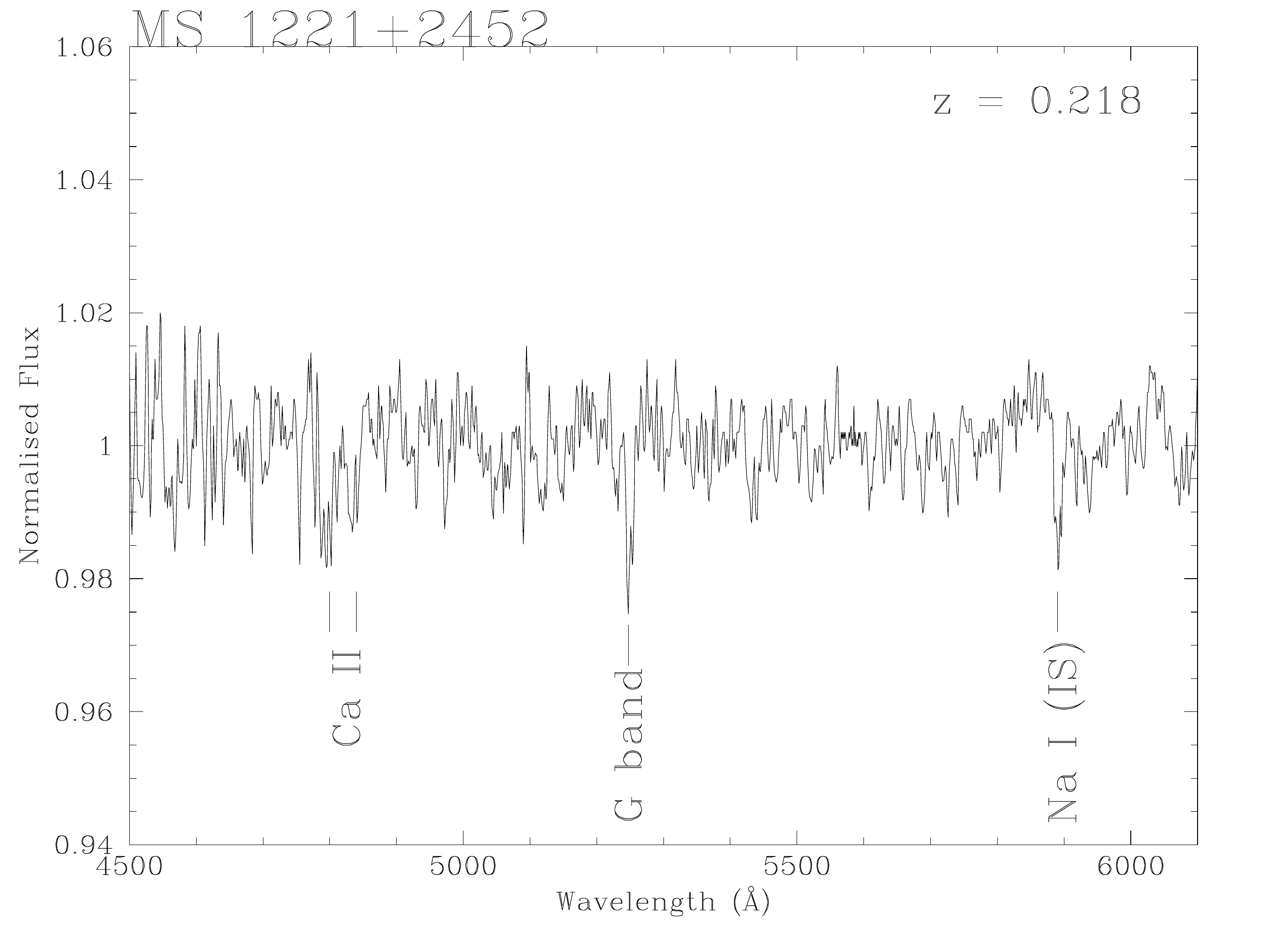}
   \includegraphics[width=0.4\textwidth, angle=-90]{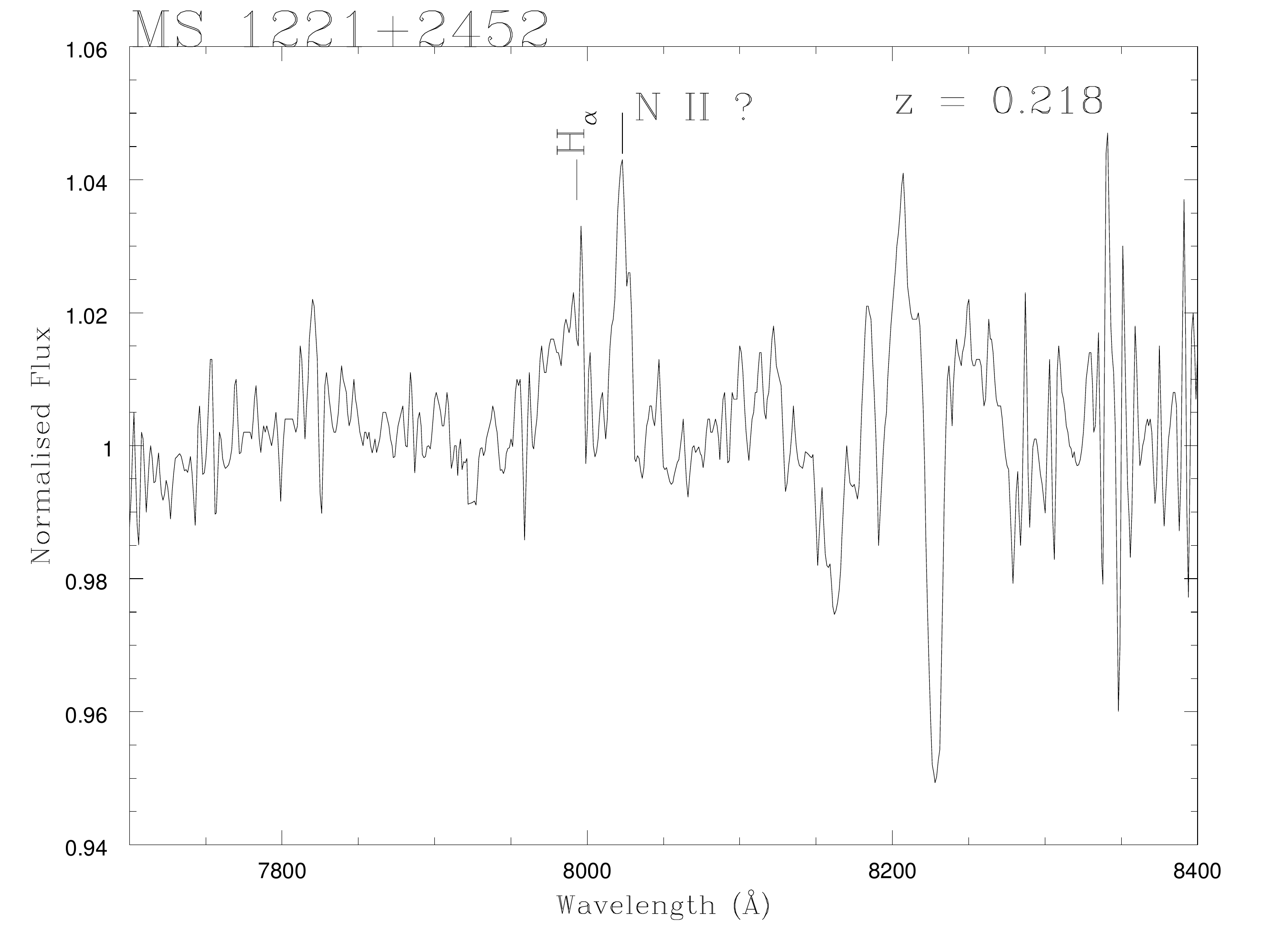}
   \includegraphics[width=0.4\textwidth, angle=-90]{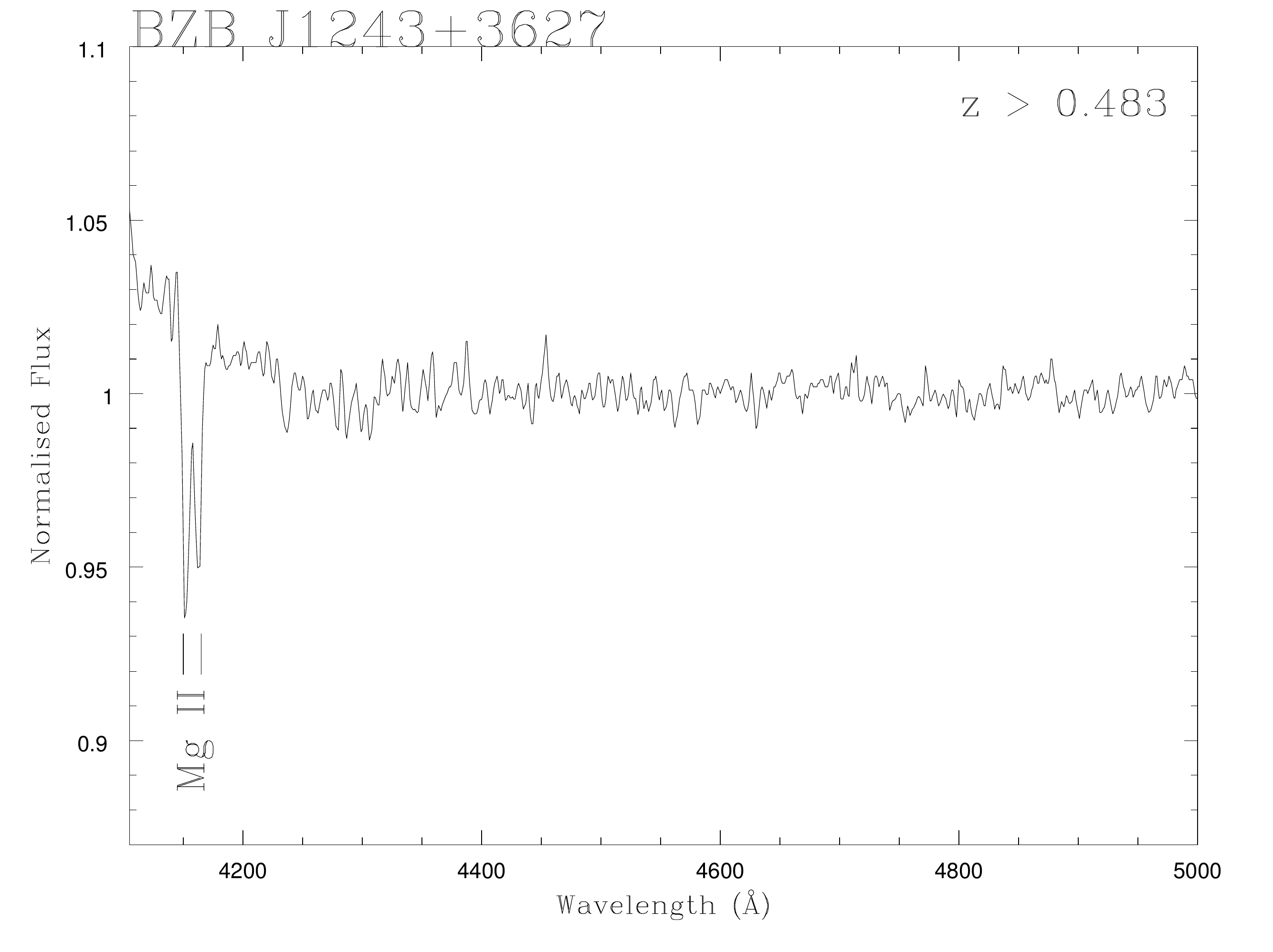}
\caption{Close-up of the normalized spectra around the detected spectral features of the TeV sources and TeV-candidates obtained at GTC. Main telluric bands are indicated as $\oplus$, spectral lines are marked by line identification. } 
   \label{fig:spectraCU}
\end{figure*}

\setcounter{figure}{4}
\begin{figure*}
    \includegraphics[width=0.4\textwidth, angle=-90]{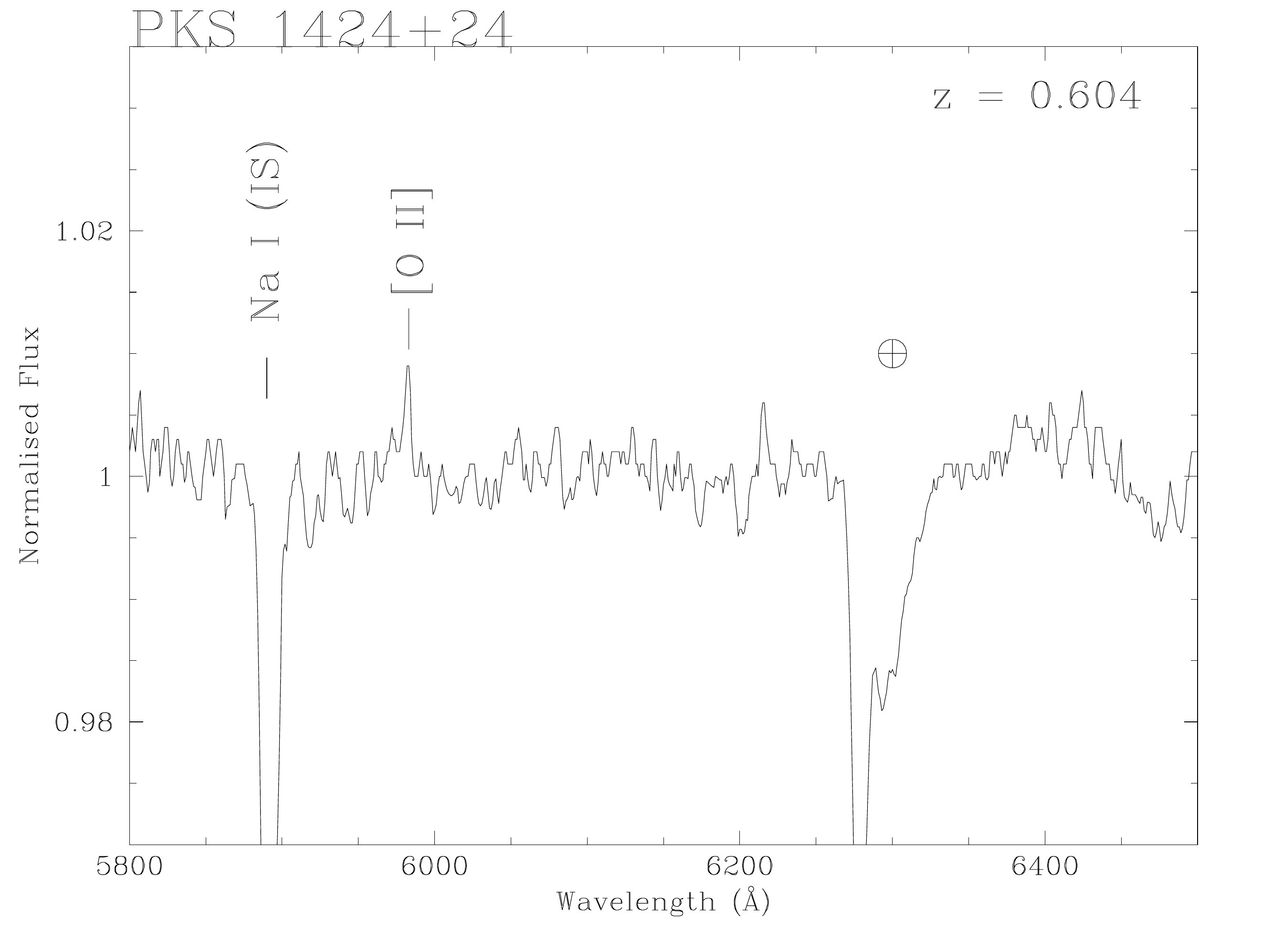}
    \includegraphics[width=0.4\textwidth, angle=-90]{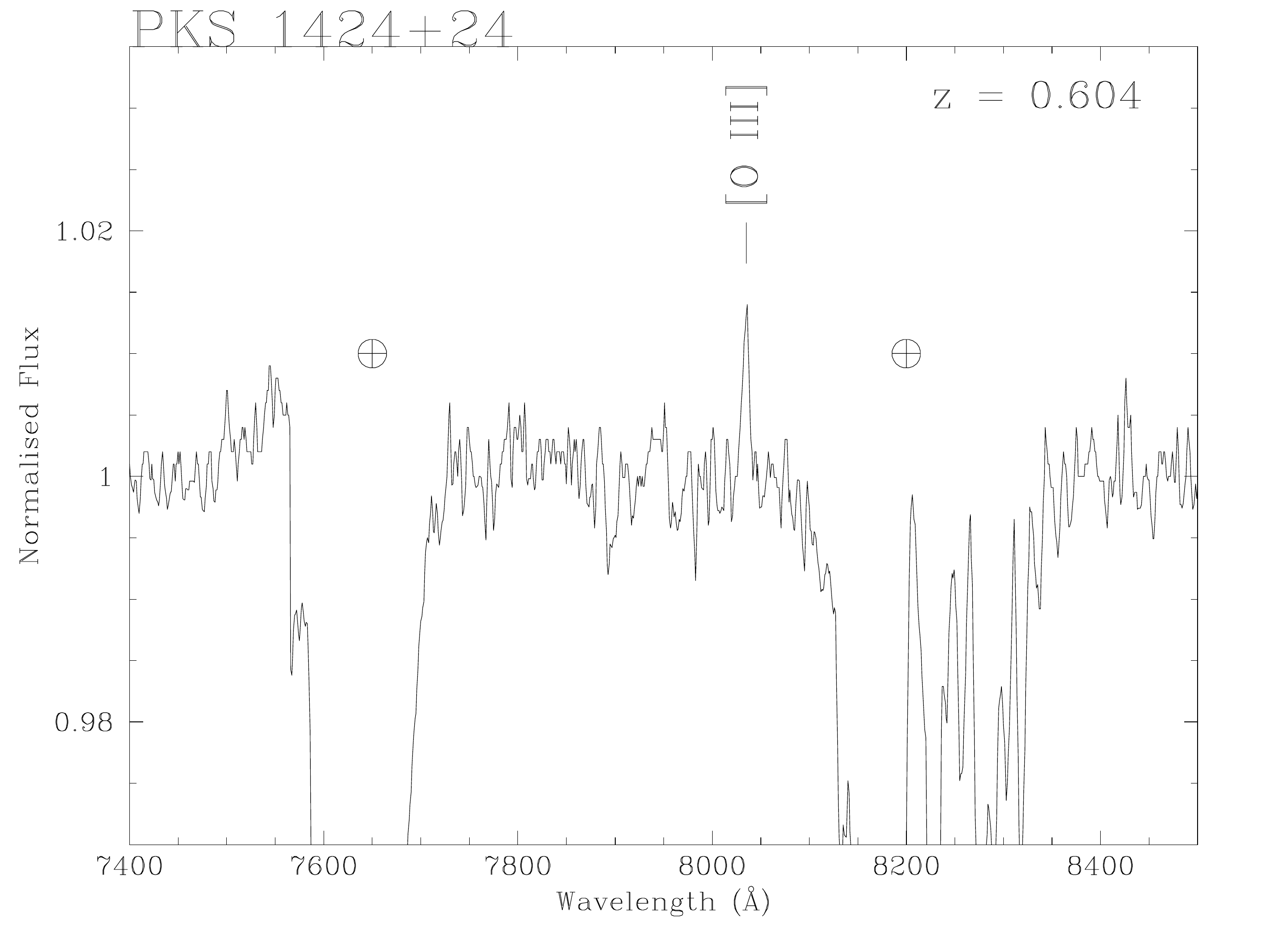}
    \includegraphics[width=0.4\textwidth, angle=-90]{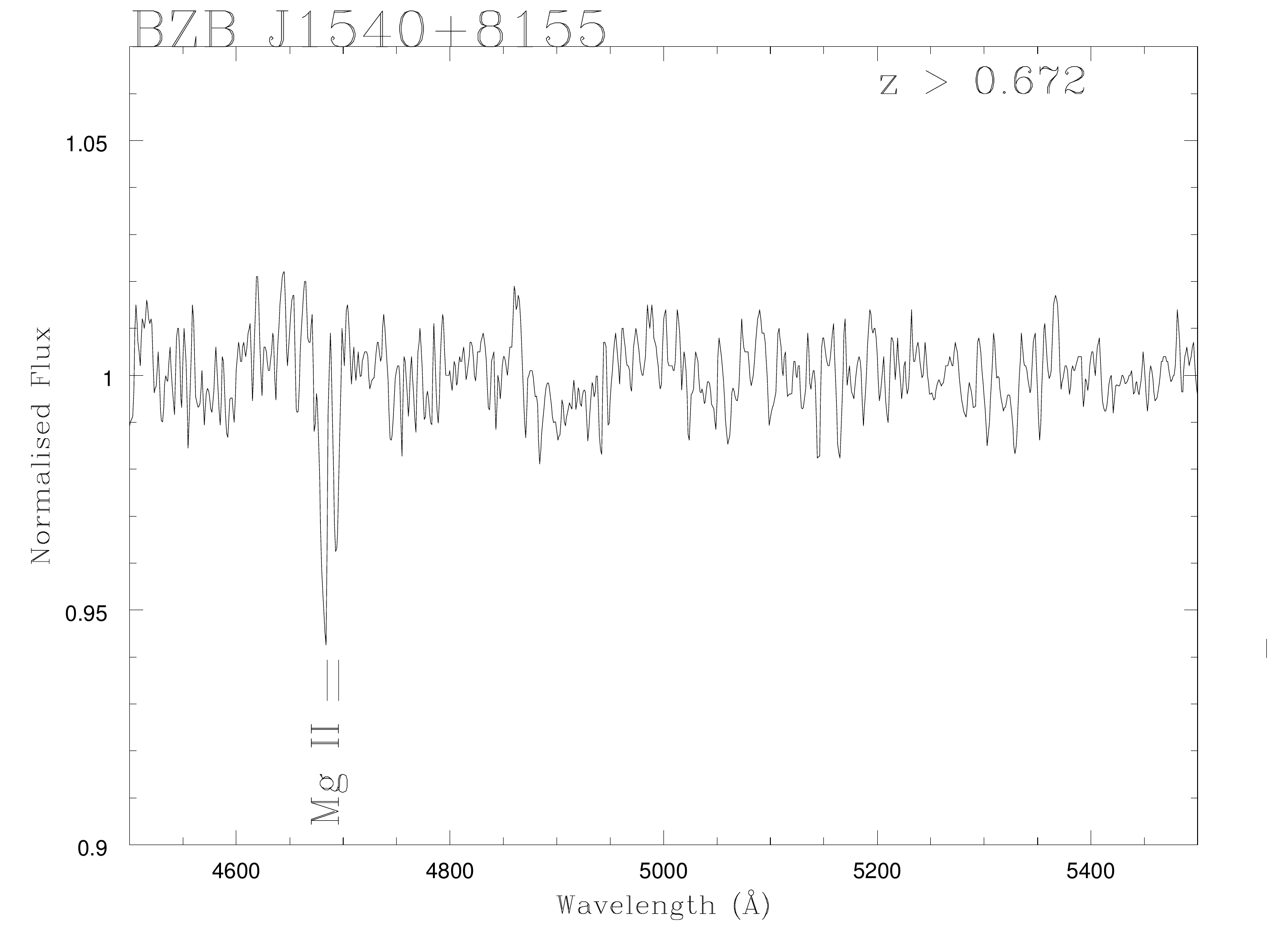}
   \includegraphics[width=0.4\textwidth, angle=-90]{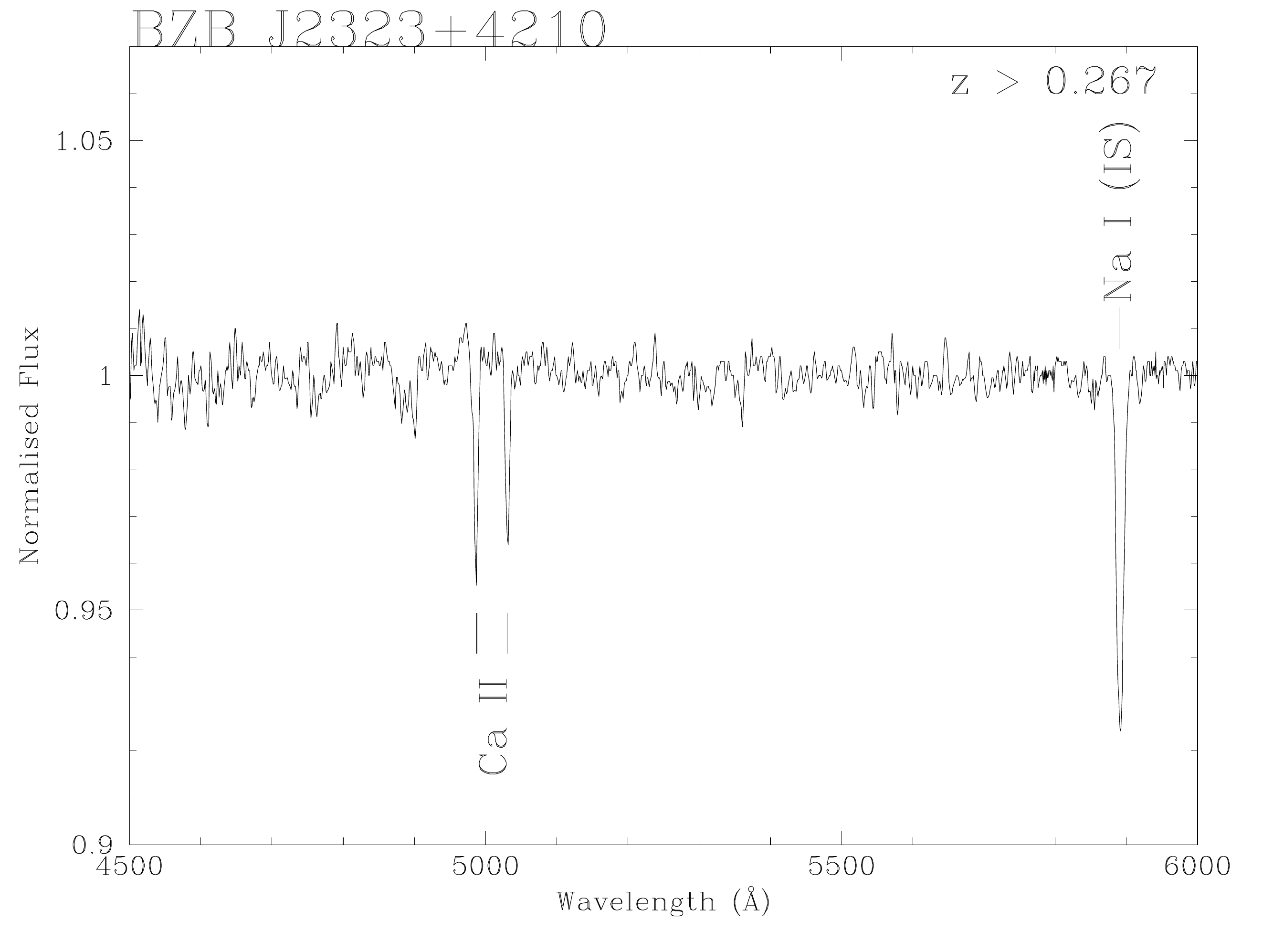}
  \caption{Continued.} 
  \end{figure*}

\clearpage
\newpage

\bibliographystyle{aasjournal}
\bibliography{tevbiblio}

\begin{thebibliography}{}
\expandafter\ifx\csname natexlab\endcsname\relax\def\natexlab#1{#1}\fi

\bibitem[{{Ahnen} {et~al.}(2016){Ahnen}, {Ansoldi}, {Antonelli}, {Antoranz},
  {Arcaro}, {Babic}, {Banerjee}, {Bangale}, {Barres de Almeida}, {Barrio},
  {Becerra Gonz{\'a}lez}, {Bednarek}, {Bernardini}, {Berti}, {Biasuzzi},
  {Biland}, {Blanch}, {Bonnefoy}, {Bonnoli}, {Borracci}, {Bretz}, {Buson},
  {Carosi}, {Chatterjee}, {Clavero}, {Colin}, {Colombo}, {Contreras},
  {Cortina}, {Covino}, {Da Vela}, {Dazzi}, {De Angelis}, {De Lotto}, {de
  O{\~n}a Wilhelmi}, {Di Pierro}, {Doert}, {Dom{\'{\i}}nguez}, {Dominis
  Prester}, {Dorner}, {Doro}, {Einecke}, {Eisenacher Glawion}, {Elsaesser},
  {Engelkemeier}, {Fallah Ramazani}, {Fern{\'a}ndez-Barral}, {Fidalgo},
  {Fonseca}, {Font}, {Frantzen}, {Fruck}, {Galindo}, {Garc{\'{\i}}a L{\'o}pez},
  {Garczarczyk}, {Garrido Terrats}, {Gaug}, {Giammaria}, {Godinovi{\'c}},
  {Gora}, {Guberman}, {Hadasch}, {Hahn}, {Hayashida}, {Herrera}, {Hose},
  {Hrupec}, {Hughes}, {Idec}, {Kodani}, {Konno}, {Kubo}, {Kushida}, {La
  Barbera}, {Lelas}, {Lindfors}, {Lombardi}, {Longo}, {L{\'o}pez},
  {L{\'o}pez-Coto}, {Majumdar}, {Makariev}, {Mallot}, {Maneva}, {Manganaro},
  {Mannheim}, {Maraschi}, {Marcote}, {Mariotti}, {Mart{\'{\i}}nez}, {Mazin},
  {Menzel}, {Miranda}, {Mirzoyan}, {Moralejo}, {Moretti}, {Nakajima},
  {Neustroev}, {Niedzwiecki}, {Nievas Rosillo}, {Nilsson}, {Nishijima}, {Noda},
  {Nogu{\'e}s}, {Paiano}, {Palacio}, {Palatiello}, {Paneque}, {Paoletti},
  {Paredes}, {Paredes-Fortuny}, {Pedaletti}, {Peresano}, {Perri}, {Persic},
  {Poutanen}, {Prada Moroni}, {Prandini}, {Puljak}, {Garcia}, {Reichardt},
  {Rhode}, {Rib{\'o}}, {Rico}, {Saito}, {Satalecka}, {Schroeder}, {Schweizer},
  {Shore}, {Sillanp{\"a}{\"a}}, {Sitarek}, {Snidaric}, {Sobczynska},
  {Stamerra}, {Strzys}, {Suri{\'c}}, {Takalo}, {Tavecchio}, {Temnikov},
  {Terzi{\'c}}, {Tescaro}, {Teshima}, {Torres}, {Toyama}, {Treves}, {Vanzo},
  {Verguilov}, {Vovk}, {Ward}, {Will}, {Wu}, {Zanin}, \&
  {Desiante}}]{ahnen2016}
{Ahnen}, {Ansoldi}, S., {Antonelli}, L.~A., {et~al.} 2016, ArXiv e-prints,
  arXiv:1609.01095

\bibitem[{{{\'A}lvarez Crespo} {et~al.}(2016){{\'A}lvarez Crespo}, {Masetti},
  {Ricci}, {Landoni}, {Pati{\~n}o-{\'A}lvarez}, {Massaro}, {D'Abrusco},
  {Paggi}, {Chavushyan}, {Jim{\'e}nez-Bail{\'o}n}, {Torrealba}, {Latronico},
  {La Franca}, {Smith}, \& {Tosti}}]{crespo2016}
{{\'A}lvarez Crespo}, N., {Masetti}, N., {Ricci}, F., {et~al.} 2016, \aj, 151,
  32

\bibitem[{{Archambault} {et~al.}(2013){Archambault}, {Arlen}, {Aune}, {Behera},
  {Beilicke}, {Benbow}, {Bird}, {Bouvier}, {Buckley}, {Bugaev}, {Byrum},
  {Cesarini}, {Ciupik}, {Connolly}, {Cui}, {Errando}, {Falcone}, {Federici},
  {Feng}, {Finley}, {Fortson}, {Furniss}, {Galante}, {Gall}, {Gillanders},
  {Griffin}, {Grube}, {Gyuk}, {Hanna}, {Holder}, {Hughes}, {Humensky},
  {Kaaret}, {Kertzman}, {Khassen}, {Kieda}, {Krawczynski}, {Krennrich},
  {Kumar}, {Lang}, {Madhavan}, {Maier}, {Majumdar}, {McArthur}, {McCann},
  {Millis}, {Moriarty}, {Mukherjee}, {O'Faol{\'a}in de Bhr{\'o}ithe}, {Ong},
  {Otte}, {Park}, {Perkins}, {Pohl}, {Popkow}, {Prokoph}, {Quinn}, {Ragan},
  {Reyes}, {Reynolds}, {Richards}, {Roache}, {Saxon}, {Sembroski}, {Smith},
  {Staszak}, {Telezhinsky}, {Theiling}, {Varlotta}, {Vassiliev}, {Vincent},
  {Wakely}, {Weekes}, {Weinstein}, {Welsing}, {Williams}, {Zitzer}, {VERITAS
  Collaboration}, {B{\"o}ttcher}, {Fegan}, {Fortin}, {Halpern}, {Kovalev},
  {Lister}, {Liu}, {Pushkarev}, \& {Smith}}]{archambault2013}
{Archambault}, S., {Arlen}, T., {Aune}, T., {et~al.} 2013, \apj, 776, 69

\bibitem[{{Bach} {et~al.}(2007){Bach}, {Raiteri}, {Villata}, {Fuhrmann},
  {Buemi}, {Larionov}, {Letog}, {Arkharov}, {Coloma}, {di Paola}, {Dolci},
  {Efimova}, {Forn{\'e}}, {Ibrahimov}, {Hagen-Thorn}, {Konstantinova},
  {Kopatskaya}, {Lanteri}, {Kurtanidze}, {Maccaferri}, {Nikolashvili},
  {Orlati}, {Ros}, {Tosti}, {Trigilio}, \& {Umana}}]{bach2007}
{Bach}, U., {Raiteri}, C.~M., {Villata}, M., {et~al.} 2007, \aap, 464, 175

\bibitem[{{Bade} {et~al.}(1998){Bade}, {Beckmann}, {Douglas}, {Barthel},
  {Engels}, {Cordis}, {Nass}, \& {Voges}}]{bade1998}
{Bade}, N., {Beckmann}, V., {Douglas}, N.~G., {et~al.} 1998, \aap, 334, 459

\bibitem[{{Browne} {et~al.}(1993){Browne}, {Patnaik}, {Walsh}, \&
  {Wilkinson}}]{browne1993}
{Browne}, I.~W.~A., {Patnaik}, A.~R., {Walsh}, D., \& {Wilkinson}, P.~N. 1993,
  \mnras, 263, L32

\bibitem[{{Cardelli} {et~al.}(1989){Cardelli}, {Clayton}, \&
  {Mathis}}]{cardelli1989}
{Cardelli}, J.~A., {Clayton}, G.~C., \& {Mathis}, J.~S. 1989, \apj, 345, 245

\bibitem[{{Carilli} {et~al.}(1993){Carilli}, {Rupen}, \& {Yanny}}]{carilli1993}
{Carilli}, C.~L., {Rupen}, M.~P., \& {Yanny}, B. 1993, \apjl, 412, L59

\bibitem[{{Cepa} {et~al.}(2003){Cepa}, {Aguiar-Gonzalez}, {Bland-Hawthorn},
  {Castaneda}, {Cobos}, {Correa}, {Espejo}, {Fragoso-Lopez}, {Fuentes},
  {Gigante}, {Gonzalez}, {Gonzalez-Escalera}, {Gonzalez-Serrano},
  {Joven-Alvarez}, {Lopez-Ruiz}, {Militello}, {Cano}, {Perez}, {Perez},
  {Rasilla}, {Sanchez}, \& {Tejada}}]{cepa2003}
{Cepa}, J., {Aguiar-Gonzalez}, M., {Bland-Hawthorn}, J., {et~al.} 2003, in
  Proc. SPIE, Vol. 4841, -, 1739--1749

\bibitem[{{Cohen} {et~al.}(2003){Cohen}, {Lawrence}, \&
  {Blandford}}]{cohen2003}
{Cohen}, J.~G., {Lawrence}, C.~R., \& {Blandford}, R.~D. 2003, \apj, 583, 67

\bibitem[{{Danforth} {et~al.}(2013){Danforth}, {Nalewajko}, {France}, \&
  {Keeney}}]{danforth2013}
{Danforth}, C.~W., {Nalewajko}, K., {France}, K., \& {Keeney}, B.~A. 2013,
  \apj, 764, 57

\bibitem[{{Falomo} \& {Kotilainen}(1999)}]{falomo1999}
{Falomo}, R., \& {Kotilainen}, J.~K. 1999, \aap, 352, 85

\bibitem[{{Falomo} {et~al.}(2014){Falomo}, {Pian}, \& {Treves}}]{falomo2014}
{Falomo}, R., {Pian}, E., \& {Treves}, A. 2014, \aapr, 22, 73

\bibitem[{{Finke} {et~al.}(2008){Finke}, {Shields}, {B{\"o}ttcher}, \&
  {Basu}}]{finke2008}
{Finke}, J.~D., {Shields}, J.~C., {B{\"o}ttcher}, M., \& {Basu}, S. 2008, \aap,
  477, 513

\bibitem[{{Fischer} {et~al.}(1998){Fischer}, {Hasinger}, {Schwope}, {Brunner},
  {Boller}, {Tr{\"u}mper}, {Voges}, \& {Neizvestnyj}}]{fischer1998}
{Fischer}, J.-U., {Hasinger}, G., {Schwope}, A.~D., {et~al.} 1998,
  Astronomische Nachrichten, 319, 347

\bibitem[{{Fleming} {et~al.}(1993){Fleming}, {Green}, {Jannuzi}, {Liebert},
  {Smith}, \& {Fink}}]{fleming1993}
{Fleming}, T.~A., {Green}, R.~F., {Jannuzi}, B.~T., {et~al.} 1993, \aj, 106,
  1729

\bibitem[{{Furniss} {et~al.}(2013){Furniss}, {Williams}, {Danforth},
  {Fumagalli}, {Prochaska}, {Primack}, {Urry}, {Stocke}, {Filippenko}, \&
  {Neely}}]{furniss2013}
{Furniss}, A., {Williams}, D.~A., {Danforth}, C., {et~al.} 2013, \apjl, 768,
  L31

\bibitem[{{Giommi} {et~al.}(2005){Giommi}, {Piranomonte}, {Perri}, \&
  {Padovani}}]{giommi2005}
{Giommi}, P., {Piranomonte}, S., {Perri}, M., \& {Padovani}, P. 2005, \aap,
  434, 385

\bibitem[{{Healey} {et~al.}(2008){Healey}, {Romani}, {Cotter}, {Michelson},
  {Schlafly}, {Readhead}, {Giommi}, {Chaty}, {Grenier}, \&
  {Weintraub}}]{healey2008}
{Healey}, S.~E., {Romani}, R.~W., {Cotter}, G., {et~al.} 2008, \apjs, 175, 97

\bibitem[{{Henstock} {et~al.}(1997){Henstock}, {Browne}, {Wilkinson}, \&
  {McMahon}}]{henstock1997}
{Henstock}, D.~R., {Browne}, I.~W.~A., {Wilkinson}, P.~N., \& {McMahon}, R.~G.
  1997, \mnras, 290, 380

\bibitem[{{Jackson} {et~al.}(2000){Jackson}, {Xanthopoulos}, \&
  {Browne}}]{jackson2000}
{Jackson}, N., {Xanthopoulos}, E., \& {Browne}, I.~W.~A. 2000, \mnras, 311, 389

\bibitem[{{Kinney} {et~al.}(1996){Kinney}, {Calzetti}, {Bohlin}, {McQuade},
  {Storchi-Bergmann}, \& {Schmitt}}]{kinney1996}
{Kinney}, A.~L., {Calzetti}, D., {Bohlin}, R.~C., {et~al.} 1996, \apj, 467, 38

\bibitem[{{Kotilainen} {et~al.}(2011){Kotilainen}, {Hyv{\"o}nen}, {Falomo},
  {Treves}, \& {Uslenghi}}]{kotilainen2011}
{Kotilainen}, J.~K., {Hyv{\"o}nen}, T., {Falomo}, R., {Treves}, A., \&
  {Uslenghi}, M. 2011, \aap, 534, L2

\bibitem[{{Landoni} {et~al.}(2014){Landoni}, {Falomo}, {Treves}, \&
  {Sbarufatti}}]{landoni2014}
{Landoni}, M., {Falomo}, R., {Treves}, A., \& {Sbarufatti}, B. 2014, \aap, 570,
  A126

\bibitem[{{Landoni} {et~al.}(2013){Landoni}, {Falomo}, {Treves}, {Sbarufatti},
  {Barattini}, {Decarli}, \& {Kotilainen}}]{landoni2013}
{Landoni}, M., {Falomo}, R., {Treves}, A., {et~al.} 2013, \aj, 145, 114

\bibitem[{{Landoni} {et~al.}(2012){Landoni}, {Falomo}, {Treves}, {Sbarufatti},
  {Decarli}, {Tavecchio}, \& {Kotilainen}}]{landoni2012}
---. 2012, \aap, 543, A116

\bibitem[{{Landoni} {et~al.}(2015){Landoni}, {Falomo}, {Treves}, {Scarpa}, \&
  {Reverte Pay{\'a}}}]{landoni2015}
{Landoni}, M., {Falomo}, R., {Treves}, A., {Scarpa}, R., \& {Reverte Pay{\'a}},
  D. 2015, \aj, 150, 181

\bibitem[{{Laurent-Muehleisen} {et~al.}(1998){Laurent-Muehleisen}, {Kollgaard},
  {Ciardullo}, {Feigelson}, {Brinkmann}, \& {Siebert}}]{laurent1998}
{Laurent-Muehleisen}, S.~A., {Kollgaard}, R.~I., {Ciardullo}, R., {et~al.}
  1998, \apjs, 118, 127

\bibitem[{{Lawrence} {et~al.}(1986){Lawrence}, {Pearson}, {Readhead}, \&
  {Unwin}}]{lawrence1986}
{Lawrence}, C.~R., {Pearson}, T.~J., {Readhead}, A.~C.~S., \& {Unwin}, S.~C.
  1986, \aj, 91, 494

\bibitem[{{Marcha} {et~al.}(1996){Marcha}, {Browne}, {Impey}, \&
  {Smith}}]{marcha1996}
{Marcha}, M.~J.~M., {Browne}, I.~W.~A., {Impey}, C.~D., \& {Smith}, P.~S. 1996,
  \mnras, 281, 425

\bibitem[{{Massaro} {et~al.}(2009){Massaro}, {Giommi}, {Leto}, {Marchegiani},
  {Maselli}, {Perri}, {Piranomonte}, \& {Sclavi}}]{bzcat2009}
{Massaro}, E., {Giommi}, P., {Leto}, C., {et~al.} 2009, \aap, 495, 691

\bibitem[{{Massaro} {et~al.}(2015){Massaro}, {Landoni}, {D'Abrusco},
  {Milisavljevic}, {Paggi}, {Masetti}, {Smith}, \& {Tosti}}]{massaro2015a}
{Massaro}, F., {Landoni}, M., {D'Abrusco}, R., {et~al.} 2015, \aap, 575, A124

\bibitem[{{Massaro} {et~al.}(2014){Massaro}, {Masetti}, {D'Abrusco}, {Paggi},
  \& {Funk}}]{massaro2014}
{Massaro}, F., {Masetti}, N., {D'Abrusco}, R., {Paggi}, A., \& {Funk}, S. 2014,
  \aj, 148, 66

\bibitem[{{Massaro} {et~al.}(2013){Massaro}, {Paggi}, {Errando}, {D'Abrusco},
  {Masetti}, {Tosti}, \& {Funk}}]{massaro2013}
{Massaro}, F., {Paggi}, A., {Errando}, M., {et~al.} 2013, \apjs, 207, 16

\bibitem[{{Meisner} \& {Romani}(2010)}]{meisner2010}
{Meisner}, A.~M., \& {Romani}, R.~W. 2010, \apj, 712, 14

\bibitem[{{Miller} {et~al.}(1978){Miller}, {French}, \& {Hawley}}]{miller1978}
{Miller}, J.~S., {French}, H.~B., \& {Hawley}, S.~A. 1978, in BL Lac Objects,
  ed. A.~M. {Wolfe}, 176--187

\bibitem[{{Morris} {et~al.}(1991){Morris}, {Stocke}, {Gioia}, {Schild},
  {Wolter}, {Maccacaro}, \& {della Ceca}}]{morris1991}
{Morris}, S.~L., {Stocke}, J.~T., {Gioia}, I.~M., {et~al.} 1991, \apj, 380, 49

\bibitem[{{Nass} {et~al.}(1996){Nass}, {Bade}, {Kollgaard},
  {Laurent-Muehleisen}, {Reimers}, \& {Voges}}]{nass1996}
{Nass}, P., {Bade}, N., {Kollgaard}, R.~I., {et~al.} 1996, \aap, 309, 419

\bibitem[{{Nilsson} {et~al.}(2003){Nilsson}, {Pursimo}, {Heidt}, {Takalo},
  {Sillanp{\"a}{\"a}}, \& {Brinkmann}}]{nilsson2003}
{Nilsson}, K., {Pursimo}, T., {Heidt}, J., {et~al.} 2003, \aap, 400, 95

\bibitem[{{Nilsson} {et~al.}(2008){Nilsson}, {Pursimo}, {Sillanp{\"a}{\"a}},
  {Takalo}, \& {Lindfors}}]{nilsson2008}
{Nilsson}, K., {Pursimo}, T., {Sillanp{\"a}{\"a}}, A., {Takalo}, L.~O., \&
  {Lindfors}, E. 2008, \aap, 487, L29

\bibitem[{{Nilsson} {et~al.}(2012){Nilsson}, {Pursimo}, {Villforth},
  {Lindfors}, {Takalo}, \& {Sillanp{\"a}{\"a}}}]{nilsson2012}
{Nilsson}, K., {Pursimo}, T., {Villforth}, C., {et~al.} 2012, \aap, 547, A1

\bibitem[{{Paiano} {et~al.}(2016){Paiano}, {Landoni}, {Falomo}, {Scarpa}, \&
  {Treves}}]{paiano2016}
{Paiano}, S., {Landoni}, M., {Falomo}, R., {Scarpa}, R., \& {Treves}, A. 2016,
  \mnras, 458, 2836

\bibitem[{{Patnaik} {et~al.}(1993){Patnaik}, {Browne}, {King}, {Muxlow},
  {Walsh}, \& {Wilkinson}}]{patnaik1993}
{Patnaik}, A.~R., {Browne}, I.~W.~A., {King}, L.~J., {et~al.} 1993, \mnras,
  261, 435

\bibitem[{{Perlman} {et~al.}(1996){Perlman}, {Stocke}, {Schachter}, {Elvis},
  {Ellingson}, {Urry}, {Potter}, {Impey}, \& {Kolchinsky}}]{perlman1996}
{Perlman}, E.~S., {Stocke}, J.~T., {Schachter}, J.~F., {et~al.} 1996, \apjs,
  104, 251

\bibitem[{{Piranomonte} {et~al.}(2007){Piranomonte}, {Perri}, {Giommi},
  {Landt}, \& {Padovani}}]{piranomonte2007}
{Piranomonte}, S., {Perri}, M., {Giommi}, P., {Landt}, H., \& {Padovani}, P.
  2007, \aap, 470, 787

\bibitem[{{Plotkin} {et~al.}(2008){Plotkin}, {Anderson}, {Hall}, {Margon},
  {Voges}, {Schneider}, {Stinson}, \& {York}}]{plotkin2008}
{Plotkin}, R.~M., {Anderson}, S.~F., {Hall}, P.~B., {et~al.} 2008, \aj, 135,
  2453

\bibitem[{{Plotkin} {et~al.}(2010){Plotkin}, {Anderson}, {Brandt},
  {Diamond-Stanic}, {Fan}, {Hall}, {Kimball}, {Richmond}, {Schneider},
  {Shemmer}, {Voges}, {York}, {Bahcall}, {Snedden}, {Bizyaev}, {Brewington},
  {Malanushenko}, {Malanushenko}, {Oravetz}, {Pan}, \& {Simmons}}]{plotkin2010}
{Plotkin}, R.~M., {Anderson}, S.~F., {Brandt}, W.~N., {et~al.} 2010, \aj, 139,
  390

\bibitem[{{Rector} \& {Stocke}(2001)}]{rector2001}
{Rector}, T.~A., \& {Stocke}, J.~T. 2001, \aj, 122, 565

\bibitem[{{Rector} {et~al.}(2000){Rector}, {Stocke}, {Perlman}, {Morris}, \&
  {Gioia}}]{rector2000}
{Rector}, T.~A., {Stocke}, J.~T., {Perlman}, E.~S., {Morris}, S.~L., \&
  {Gioia}, I.~M. 2000, \aj, 120, 1626

\bibitem[{{Ricci} {et~al.}(2015){Ricci}, {Massaro}, {Landoni}, {D'Abrusco},
  {Milisavljevic}, {Stern}, {Masetti}, {Paggi}, {Smith}, \&
  {Tosti}}]{ricci2015}
{Ricci}, F., {Massaro}, F., {Landoni}, M., {et~al.} 2015, \aj, 149, 160

\bibitem[{{Rovero} {et~al.}(2016){Rovero}, {Muriel}, {Donzelli}, \&
  {Pichel}}]{rovero2016}
{Rovero}, A.~C., {Muriel}, H., {Donzelli}, C., \& {Pichel}, A. 2016, \aap, 589,
  A92

\bibitem[{{Sbarufatti} {et~al.}(2008){Sbarufatti}, {Ciprini}, {Kotilainen},
  {Decarli}, {Treves}, {Veronesi}, \& {Falomo}}]{sbarufatti2008}
{Sbarufatti}, B., {Ciprini}, S., {Kotilainen}, J., {et~al.} 2008, ArXiv
  e-prints, arXiv:0810.3563

\bibitem[{{Sbarufatti} {et~al.}(2006{\natexlab{a}}){Sbarufatti}, {Falomo},
  {Treves}, \& {Kotilainen}}]{sbarufatti2006b}
{Sbarufatti}, B., {Falomo}, R., {Treves}, A., \& {Kotilainen}, J.
  2006{\natexlab{a}}, \aap, 457, 35

\bibitem[{{Sbarufatti} {et~al.}(2005{\natexlab{a}}){Sbarufatti}, {Treves}, \&
  {Falomo}}]{sbarufatti2005a}
{Sbarufatti}, B., {Treves}, A., \& {Falomo}, R. 2005{\natexlab{a}}, \apj, 635,
  173

\bibitem[{{Sbarufatti} {et~al.}(2005{\natexlab{b}}){Sbarufatti}, {Treves},
  {Falomo}, {Heidt}, {Kotilainen}, \& {Scarpa}}]{sbarufatti2005b}
{Sbarufatti}, B., {Treves}, A., {Falomo}, R., {et~al.} 2005{\natexlab{b}}, \aj,
  129, 559

\bibitem[{{Sbarufatti} {et~al.}(2006{\natexlab{b}}){Sbarufatti}, {Treves},
  {Falomo}, {Heidt}, {Kotilainen}, \& {Scarpa}}]{sbarufatti2006a}
---. 2006{\natexlab{b}}, \aj, 132, 1

\bibitem[{{Scarpa} {et~al.}(1999){Scarpa}, {Urry}, {Falomo}, {Pesce},
  {Webster}, {O'Dowd}, \& {Treves}}]{scarpa1999}
{Scarpa}, R., {Urry}, C.~M., {Falomo}, R., {et~al.} 1999, \apj, 521, 134

\bibitem[{{Scarpa} {et~al.}(2000){Scarpa}, {Urry}, {Padovani}, {Calzetti}, \&
  {O'Dowd}}]{scarpa2000}
{Scarpa}, R., {Urry}, C.~M., {Padovani}, P., {Calzetti}, D., \& {O'Dowd}, M.
  2000, \apj, 544, 258

\bibitem[{{Schachter} {et~al.}(1993){Schachter}, {Stocke}, {Perlman}, {Elvis},
  {Remillard}, {Granados}, {Luu}, {Huchra}, {Humphreys}, {Urry}, \&
  {Wallin}}]{schachter1993}
{Schachter}, J.~F., {Stocke}, J.~T., {Perlman}, E., {et~al.} 1993, \apj, 412,
  541

\bibitem[{{Shaw} {et~al.}(2013){Shaw}, {Romani}, {Cotter}, {Healey},
  {Michelson}, {Readhead}, {Richards}, {Max-Moerbeck}, {King}, \&
  {Potter}}]{shaw2013a}
{Shaw}, M.~S., {Romani}, R.~W., {Cotter}, G., {et~al.} 2013, \apj, 764, 135

\bibitem[{{Stickel} \& {Kuhr}(1993)}]{stickel1993}
{Stickel}, M., \& {Kuhr}, H. 1993, \aaps, 101, 521

\bibitem[{{Tavecchio} {et~al.}(2015){Tavecchio}, {Roncadelli}, \&
  {Galanti}}]{tavecchio2015}
{Tavecchio}, F., {Roncadelli}, M., \& {Galanti}, G. 2015, Physics Letters B,
  744, 375

\bibitem[{{Urry} {et~al.}(2000){Urry}, {Scarpa}, {O'Dowd}, {Falomo}, {Pesce},
  \& {Treves}}]{urry2000}
{Urry}, C.~M., {Scarpa}, R., {O'Dowd}, M., {et~al.} 2000, \apj, 532, 816

\bibitem[{{van Dokkum}(2001)}]{lacos}
{van Dokkum}, P.~G. 2001, \pasp, 113, 1420

\bibitem[{{Wei} {et~al.}(1999){Wei}, {Xu}, {Dong}, \& {Hu}}]{wei1999}
{Wei}, J.~Y., {Xu}, D.~W., {Dong}, X.~Y., \& {Hu}, J.~Y. 1999, \aaps, 139, 575

\bibitem[{{Weistrop} {et~al.}(1985){Weistrop}, {Shaffer}, {Hintzen}, \&
  {Romanishin}}]{weistrop1985}
{Weistrop}, D., {Shaffer}, D.~B., {Hintzen}, P., \& {Romanishin}, W. 1985,
  \apj, 292, 614

\bibitem[{{White} {et~al.}(2000){White}, {Becker}, {Gregg},
  {Laurent-Muehleisen}, {Brotherton}, {Impey}, {Petry}, {Foltz}, {Chaffee},
  {Richards}, {Oegerle}, {Helfand}, {McMahon}, \& {Cabanela}}]{white2000}
{White}, R.~L., {Becker}, R.~H., {Gregg}, M.~D., {et~al.} 2000, \apjs, 126, 133

\bibitem[{{Wills} \& {Wills}(1974)}]{wills1974}
{Wills}, B.~J., \& {Wills}, D. 1974, \apjl, 190, L97

\bibitem[{{Wills} \& {Wills}(1976)}]{wills1976}
{Wills}, D., \& {Wills}, B.~J. 1976, \apjs, 31, 143

\bibitem[{{York} {et~al.}(2005){York}, {Jackson}, {Browne}, {Wucknitz}, \&
  {Skelton}}]{york2005}
{York}, T., {Jackson}, N., {Browne}, I.~W.~A., {Wucknitz}, O., \& {Skelton},
  J.~E. 2005, \mnras, 357, 124

\bibitem[{{Zhu} \& {M{\'e}nard}(2013)}]{zhu2013}
{Zhu}, G., \& {M{\'e}nard}, B. 2013, \apj, 770, 130

\end{thebibliography}






\clearpage
\newpage
\clearpage

\appendix
\section{Redshift lower limit of BL Lac objects from host galaxy absorption lines} \label{App}

Given the featureless nature of many BL Lac sources it is of great interest to estimate lower limits of the redshift for these kind of targets and in particular for those that are also emitters (or candidates) at $\gamma$ frequencies.
In the cases where no spectral features are detected the only way to estimate the redshift or a lower limit from optical data is to use the characteristics of the host galaxies. 
A direct method is to use high quality images to detect the surrounding nebulosity, or to asses upper limits of the host brightness and  then derive a redshift lower limit.
Alternatively one can search for the host galaxy features that are heavily obscured by the dominant non thermal emission \citep{sbarufatti2006b} .
In the first case high resolution and deep images are required while in the second one very high SNR spectra of adequate resolution are needed. Here we focus  on this second approach.

Assuming the observed targets have giant elliptical host galaxies of similar luminosity distribution of the  population of BLLs for which the host galaxy has been resolved  it is plausible to assume that they are hosted by massive early type galaxies (see e.g. \cite{falomo2014} and references therein). 
Since the luminosity distribution of these host galaxies is relatively narrow \citep[see e.g.][]{urry2000} it is possible to use the host galaxy luminosity as a sort of {\t standard candle} to evaluate the distance of the objects or to set lower limits in the cases where no signature from the starlight is found (\citet{sbarufatti2006b} ). 
 
\begin{figure}[htbp]  
   \includegraphics[width=0.5\textwidth]{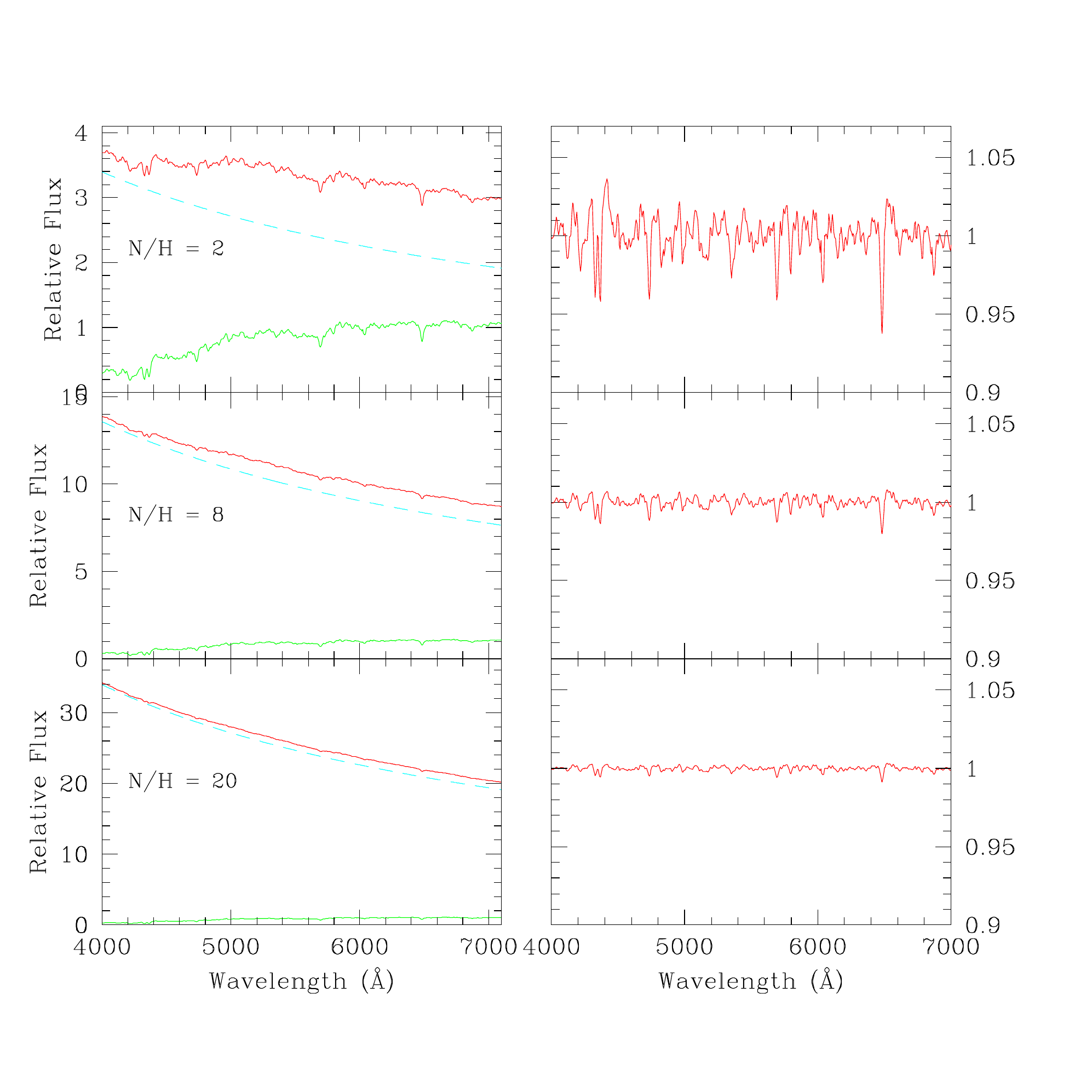}
   \caption{ LEFT: Simulation of the optical spectrum of a BL Lac object at z~$=$~0.2 for different nucleus to host galaxy ratio (N/H). The values of N/H refer to r band. The non thermal component is assumed a power law with spectral index -1. No noise is included. RIGHT: The normalized spectrum of the source (non thermal plus   starlight). The figure 
   illustrates the effect of the non thermal component to dilute the equivalent width of the spectral features.}
   \label{fig1_app}
\end{figure}

In order to estimate lower limits to the redshift of line-less objects we follow and extend the procedure proposed by 
\cite{sbarufatti2006b} for good SNR optical spectra. The basic idea is that under the assumption that the observed spectrum is due to the contribution of a (often dominant) non thermal component, usually described by a power law, and to starlight component from a {\it standard} host galaxy (see example Fig. \ref{fig1_app}), it is possible to set lower limits to their redshift

In fact the effect of the strong non thermal emission  is to dilute the equivalent width of the absorption features of the host galaxy depending on the flux ratio of the two components (non thermal and starlight). 
Using high SNR spectra it is thus possible to set suitable upper limits to the equivalent width of the absorption features from the host galaxy. These limits depend on the SNR and the spectral resolution of the observations  and on the brightness of the source. 

We assume that the underlying host galaxy is a giant elliptical of absolute magnitude M(R)~$=$~-22.9 \cite{sbarufatti2005a} and as spectrum template that of \citet{kinney1996}.

For each observed spectrum we then evaluate the dilution factor of an absorption line (namely H,K of  Ca II 3924, 3970 \AA, G band 4304 \AA , Mg I 5175 \AA )  of this host galaxy as a function of the redshift.
To perform this we took into account both k-correction (using the host galaxy template spectrum) and the starlight flux lost through the slit. The latter term was computed by assuming the host galaxy has a de Vaucouleurs brightness profile and an effective radius Re~$=$~8 kpc. 

\begin{figure}[htbp]  
   \includegraphics[width=0.5\textwidth]{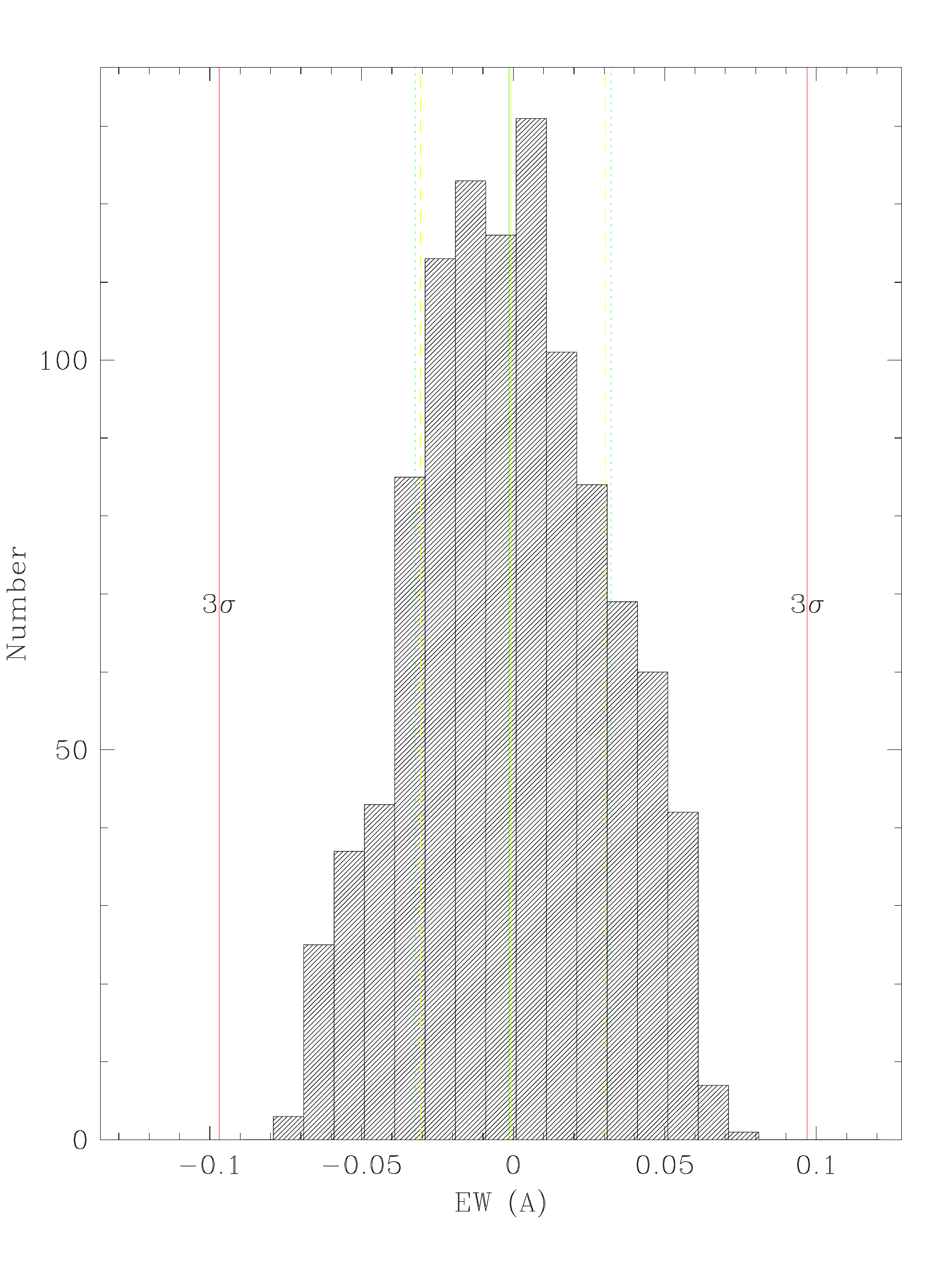}
   \caption{Distribution of all measurements of EW computed in a defined spectral interval (avoiding the telluric bands) adopting a running window of fixed size.  The EW$_{min}$ is defined as 3 times $\sigma_{EW}$ where  $\sigma_{EW}$ is the standard deviation of the distribution (see text for more details).}
   \label{figSP_app}
\end{figure}

\begin{figure}[htbp]  
   \includegraphics[width=0.5\textwidth]{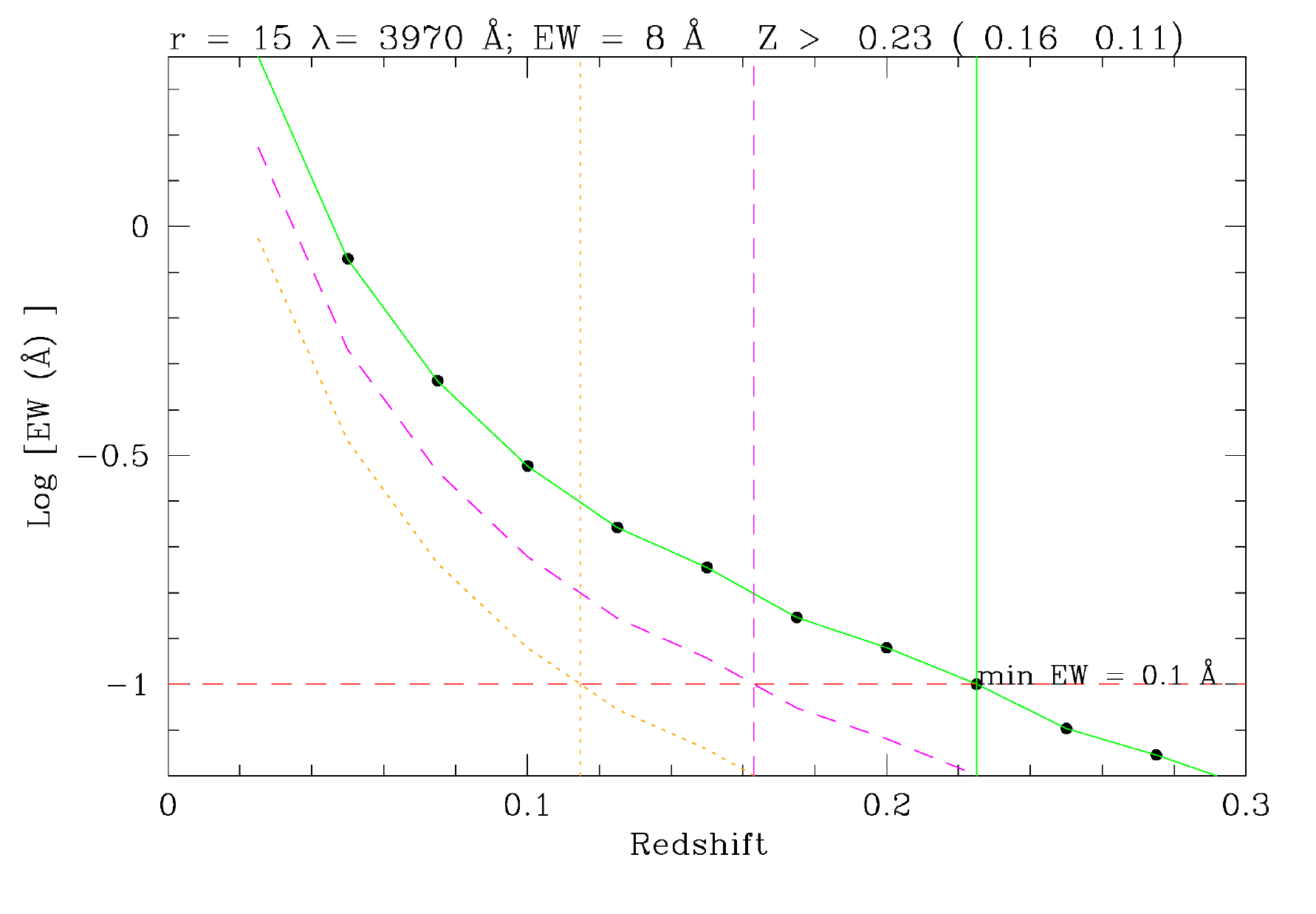}
   \caption{ The relation between the EW of Ca II absorption feature (assuming EW~$=$~8 \AA) and the redshift for the spectrum of a BL Lac object of magnitude r~$=$~15. The relation assumes that the host galaxy has M(R)= -22.9 (green solid line and filled circles).
   The other two similar relations are for M(R)~$=$~-22.4 (magenta dashed lines) and M(R)~$=$~-21.9 (orange dotted line).
   The dashed horizontal line gives the assumed EW$_{min}$ level (0.1 \AA) and the vertical lines represent the intersection with the above relationships with EW$_{min}$ level. In this case the redshift is : z~$>$~0.23 (for the average host galaxy; other values in parenthesis)   }
   \label{fig3_app}
\end{figure}

\begin{figure*}[htbp]  
   \includegraphics[width=1.0\textwidth]{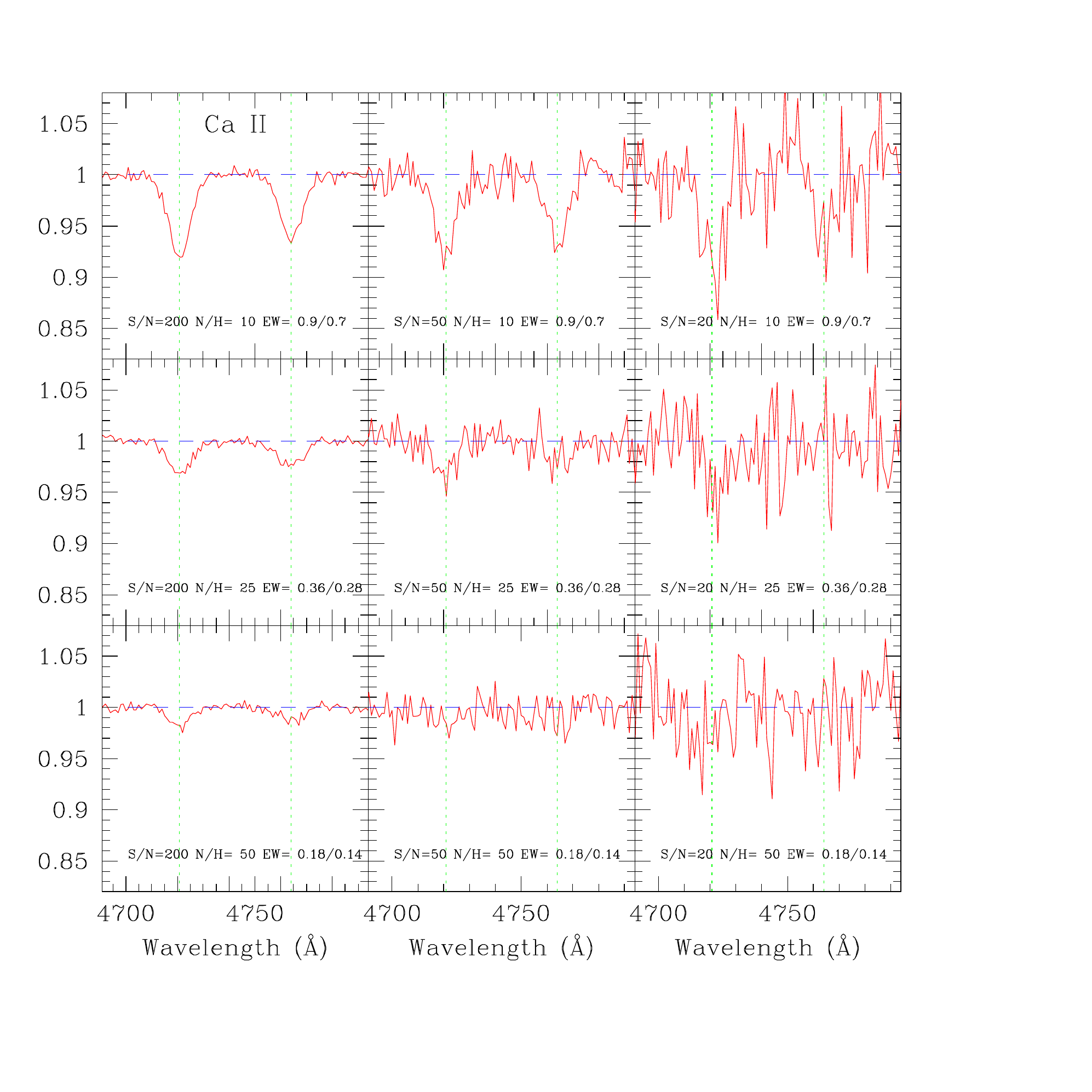}
   \caption{ Simulation of the normalized optical spectrum of a BL Lac object at z~$=$~0.2  in the region of H, K Ca II lines. 
   The simulation assume three different nucleus to starlight flux  ratio  at the observed wavelength of Ca II lines (from top to bottom) and 
   three levels of SNR (from left to right). 
   The two dotted vertical lines indicate the position of H, K features for reference. The horizontal dashed line gives the 
   normalized continuum. The simulation include the statistical noise. The figure illustrate how the detectability of CaII lines depend on N/H and SNR (see also text). 
In each panel we give the SNR of the spectrum, the nucleus to host ratio (N/H) and the minimum EW assuming standard or 1 mag fainter host galaxy luminosity.
     }
   \label{fig2_app}
\end{figure*}

From the observed magnitude of the source and the assumption that the underlying host galaxy is a giant elliptical we derive the minimum redshift of the target from the minimum detectable equivalent width (EW$_{min}$) of a specific absorption feature (see example in Fig. \ref{fig3_app}).
This depends on the SNR of the spectrum and the brightness of the object during the observations.
To estimate EW$_{min}$ we computed the nominal EW adopting a running window of fixed size (typically 15 $\textrm{\AA}$)  for a number of intervals where the SNR is approximately constant and 
 avoiding the prominent telluric absorption bands. For each interval we define EW$_{min}$~$=$~3 $\times$ $\sigma_{EW}$ 
 where $\sigma_{EW}$  is the standard deviation of the distribution of all measurements of EW (see Fig. \ref{figSP_app})
For each spectral interval a given feature (for instance Ca II absorption) is considered  detected only if the SNR is 
 sufficiently high to measure the absorption feature with EW~$>$~EW$_{min}$. 
 In Figure \ref{fig2_app}  we show an example of simulated optical spectra of BLLs in the region of Ca II absorption lines assuming different values for the N/H and SNR.

We used the EW$_{min}$ of CaII and MgI to derive lower limits to the redshift of featureless BLLs in our sample (see Tab.\ref{tab:table3} and Tab.\ref{tab:table4}). These limits range from 0.1 to 0.55 \AA.
For a number of sources other authors have derived redshift limits based on a similar approach but with somewhat different results. In particular \citet{shaw2013a} propose redshift limits for 15 objects of our sample.
For about half of them the difference of redshift limit is small and could be explained by some differences in the adopted method. \citet{shaw2013a} adopt the method proposed by \citet{plotkin2010} who perform a best fit of the observed spectrum with a host galaxy template and a power law and set a 2$\sigma$ level threshold to asses the redshift 
(compared with 3$\sigma$ assumed in our work). However, in some cases \citet{shaw2013a} claim 
redshift limits that are significantly higher than those derived in this work.
For instance in the case of RGB J0136+391 we found z~$>$~0.27 while \citet{shaw2013a}  gives z~$>$~0.88. 
The brightness level of the source was similar during the  two observations and also the SNR (estimated roughly only from the figures presented by Shaw et al.)
We believe that the latter value is implausible since at this redshift the bluest starlight feature (the CaII absorption lines) should be appearent at $\lambda>$7450$\textrm{\AA}$. 
Since the observed flux at this wavelength is $\sim$15$\times$10$^{-16}$, the contribution of the host galaxy should be one hundredth of this observed flux. This would imply the ability to detect spectral absorption with EW$\sim$0.01$\textrm{\AA}$ that  is inconsistent with the spectral resolution and SNR of the presented data.  
Another clear discrepant case is that of BZB J0915+2933 for which we give z~$>$~0.13 while Shaw et al. reports z~$>$~0.53. In this case the source was a factor $\sim$ 5 fainter, that could help to detect the starlight component, however their SNR is visually worse ($\sim$ 30-50 ) than that of our spectra ($\sim$ 100-200) therefore the detection of the host galaxy signature at z$\sim$0.5 is very unlikely.

\end{document}